# A constant potential reactor framework for electrochemical reaction simulations

Letian Chen[1], Yun Tian[2], Xu Hu[1], Suya Chen[1], Huijuan Wang[1], Xu Zhang[2]*, Zhen Zhou[1,2]*

[1] School of Materials Science and Engineering, Institute of New Energy Material Chemistry, Renewable Energy Conversion and Storage Center (ReCast), Key Laboratory of Advanced Energy Materials Chemistry (Ministry of Education), Nankai University, Tianjin 300350, China

[2] School of Chemical Engineering, Zhengzhou University, Zhengzhou 450001, Henan, China

* Corresponding author. E-mail: zzuzhangxu@zzu.edu.cn; zhouzhen@nankai.edu.cn

## Abstract

Understanding the evolution of electrified solid-liquid interfaces during electrochemical reactions is crucial. However, capturing the dynamic behavior of the interfaces with high temporal resolution and accuracy over long timescales remains a major challenge for both experimental and computational techniques. Here, we present a constant potential reactor framework that enables the simulation of electrochemical reactions with *ab initio* accuracy over extended timescales, allowing for real-time atomic scale observations for the electrified solid-liquid interface evolution. By implementing an enhanced sampling active learning protocol, we develop fast, accurate, and scalable neural network potentials that generalize across systems with varying electron counts, based on high-throughput density functional theory computations within an explicit-implicit hybrid solvent model. The simulation of reactions in realistic electrochemical environments uncovers the intrinsic mechanisms through which alkali metal cations promote $CO_2$ adsorption and suppress the hydrogen evolution reaction. These findings align with previous experimental results and clarify previously elusive observations, offering valuable computational insights. Our framework lay the groundwork for future studies exploring the dynamic interplay between interfacial structure and reactivity in electrochemical environments.

## Introduction

Despite the crucial role of the electrical double layer (EDL) in numerous electrochemical processes[1-3], the dynamic evolution of the electrified solid-liquid interface (ESLI) during reactions remains shrouded in mystery[4]. Even for the simplest hydrogen evolution reaction (HER) occurring at the platinum-water electrochemical interface, the kinetic mechanism remains a subject of ongoing debate[5-7]. Additionally, alkali metal cations, which have been found to uniquely enhance $CO_2$ reduction reactions ($CO_2$RR) beyond their inherent properties[8], add further complexity. While global efforts have advanced theoretical and experimental insights, a comprehensive understanding of the underlying mechanisms remains elusive[9]. Some studies suggested a potential link between catalytic performance and cation coverage on the surface of catalysts[10], while others proposed that cations modify the electric field distribution at the electrochemical interface, strengthening electrostatic interactions with adsorbed intermediates and thereby stabilizing and accelerating the reaction[2,11,12]. Moreover, the direct interaction of cations with intermediates[13], and their influence on the hydrogen bonding network at the interface[14,15], are also suggested as critical factors to impact the reaction. This knowledge gap stems from challenges in *in situ* probing of the ESLI and the computational restrictions of existing first-principles methods, which cannot simulate reactive interfaces over long timescale under realistic operating conditions.

Indeed, numerous advanced *in situ* techniques, including infrared absorption spectroscopy[16], Raman spectroscopy[14,17], UV-Vis spectroscopy[18], and scanning tunneling microscopy[19], have been employed to investigate ESLI during electrochemical reactions. Regrettably, the techniques available for studying ESLI at the atomic level are constrained by high cost, complexity and spatial resolution limitations, restricting their ability to cover the wide range of experimental conditions[20,21]. Remarkably, ab initio molecular dynamics (AIMD) simulations based on electronic structure calculations offer atomistic resolution images[6]. To precisely capture electrochemical interface dynamics, researchers have developed various AIMD based approaches. Chan et al.[22,23] introduced a charge extrapolation approach for correcting reaction free energies, while it overlooks microstructural evolution at the interface under electric bias and the corresponding influence on the reaction mechanisms. Leveraging the computational standard hydrogen electrode (cSHE) method developed by Cheng's group[24] and the constraint potential-hybrid solvent-dynamics (CP-HS-DM) method developed by Liu's group[25], the structural evolution at the ESLI can be observed. These methods enable precise calculations of the potential of zero charge (PZC)[24] and the differential capacity[26] for electrocatalysts, as well as the simulation of electrochemical reaction processes[27,28]. However, AIMD under constant potential conditions come with a substantial computational cost, frequently restricting simulations to only a few hundred atoms over a picosecond (ps) timescale. This limitation hinders the comprehensive exploration of the complicated evolution and reaction behavior at realistic electrochemical interfaces under operating conditions.

Fortunately, the rapid advancements in neural network potential (NNP) pave the way for achieving long-duration simulations with first-principles accuracy for large-scale systems[29,30]. Leveraging this opportunity, simulations can now capture surface reconstruction phenomena[31] and solid-liquid interfaces[32] over extended timeframes, achieving equilibria that were previously unattainable. For example, Zhu et al.[33] revealed the unusual superdiffusive rotational motion of interfacial water molecules on a Pt surface through long-timescale MD simulations based on NNP. Moreover, long-timescale MD simulations have facilitated comprehensive sampling of intricate interfacial reactions, elucidating kinetic mechanisms of important electrocatalytic processes such as oxygen reduction reaction (ORR)[34,35], HER[36], and formic acid decomposition[37] at electrochemical interfaces. While computational studies have partially explained experimental observations, these simulations remain limited to solid-liquid interfaces under zero electric bias. In practical scenarios, ESLI is often subjected to varying operating applied potentials, which can trigger distinct surface reconstructions and hydrogen bond network rearrangements, potentially altering reaction mechanisms[28,38]. Therefore, accounting for the applied electric bias is essential for a comprehensive understanding of the kinetics and mechanisms involved.

In this study, we propose a novel constant potential reactor framework (as shown in Fig. 1) that integrates a NNP capable of handling systems with varying charge states, alongside constant potential MD simulations. This framework facilitates efficient sampling of large-scale systems within the grand canonical ensemble. By combining this framework with enhanced sampling methods such as metadynamics[39], Blue Moon[40], and slow-growth[41] techniques, we can accelerate the sampling of rare events, thereby expediting the acquisition of accurate potential energy surface (PES). The precise obtained PES ensures the long-timescale simulations of ESLI with density functional theory (DFT) accuracy under realistic working conditions. Utilizing the framework, we systematically investigated the Au-water interface, revealing the impact of applied potential and cations on the hydrogen-bond network in the solution and elucidating the significant role of cations in $CO_2$RR. Our results demonstrate that cations promote $CO_2$ activation by facilitating surface charge accumulation, which stabilizes the strong dipole formed during $CO_2$ activation through an intensified electric field, maintaining the stability of the activated $CO_2$ state. Furthermore, our findings provide theoretical evidence for the effects of surface reconstruction on promoting reduction reactions and the cation mediated suppression of $H^+$ mass transfer, which align well with previous experimental results[16,42,43]. This framework enables rapid simulations of electrochemical reactions and ESLI evolution under varying applied potentials, establishing a powerful tool for advancing our understanding of electrochemical systems.

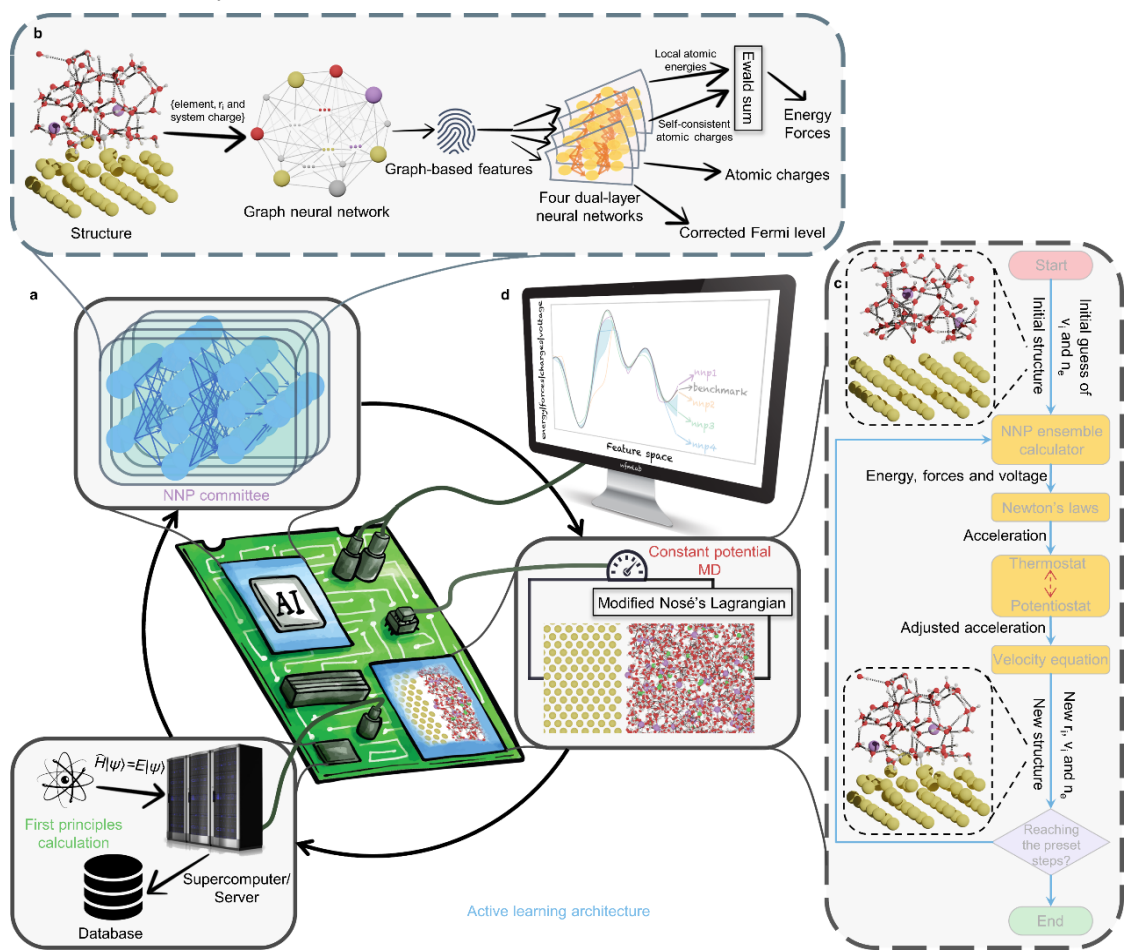

**Fig. 1. Schematic Overview of the Electrochemical Reaction Simulation within a Constant Potential Reactor. a**, Architecture of active learning (AL). The automated iterative AL cycle expedites crafting a NNP attuned to variant system charges. The training dataset, assembled by the AL process, selectively encompasses structures exhibiting high-uncertainty in characteristics (atomic energies, forces, atomic charges, and corrected Fermi level), pinpointed

with a purpose-built ensemble calculator formed by a committee of NNPs established herein. **b**, Generalized NNP framework. The framework prognosticates energies, forces, atomic charges, and corrected Fermi level by extracting elemental, positional, and systemic charge features as graph-based features via a Graph Neural Network. **c**, MD program operating under constant electrode potential. The program utilizes a modified Nosé's Lagrangian and adhering to Newton law of motion, as detailed in the section of Methods. **d**, A schematic of identifying highly uncertain structures during structure evolution. The graphical representation uses purple, yellow, green, and blue curves to denote the predictions of target properties in the structural space by four different NNP models. These models are uniformly trained with identical hyperparameters but distinct random seeds within the NNP committee. The blue shaded region on the graph signifies the standard deviation, illustrating the variance in the predictions from the four NNPs and serving as a quantifier for the uncertainty linked to the target properties of the evolved structures. The gray curve represents the actual target values. The notation $r_i$ represents the position of the *i*th atom, $v_i$ indicates the velocity of the *i*th atom, and $n_e$ denotes the electron count in the system.

## Results

**Workflow of the constant potential reactor.** Fig. 1 illustrates the workflow of our constant potential reactor algorithm, designed for efficient and accurate simulation of electrochemical reactions. The algorithm iteratively refines a machine-learned PES, guided by uncertainty quantification and targeted DFT calculations. The process begins with training an ensemble of *n* variable electronic neural network potentials (veNNPs) on an initial dataset. Unlike traditional NNPs, veNNPs incorporate structural fingerprints derived from graph neural networks and feature multiple output layers, enabling simultaneous predictions across multiple target properties (Fig. 1b), which we refer to as multi-objective learning. The ensemble of veNNPs provides multiple predictions per configuration, enabling the estimation of model uncertainty based on the prediction standard deviations (Fig. 1d). Crucially, the veNNPs are designed to explicitly account for system charge variations, leveraging neural network fitting capability to capture the interplay between system charge, structure, and applied potential accurately. During MD simulations, this charge-structure-potential relationship is utilized through a modified Nosé-Hoover Lagrangian formalism[44], which dynamically adjusts the system charge to maintain a constant applied potential (Fig. 1c, see Methods for derivation and implementation details). Enhanced sampling techniques are integrated within the MD simulations to expedite PES exploration and ensure comprehensive sampling of significant configurations. Regions of high uncertainty, as identified by the veNNP ensemble, are flagged for targeted DFT calculations. These high-accuracy DFT data points are then incorporated into the training set, and the veNNP ensemble is retrained. It is worth noting that the system applied potential determination in this process is based on an explicit-implicit solvent model, as shown in Supplementary Fig. 1. For more details, please refer to the Methods section. The iterative process of uncertainty-guided refinement progressively expands the domain of applicability of the NNP towards the target electrochemical reaction. Once sufficient coverage of the relevant configuration space is achieved, as assessed by convergence in the predicted properties and reduced uncertainty, a single, highly accurate veNNP is trained. The refined veNNP, coupled with the modified Nosé-Hoover MD and enhanced sampling, provides an efficient and reliable framework for simulating the target electrochemical reaction under applied potentials.

**Evaluating multi-objective learning performance.** To assess the performance of the multi-objective learning-based veNNP, we selected the Au-water interface system as a case study. After generating a representative dataset of configurations through an active learning framework and ensuring its convergence, we divided the data into training, validation, and test sets with an 8:1:1 ratio. A veNNP model was trained on this dataset, and its predictions of energy, atomic forces, Bader charges[45], and applied potential were compared against DFT calculations (Fig. 2). The model demonstrates remarkable accuracy for energy, forces, and Fermi level, with root mean square errors (RMSEs) of 0.7 meV/atom, 14 meV/Å, and 4 meV, respectively, indicating that simulations conducted with the veNNP retain DFT-level precision. Bader charges exhibit a comparatively higher RMSE of 0.05 |e|, with the discrepancies primarily appearing within the charge ranges of -2 to -1.4 |e| and 0.7 to 1 |e| (Fig. 2c), corresponding to oxygen and hydrogen atoms in water molecules. This observation is further corroborated by a separate comparison of Bader charges for water molecules (Supplementary Fig. 2a). Notably, the calculated charge values for hydrogen, ranging from approximately 0.7 |e| to -0.1 |e|, include contributions from hydrogen atoms involved in the Volmer step, demonstrating the accuracy of our model in predicting the charge of hydrogen during the Volmer process, although the accuracy diminishes for hydrogen within bulk water. The veNNP model maintains solid performance for other atomic species in the system (Supplementary Fig. 2b). We attribute the discrepancies in Bader charges to inherent limitations in DFT and Bader charge analysis[45], as the partial charges of hydrogen and oxygen in water are expected to fluctuate within a narrower range than observed in the DFT calculations. However, since the overall charge of water molecules remains close to neutral, we believe that this discrepancy has a negligible impact on our subsequent analyses.

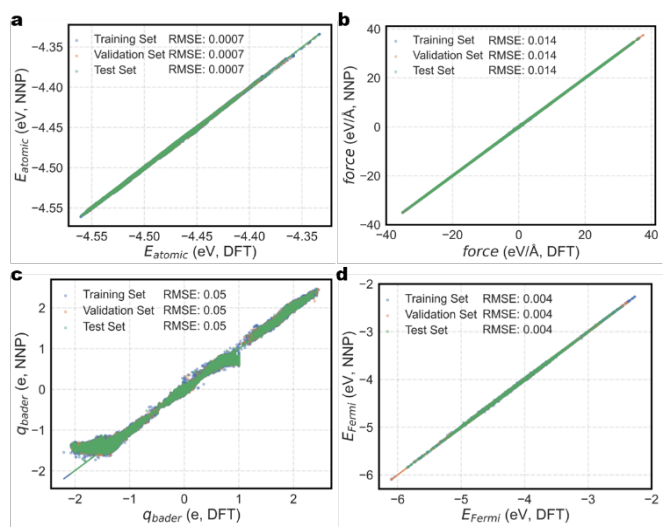

**Fig. 2. Performance comparison between veNNP and DFT for multi-objective learning.** The plots compare the accuracy of two methods for **a**, energy per atom, **b**, forces, **c**, atomic charges, and **d**, Fermi level.

**Nanosecond-scale constant potential MD simulations.** With the veNNP achieving high accuracy, we apply it to conduct nanosecond (ns)-scale MD simulations on eight distinct Au-water interfacial systems to explore the effects of surface structure, $CO_2$ adsorption and $K^+$ ion on electrocatalytic reaction processes. Specifically, we simulate both Au(110) and Au(111) surfaces under four specific conditions: (1) pure water, (2) water with dissolved $CO_2$, (3) water containing 2 $K^+$ ions, and (4) water with both dissolved $CO_2$ and 2 $K^+$ ions. The simulations are executed under both constant potential and constant charge conditions. Constant potential molecular dynamics (CPMD) simulations are conducted at -0.2 V, 0 V, -0.2 V, -0.4 V, and -0.6 V versus the standard hydrogen electrode (SHE).

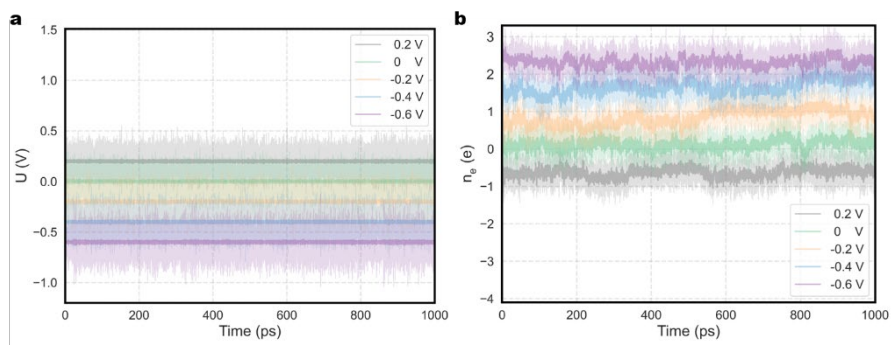

**Fig. 3. Applied potential and net electron count of Au(110)-water interface with 2 $K^+$ ions and a $CO_2$ molecule. a**, The applied potential and **b**, the net electron count of Au(110)-water interface with 2 $K^+$ ions and a $CO_2$ molecule as functions of simulation time. All results were obtained under a 1 ns constant potential (0.2, 0, -0.2, -0.4, -0.6 V vs SHE) MD simulation. The light-colored line represents the results based on each frame, while the dark-colored line is the average obtained with a 200 fs moving window.

Supplementary Figs. 3-18 illustrate the energy and temperature fluctuations of eight systems during ns scale MD simulations. The system temperature remains well controlled around the target value of 300 K. Energy fluctuations are more pronounced in CPMD compared with constant-charge MD, even when averaged over a 200 fs moving window. This phenomenon likely stems from variations in total charge of the system to maintain a constant potential, which directly impact the system energy, leading to larger energy fluctuations in CPMD. Fig. 3 illustrates the applied potential and net electron count fluctuations in an Au(110)-water interface system containing $CO_2$ molecules and 2 $K^+$ ions. The net electron count is defined as the difference between the total electron number and total nuclear charge. Corresponding data for other systems are presented in Supplementary Figs. 19 and 20. The potentiostat control maintains the applied potential near the target range within ±0.3 V fluctuation vs SHE. Furthermore, the net electron number generally increases as the target applied potential decreases, consistent with a direct relationship between applied potential and system charge. This demonstrates the effectiveness of our method in regulating the applied potential of the system during MD simulations, enabling electrochemical reactions to be simulated at the desired potential. The relationship between net electron number and energy fluctuation over time reveals that these quantities reach a relatively stable state within the first few to tens of ps, suggesting that the system equilibrates into a set of relatively stable and metastable configurations. Importantly, this stability is relative, rather than oscillating around a fixed value, the parameters continue to fluctuate dynamically, maintaining a realistic model of the system's electrochemical environment.

Similar to CPMD, the energy and applied potential in constant charge MD reach a relatively stable state within the first few to tens of ps (Supplementary Figs. 21 and 22). However, the applied potential fluctuates within a range instead of converging to a single representative value. Therefore, averaging the applied potential over the simulation trajectory under the canonical ensemble is not a reliable approach for characterizing electrochemical potential of the system. Nonetheless, the prevailing computational method for determining the PZC at solid-liquid interfaces still relies on averaging the potential under the canonical ensemble. This approach generally works well for metal-pure water interfaces, as the canonical ensemble ensures zero net charge on the metal electrode. While averaging over different simulation time intervals might yield slightly different PZC values, the potential fluctuations are typically small, leading to acceptable errors. In this study, the calculated PZCs for Au(111) and Au(110) are approximately 0.2 V and 0 V, respectively. Although these values deviate slightly from experimental measurements, the discrepancy could be due to using 4.44 V as the computational reference for the SHE potential ($U_{SHE}$). The accepted $U_{SHE}$ value ranges between 4.2 and 4.7 V[46], which introduces uncertainty in the precise computational alignment of calculated PZCs. More accurate computational determination of U and PZC requires methods such as the cSHE approach developed by Cheng et al.[24], or fitting a computational reference $U_{SHE}$ value by comparing calculated work functions of multiple surfaces with experimental PZCs[28]. For instance, Yu et al. obtained a fitted $U_{SHE}$ value of 4.2 V[28], which includes Au(110) and Au(111) surfaces. Using this reference, our calculated PZCs are much closer to experimental values[24,47]. Notably, the difference between our calculated PZCs for Au(111) and Au(110) agrees well with experimental measurements[47], supporting the validity of our approach for calculating applied potential.

**Structure of interface water.** Following ns scale MD simulations, we analyzed the structural arrangement of water at the Au-water interface. Supplementary Figs. 23 and 24 show the water density profiles along the surface normal (z-axis) at various applied potentials (PZC, 0.2, 0, -0.2, -0.4, and -0.6 V vs. SHE). Here, the PZC refers to the simulations performed under canonical ensemble conditions with a constant net charge. Unlike previous findings by Cheng et al.[48], which showed an increase in the first peak intensity at increasingly negative potentials, our results do not reveal a clear trend correlating density peak heights with applied potential This discrepancy likely arises from their use of a fully explicit ESLI model, where explicit ions exchange charge with the electrode, thus altering the electrode potential and creating a coupling between ion concentration and applied potential. Our implicit/explicit model decouples this relationship. Nevertheless, the applied potential does influence the density profiles. In systems without cations, the position of the first peak of the water density profile exhibits a dependence on the electrode potential. For both Au(110) and Au(111), with and without $CO_2$, this first peak shifts away from the Au surface with decreasing potential. This shift likely arises from the increasing negative charge on the Au surface at lower potentials, attracting the positively charged hydrogen atoms of the interfacial water molecules. As a result, the water molecules reorient with their hydrogen atoms facing the surface, and the oxygen atoms displace outward. Since our density profiles are based on the oxygen atom positions, this reorientation manifests as a shift of the first peak away from the surface as the potential decreases. Furthermore, the addition of cations significantly enhances the intensity of the first peak in the density profiles for both Au(110) and Au(111) surfaces, indicating a strong correlation between cation presence and peak intensity, while eliminating the clear dependence of the first peak position on applied potential. It likely arises from cations coordinating with the interfacial water molecules, causing oxygen atoms to orient towards the cations and the hydrogen atoms towards the Au surface. It also suggests that the conclusion reached by Cheng and colleagues[48] may be due to the influence of cations rather than applied potentials. This cation-water interaction diminishes the influence of the applied potential on the orientation of the first water layer, effectively decoupling the first peak position from applied potential. To further support this observation, we analyzed the radial distribution functions (RDFs) of interfacial water at different potentials, including H-H, O-H, and O-O pairs (Supplementary Figs. 25 and 26). For pure water interface, the applied potential significantly influences the RDFs. However, the addition of 2 $K^+$ ions markedly diminishes this influence, corroborating our previous analysis.

Furthermore, we analyzed the probability distributions of the interfacial water dipole orientation ($\varphi$) and the angle ($\theta$) between the O-H bond and the surface normal at different applied potentials. The insets of Fig. 4a and 4b illustrate the definitions of $\varphi$ and $\theta$, respectively, while Supplementary Fig. 27a and 27b depict the corresponding configurations at the Au(110)-water interface. As shown in Fig. 4a and 4b, at the PZC and in the absence of cations, interfacial water dipoles predominantly orient between 50° and 120° relative to the surface normal, while $\theta$ peaks around 90°. This indicates that the plane of the water molecule lies predominantly parallel or at a small angle to the Au(110) surface, consistent with the previous study[17]. As the potential increases, the positively charged surface attracts the oxygen atoms of the water molecules, causing them to reorient towards the surface, while the hydrogen atoms are repelled. Consequently, both $\varphi$ and $\theta$ exhibit higher probability densities at smaller angles. Conversely, as the potential decreases, the probability density peaks shift towards larger angles. At -0.6 V, a pronounced peak in the $\varphi$ distribution appears between 125° and 130°, indicating that the water dipoles are largely oriented towards the surface. These trends are consistent with those observed in earlier studies with fully explicit models[17,48]. Upon the introduction of $K^+$ ions, a peak emerges in the $\varphi$ distribution above 120° for both Au(110) and Au(111) (Supplementary Figs. 28 and 29). This suggests that the coordination of water molecules with cations, combined with the strong electric field near the surface, significantly diminishes the influence of the applied potential on interfacial water structure, further supporting our previous conclusions.

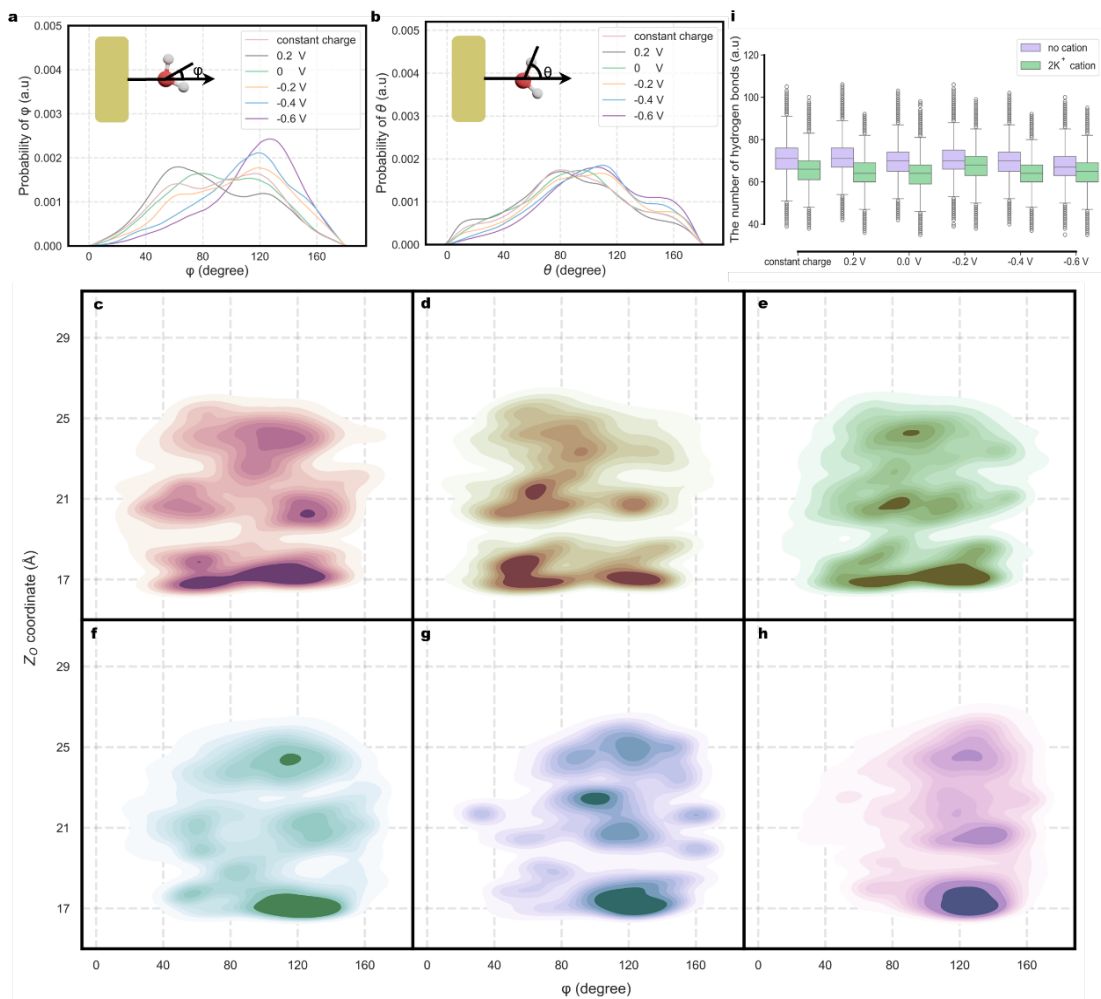

**Fig. 4. Structure of Water at the Au(110)-Water Interface.** Probability distribution profiles for **a**, angle φ and **b**, angle θ of the interfacial water at the Au(110)-water interface. Angle φ represents the angle between the bisector of the water molecule and the surface normal, while angle θ indicates the angle between the O-H bond of the water molecule and the surface normal. Both angles are illustrated in the inset, with a detailed schematic of the interfacial structure provided in Supplementary Fig. 27a and 27b. **c-h**, Probability distribution profiles for angle φ and the Z-coordinate of the corresponding oxygen atom in the water molecule at the Au(110)-water interface were obtained from 1 ns MD simulations under different conditions (**c**, constant charge, **d**, 0.2 V vs SHE, **e**, 0 V vs SHE, **f**, -0.2 V vs SHE, **g**, -0.4 V vs SHE, and **h**, -0.6 V vs SHE). **i**, Statistics of the number of hydrogen bonds in interfacial water at the Au-water interface. The hydrogen bond network was compared under different simulation conditions (constant charge, 0.2 V vs SHE, 0 V vs SHE, -0.2 V vs SHE, -0.4 V vs SHE, -0.6 V vs SHE) and different solution environments (with and without $K^+$ ions). In the box plot, the green and purple colors represent the differences between solutions with and without $K^+$ ions, respectively.

To elucidate the influence of applied potential and cations on different water layers, we analyzed the probability distributions of φ and θ for water molecules at various distances from the surface (Fig. 4c-h and Supplementary Figs. 30-44). The influence of the applied potential on water structure clearly diminishes with increasing distance from the surface, with the first water layer exhibiting the most pronounced response. Across different applied potentials, a decrease generally shifts both φ and θ distributions towards higher angles, at least up to a distance of approximately 10 Å from the interface. With the addition of cations, the distributions of both φ and θ for the first water layer become virtually identical across all applied potentials, indicating a diminished response to potential in this layer. In contrast, the diminished influence of the cations on subsequent water layers allows these layers to exhibit a dependence on the applied potential. Notably, the influence of the applied potential on water structure appears to initially increase with distance from the surface before decreasing further out.

We further examined the angle ψ at different applied potentials, defined as the angle between the bisector of the water molecule and the vector connecting the water oxygen atom to the $K^+$ ion (illustrated in Supplementary Fig. 27c). This analysis considered the differences among water layers in their coordination with $K^+$ ions (Supplementary Fig. 45-48). Across the four systems containing $K^+$ ions, the ψ distributions are remarkably consistent. In the first water layer, located approximately 3 Å from the $K^+$ ion, the water molecules are strongly bound to the ion, with ψ predominantly distributed between 100° and 150°. This indicates that the oxygen atoms orient towards the central $K^+$ ion, while the water dipoles point outwards. This preferential orientation weakens with increased distance from $K^+$ ion. Notably, the first water layer, due to its strong interaction with $K^+$ ion, remains largely unaffected by the applied potential. In contrast, the more distant water layers exhibit a clear potential dependence, with ψ decreasing as the potential decreases. Supplementary Fig. 49 shows the vertical positioning of $K^+$ ion during the ns scale MD simulations, with the z-coordinate in the cell model corresponding to the same spatial dimension as depicted in Supplementary Fig. 1. The $K^+$ ion tends to reside near the surface, with its height decreasing as the applied potential becomes more negative. It is reasonable, as more negative potentials likely facilitate the partial dehydration of the cation and its specific adsorption onto the surface[17,26,48]. The decrease in ψ with decreasing potential indicates that the water dipoles increasingly point towards the $K^+$ ion. Given the height of the $K^+$ ion, these potential-sensitive water molecules are primarily located above the cation. Therefore, the decrease in ψ with decreasing potential is due to the progressive reorientation of these water molecules, aligning their dipoles towards the Au surface as the potential becomes more negative.

Our results clearly demonstrate the influence of applied potential and cations on the interfacial water structure. Both the applied potential and the presence of cations encourage the reorientation of interfacial water dipoles towards the Au surface. Furthermore, the strong coordination of water molecules within the first solvation shell of the cations significantly disrupts the hydrogen-bonding network of the interfacial water, as illustrated in Fig. 4i. At the Au(110)-water interface, the number of hydrogen bonds generally decreases with decreasing the potential. The introduction of $K^+$ ions further reduces hydrogen bonds, indicating that both the applied potential and cations, particularly the latter, disrupt the connectivity of the interfacial water network. This trend, also evident across other systems (Supplementary Figs. 50 and 51), aligns well with previous experimental findings[6]. This disruption of the hydrogen-bond network can hinder proton transfer, thereby inhibiting the HER[42]. Additionally, the number of hydrogen bonds forming between the oxygen atoms of $CO_2$ and interfacial water molecules was also examined and is shown in Supplementary Fig. 52. In the pure water interface at -0.6 V without cations, the oxygen atoms of $CO_2$ participate in almost no hydrogen bonding, likely due to the expanded water cavity surrounding $CO_2$ under more negative potentials. Interestingly, at the Au(110)-water interface, the oxygen atoms of $CO_2$ form hydrogen bonds with multiple interfacial water hydrogen atoms at -0.4 V and -0.6 V, which is attributed to the dynamic adsorption of $CO_2$ observed in our ns scale MD simulations at these potentials (Supplementary Videos 1 and 2). The activated $CO_2$ attracts hydrogen atoms from nearby water molecules, leading to an increased number of hydrogen bonds. These strong hydrogen bonds stabilize the adsorbed $CO_2$, consistent with our previous work[49].

**Atomic charge analysis.** Fig. 5a shows the atomic charge distribution for the Au(110)-water interface with a $CO_2$ molecule and 2 $K^+$ ions, obtained from a ns scale MD trajectory. The probability distributions of atomic charges are compared at different potentials. It reveals that as the potential becomes more negative, the added electrons primarily accumulate on the Au surface. The O, H, and K elements in the solution experience a negligible impact. This behavior is also observed in other systems, as shown in Supplementary Fig. 53 and 54. This validates our approach of modulating the system charge to control the applied potential, effectively mimicking the experimental scenario where electrons are supplied to the Au cathode during reduction reactions. The increasingly negative charge on the Au surface at lower potentials is also physically sound. The Au-water interface represents a conductor-insulator interface, with the interfacial water layer possessing a substantial band gap. The Au conduction band lies above the valence band of the water layer but well below its conduction band. Consequently, variations in the system electron population primarily affect the Au electrode, further supporting the reliability of our model. By comparing the atomic charge distributions of systems with and without $CO_2$, we identify two distinct peaks associated with oxygen atoms, with the smaller peak attributed to the oxygen atoms of $CO_2$. Notably, in the presence of both $CO_2$ and 2 $K^+$ ions at the Au(110)-water interface, additional peaks emerge in the oxygen and carbon charge distributions specifically at -0.4 V and -0.6 V, corresponding to $CO_2$ adsorption onto Au(110) surface at these potentials, as shown in Fig. 5a. These new peaks, alongside shifts in carbon and oxygen charges, reveal electron transfer from the Au surface to the $CO_2$ molecule indicating its activation. $CO_2$ adsorption was not observed in the MD simulations performed under constant charge conditions or at higher potentials. In contrast, $CO_2$ adsorption occurred spontaneously at -0.4 V and -0.6 V[2], further validating the reliability of our constant potential reactor model.

Since the changes in the applied potential primarily affect the charge distribution on the Au catalyst, we analyzed the distribution of Bader charges on the Au atoms at the interface for the four systems without $CO_2$ at -0.6 V (Fig. 5b and 5c, and Supplementary Fig. 55). The results illustrate the charge distribution on both Au(110) and Au(111) surfaces, clearly demonstrating a preferential accumulation of electrons at protruding surface sites. The protruding sites refers to locations that exhibit significant protrusions from the original surface during the simulation. This surface reconstruction which is often overlooked in previous studies[50-52] can significantly impact electrocatalytic processes. Traditional AIMD simulations often lack the necessary timescale to adequately sample these metastable reconstructed states. These metastable states introduce a diversity of adsorption sites[53], highlighting the importance of surface reconstruction in electrocatalysis. Our observed Au(110) surface reconstruction is consistent with previous experimental observations[54] and computational results[55]. While the Au(110) simulations show pronounced reconstruction, the Au(111) surface retains its overall planarity despite some structural rearrangements. This disparity may be due to the limitations in initial structure model size and simulation time. Nevertheless, even minor protrusions on Au(111) surface exhibit enhanced electron accumulation. Furthermore, as shown in Supplementary Videos 3-10, more negative potentials tend to promote surface reconstruction, a trend also supported by experiments[21]. Importantly, this preferential accumulation of electrons at protruding sites is not an artifact of individual snapshots from the MD trajectory, but a general trend observed throughout the ns scale MD simulations (Supplementary Videos 15-18). While a full investigation into the reconstruction phenomenon is beyond the scope of this work, further research could utilize structural descriptors and unsupervised learning methods to compare the different reconstructed surface structures[56]. However, this area warrants further study. This underscores the need for constant potential reactor models capable of both long-timescale and large-scale simulations of ESLI. Further analysis of the Au atomic charge distribution on the Au(110) and Au(111) surfaces with and without cations reveals that the presence of nearby cations enhances this localized accumulation of electrons at the interface.

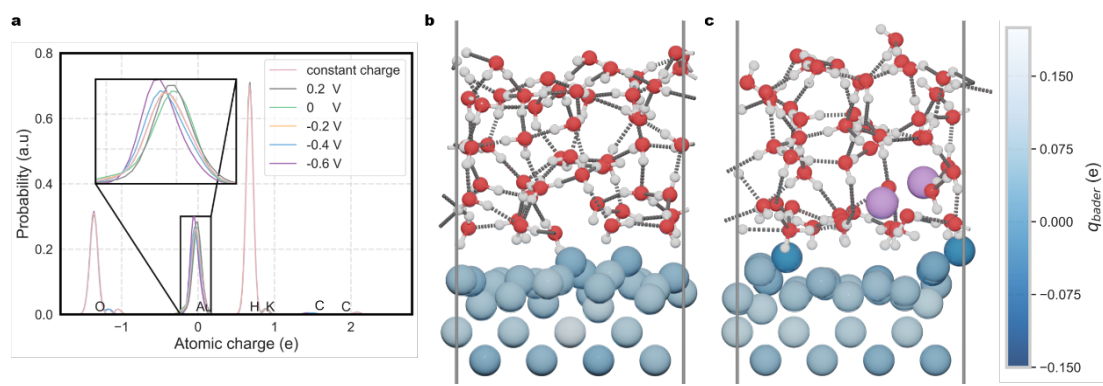

**Fig. 5. Atomic charge analysis. a**, Probability density distribution of atomic charges in the Au(110)-water interface with a $CO_2$ molecule and 2 $K^+$ ions. Atomic charges were calculated with a neural network potential (NNP) trained on Bader charges. **b**, Snapshot of the Au(110)-water interface at -0.6 V vs SHE. **c**, Snapshot of the Au(110)-water interface with 2 $K^+$ ions at -0.6 V vs SHE. The color of the Au atoms in the Au(110) surface corresponds to the Bader charge value, as indicated by the color bar. Light purple, white, and red spheres represent K, H, and O atoms, respectively.

**Volmer Step and $CO_2$ Adsorption.** To obtain reliable free energies for Volmer step and $CO_2$ adsorption, five independent slow-growth simulations are performed for each reaction (see Supplementary Note 1 for further discussion). To maintain independence between these simulations, a 20 ps MD equilibration with a fixed initial value of the collective variable (CV) preceded each slow-growth run, analogous to a single step of the blue moon ensemble method[40]. In the Volmer step, the reaction occurring at the electrode surface is expressed as $H_2O + e^- + *$ → $H* + OH^-$, where * represents an active site in the electrode surface. The corresponding reaction coordinate is defined as the difference between the distance from the reacting hydrogen atom to its oxygen atom ($d_{O-H}$) and the distance from the hydrogen atom to the target Au site ($d_{H-Au}$), *i.e.*, $d_{O-H} - d_{H-Au}$. As the simulation progresses, the reaction coordinate increases. This definition is consistent with the CV commonly used in slow-growth simulations of the Volmer step[49]. The choice of reactive Au sites on the Au(110) and Au(111) surfaces is discussed in Supplementary Note 1.

Supplementary Fig. 73 shows the free energy profiles for the Volmer step on Au(110) at both flat and protruding sites, highlighting how potential and surface structure affect reaction barriers, both with and without cations. Generally, the Volmer free energy barrier decreases with decreasing potential. However, some deviations from this trend are observed. For instance, in Supplementary Fig. 73a-c, the barrier at -0.6 V is higher than that at less negative potentials, which can be attributed to variations in reactive site geometries due to reconstructed surface structures that emerge under different potentials in ns scale MD simulations. Besides, while the slow-growth method provides a qualitative assessment of reaction favorability, it does not necessarily represent the minimum energy pathway[57,58]. This effect is less pronounced on the flatter Au(111) surface (Supplementary Fig. 84). Comparing Au(110) flat and protruding sites clearly shows that the Volmer step is more favorable at protruding surface sites in the absence of cations, with barriers 0.1-0.25 eV lower than those at flat sites at the same potential. However, the presence of 2 $K^+$ ions significantly diminishes this promoting effect of the protruding sites. The introduction of cations also generally facilitates the Volmer step, lowering the barriers by 0.1-0.4 eV at the same potential, which agrees with the results of Bender et al[59]. The same trend regarding the influence of $K^+$ ions is observed on the Au(111) surface.

We further analyzed the charge transfer between the reacting hydrogen atom and the Au surface during the Volmer step, as shown in Supplementary Figs. 74 and 85. The flattening or plateauing of the electron transfer and free energy profiles near the end of the reaction coordinate indicates that the chosen range of the reaction coordinate is appropriate. A consistent trend is observed across all systems, regardless of the Au surface (Au(110) or Au(111)), the presence of $K^+$ ions, the type of adsorption site, or the applied potential. As the hydrogen atom approaches the Au surface, it gradually gains electrons, becoming reduced, while the Au surface becomes more positively charged. Notably, under constant charge conditions, the Au surface loses more electrons, while the hydrogen atom does not gain proportionally more. Analysis of the change in the system net charge (Supplementary Figs. 75 and 86) reveals that the net charge increases as the reaction progresses under constant potential conditions. It indicates that the Au surface draws electrons from the potentiostate to maintain the applied potential, driving the reaction process, analogous to experimental electrocatalytic reduction processes. Supplementary Figs. 76, 77, 87, and 88 show the potential profiles for these systems during the reaction, further validating the ability of our constant potential reactor to maintain a fixed potential, consistent with experimental conditions. In contrast, the system potential changes significantly under constant charge conditions. Therefore, accurate electrocatalysis simulations must break from the traditional constant charge paradigm.

Several anomalous trajectories are identified during the Volmer step simulations, specifically for the Au(110) flat surface site at -0.6 V in pure water (Supplementary Fig. 74a) and the Au(111) surface at -0.4 V in pure water (Supplementary Fig. 85a). In these trajectories, an unexpected increase in the hydrogen charge can be observed. DFT calculations performed on structures extracted from these anomalous trajectories near the point of divergence confirmed that these deviations were not due to inaccuracies in our model. Further analysis of the anomalous trajectories, frame-by-frame, reveals the underlying cause. In the case of Au(110) surface at -0.6 V (Supplementary Fig. 97a), the initially adsorbed hydrogen atom is unstable (as also reflected in the free energy profile, shown in Supplementary Fig. 73a) and can be abstracted by a water molecule from the solution. The hydrogen atom originally belonging to the water molecule is subsequently adsorbed onto the other site of the surface, effectively resulting in an exchange of adsorbed hydrogen atoms. For the Au(111) surface at -0.4 V, the adsorbed hydrogen atom simply returns to the solution due to its instability (Supplementary Fig. 84b). Additionally, in the simulation of the Volmer step on the Au(110) flat surface site at -0.6 V with 2 K$^+$ ions, the adsorbed hydrogen atom can migrate to a nearby protruding surface site (Supplementary Fig. 96), which offers greater stability. It highlights a limitation of the slow-growth method, which is that the prescribed reaction coordinate may not always correspond to the actual reaction pathway. Importantly, these anomalous events occur after the initial hydrogen adsorption event, so the calculated reaction barriers remain reliable.

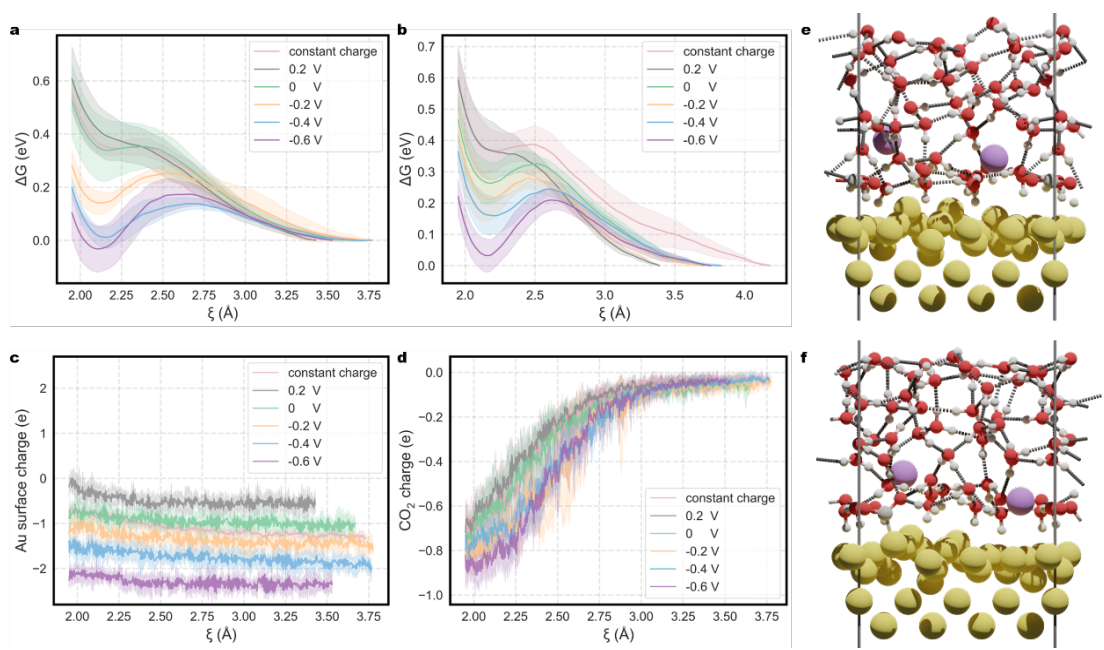

**Fig. 6. CO$_2$ adsorption.** Free energy profiles for CO$_2$ adsorption at **a**, a protruding surface site at Au(110)-water and **b**, an Au site at the Au(111)-water interfaces in the presence of 2 K$^+$ ions at various applied potentials (0.2, 0, -0.2, -0.4, and -0.6 V vs. SHE). **c**, Change in total charge on the Au(110) surface for Au(110)-water interface with 2 K$^+$ ions during CO$_2$ adsorption at a protruding surface site under different potentials. **d**, Change in the charge of CO$_2$ for Au(110)-water interface with 2 K$^+$ ions during adsorption at a protruding surface site under different potentials. Snapshots from a representative slow-growth trajectory at -0.6 V vs. SHE, showing **e**, the structure before CO$_2$ adsorption and **f**, the structure after CO$_2$ adsorption. Light purple, light gray, yellow, white, and red spheres represent K, C, Au, H, and O atoms, respectively.

For CO$_2$ activation, the reaction occurring at the electrode surface is defined as CO$_2$ + $\delta$e$^-$ + * → CO$_2^{\delta-}$*, where * represents an active site in the electrode surface. The corresponding reaction coordinate is defined as the distance ($d_{C-Au}$) between the carbon atom of CO$_2$ and the adsorbing Au site. As the simulation progresses, the reaction coordinate decreases. The determination of the initial position for CO$_2$ chemisorption is discussed in Supplementary Note 1. To validate the effectiveness of our chosen initial position for CO$_2$ chemisorption, we extracted the potential of mean force (PMF) values from a 20 ps MD simulation conducted before the reaction began.

The corresponding probability density distribution is shown in Supplementary Fig. 95. The peak values are centered around zero, and the calculated average value is also approximately zero. This indicates that our selected initial position does not contribute to the free energy of the reaction, confirming its suitability as the starting point for the reaction.

Fig. 6e and 6f illustrate the adsorption of $CO_2$ onto the flat surface site of the Au(110) surface in the presence of $K^+$ ions in the solution. Supplementary Figs. 78a and 89 illustrate the free energy changes during $CO_2$ activation at the flat surface sites of Au(110) and Au(111) in a pure water interface, respectively. Both surfaces exhibit inertness towards $CO_2$ activation, as they are unable to stably adsorb $CO_2$ regardless of the applied potential, consistent with previous experimental findings[2]. In contrast to the flat surface site, the protruding surface site of Au(110) shows an enhanced ability to activate $CO_2$ as the potential is gradually lowered, reaching -0.6 V (Supplementary Fig. 78b). This finding can explain the conflicting conclusions reached by Hicks et al.[42] and Monteiro et al.[2] in their experiments. It is possible that the catalyst surfaces tested by Hicks et al.[42] underwent significant surface reconstruction, allowing them to observe $CO_2$ reduction even in the absence of cations in the solution. This further underscores the importance of constant-potential reaction simulations, as constant-charge simulations would fail to capture the stabilization of $CO_2$ adsorption at the protruding surface site.

Our computational results align with experimental observations that cations can significantly enhance $CO_2$ reduction performance[2]. Supplementary Fig. 78c, along with Fig. 6a and 6b, demonstrates that the presence of cations significantly improves the $CO_2$ activation capability of both Au(110) and Au(111) surfaces, thereby enhancing the overall performance of $CO_2RR$. Since $CO_2$ activation is generally considered the rate-limiting step in $CO_2RR$[60], our findings are consistent with previous computational conclusions[50-52].

The electron transfer between $CO_2$ and the Au surface during the $CO_2$ adsorption process is analyzed based on the reaction coordinates. Fig. 6c and 6d illustrate the changes in the charge distribution of the Au surface and the total charge of $CO_2$ during its adsorption onto a protruding surface site in the Au(110) surface with the presence of $K^+$ ions, respectively. As $CO_2$ adsorbs, it acquires electrons from the Au surface, gradually becoming activated to form $CO_2^{\delta-}$. Under constant charge conditions, the Au surface loses more electrons, whereas under constant potential conditions, the Au surface gains electrons from the potentiostat, resulting in a smaller net electron loss. This behavior is consistent across other systems (Supplementary Figs. 79 and 90). Supplementary Figs. 81 and 92 illustrate the changes in voltage along the reaction coordinates, demonstrating that our reactor can maintain a constant system applied potential throughout the entire reaction process. In contrast, traditional constant-charge simulations show that $CO_2$ adsorption leads to an increase in the system applied potential (Supplementary Fig. 82 and 93). Additionally, these findings align with the conclusions drawn from the Volmer process, validating the consistency of our constant potential reactor in simulating electrochemical reactions.

## Discussion

We have validated the high accuracy of veNNP, enabling us to efficiently simulate the ESLI with first-principles precision. Our model also allows for the decoupling of applied potentials and cations, enabling us to systematically investigate the impact of cations on solution structure and electrochemical reactions under the same potential. We found that at more negative potentials or in the presence of cations, the dipoles of interfacial water molecules tend to orient towards the surface. It was not possible to study with previous purely explicit models[17,26]. Overall, our conclusions remain consistent with the findings from prior work[2,16,42].

Moreover, our calculations of the $CO_2$ adsorption and the Volmer step reaction barriers indicate that cations promote both reaction steps. Specifically, the promotion effect on the $CO_2$ adsorption is more pronounced, significantly lowering the reaction energy barrier and stabilizing the adsorption state. In some cases, the adsorption state is even more stable than the initial state at more negative potentials. For the Volmer process, although the reaction energy barrier is reduced, the adsorbed H atoms remain unstable and easily desorb, as previously discussed (see Supplementary Fig. 97b). Furthermore, the addition of cations significantly disrupts the connectivity of the hydrogen-bond network in the aqueous solution, hindering $H^+$ mass transfer and severely inhibiting HER performance[6,42]. This provides an explanation for the experimental observation that cations enhance $CO_2RR$ while suppressing HER, thereby improving the selectivity of $CO_2RR$.

Previous studies suggested that the promotion of $CO_2RR$ by cations might stem from direct coordination interactions between cations and $CO_2$[13,52,61]. To assess this hypothesis, we analyzed all simulated $CO_2$ adsorption reaction trajectories and found that the distances between $K^+$ ions and the two O atoms of $CO_2$ were consistently greater than 4 Å. The most probable distances are around 5 Å or even greater (see Supplementary Figs. 98 and 99). Such large $d_{K-O}$ distances make direct coordination between $K^+$ and O atoms unlikely. However, our computations indicate that cations do indeed promote $CO_2$ activation. Therefore, we can conclude that the promotion of $CO_2RR$ by cations is not achieved through direct coordination with $CO_2$, at least on Au catalysts.

To understand the underlying reasons, we analyzed the structural changes of $CO_2$ during the reaction (Supplementary Figs. 83 and 94). As $CO_2$ is gradually adsorbed, its initially linear geometry bends significantly, with the bending angle ranging between 120° and 140° upon complete adsorption. Remarkably, this bending trend corresponds to the number of electrons that $CO_2$ acquires from the Au surface during adsorption, *i.e.*, the charge on $CO_2$. A detailed comparison reveals that the more negative the charge on $CO_2$, the greater the degree of bending. By comparing the charge on $CO_2$ with and without cations (Supplementary Figs. 79 and 90), we observed that in the presence of $K^+$ ions, $CO_2$ carries a more negative charge under the same potential and reaction coordinates. Similarly, during the Volmer process, the charge on H atoms changes in a comparable manner, though to a less extent. These findings suggest that cations facilitate the reduction reaction by making it easier for the reactant molecules to acquire electrons from the catalyst surface.

Additionally, we found that at increasingly negative applied potentials, $CO_2$ more readily acquires electrons from the Au surface. Combined with previous analyses of atomic charge distribution (Fig. 5a and Supplementary Figs. 53 and 54), we know that as the potential becomes more negative, more electrons accumulate on the Au surface. Previous studies showed that electron accumulation on the catalyst surface facilitates electron transfer to reactant molecules[49,62]. Considering the frame-by-frame analysis of Bader charge in molecular dynamics trajectories (Supplementary Videos 15-18), which concluded that cations promote electron accumulation on the Au surface, we can infer that the introduction of cations enhances electron transfer between the catalytic sites and reactant molecules, thereby lowering the reaction energy barrier. As $CO_2$ acquires more electrons, the bent adsorption molecule forms a dipole, which increases with the bending angle and electron acquisition, and is stabilized by the strong electric field of $K^+$ ions[2,11]. This provides a fundamental explanation for how alkaline cations promote $CO_2RR$. Furthermore, the Bader charge analysis also provides evidence that the protruding surface site can accumulate electrons, explaining its strong catalytic activity. Essentially, both the cation effect and surface reconstruction have a similar impact to lowering the system potential, thereby promoting the reduction reaction.

Importantly, our study focused on a specific site on the Au surface through the slow-growth method. To thoroughly investigate the relationship between catalytic sites in the Au surface and $CO_2RR$ activity, more advanced tools are needed, such as the on-the-fly probability-enhanced sampling (OPES) metadynamics method[63] used by Bonati et al.[56] for studying $N_2$ activation on the Fe(111) surface. It will be addressed in our future research.

In summary, we have proposed a constant potential reactor model to probe the ESLI and the electrochemical reactions occurring on it with first-principles accuracy. The precision and efficiency of this model have been rigorously validated in this work. Using this tool, we investigated the $CO_2$ adsorption and Volmer step on the Au surface, demonstrating that cations promote $CO_2$ activation by facilitating surface charge accumulation. The dipole formed during $CO_2$ activation is stabilized by the strong electric field of the cations, ensuring the stability of the activated $CO_2$ state. Our findings provide theoretical evidence for the role of surface reconstruction in promoting reduction reactions and the hindrance of $H^+$ mass transfer by cations. Our results offer new insights into the effects of alkali metal cations and surface reconstruction on $CO_2RR$. Furthermore, the general framework of our constant potential reactor provides a powerful tool for studying ESLI structures and the electrochemical reactions occurring within them.

## Methods

**DFT calculations and establishment of an initial database**. All DFT calculations were conducted through the projector augmented wave (PAW) method[64] as implemented in the Vienna Ab-Initio Simulation Package (VASP)[65,66]. The exchange-correlation effects were described with the revised Perdew-Burke-Ernzerhof (rPBE) functional[67], and van der Waals interactions were incorporated by supplementing with the D3 dispersion correction scheme[68]. The plane wave cut-off energy was set to 400 eV for all calculations. We first optimized the crystal structure of Au bulk. For structure optimization, the energy convergence criterion was set to $10^{-5}$ eV and the force convergence criteria was set to 0.01 eV Å$^{-1}$. A gamma-centered k-point mesh with a density of 0.03 π Å$^{-1}$ and single gamma point were employed to sample the Brillouin zone for structural relaxation and molecular simulations, respectively. The Au(111) surface was subjected to lattice redirection to ensure that the lattice plane corresponding to the crystal surface is rectangular than rhombic, which facilitates the subsequent construction of the Au-water interface model. Subsequently, the Au(111) surface was expanded into a supercell with dimensions of 10.18 Å × 11.76 Å while Au(110) surface was also expanded into a supercell with dimensions of 12.47 Å × 11.76 Å. The Au surface model consisted of four atomic layers, with the bottom two layers fixed to simulate the bulk phase structure. 48 $H_2O$ molecules with a density of 1 g cm$^{-3}$ were incorporated into the Au surface system to simulate the Au-water interface. A 20 Å vacuum layer is intentionally set to isolate and minimize any periodic influences in the z direction.

The electrode potential was calculated by utilizing the VASPsol[69,70] implicit solvation model. In numerous studies, it has been demonstrated that the hybrid method of combining explicit and implicit solvent (as show in Supplementary Fig. 1) is effective for computing electrode potential in MD[27,28]. The specific electrode potential with respect to the standard hydrogen electrode (SHE) was calculated using

$$U_{SHE} = -\frac{E_{Fermi}}{e} - 4.44 \ V. \tag{1}$$

Here, $E_{Fermi}$ represents corrected Fermi level of the system which can be adjusted by the Fermi shift obtained from VASPsol and $e$ represents the elementary charge. We employed the IUPAC-recommended value of 4.44 V for SHE[46].

Subsequently, we conducted 1 ps (one frame per fs) CPMD at two randomly selected applied potentials (0.2 V, 0 V, -0.2 V, -0.4 V, and -0.6 V) for a total of 24 systems in order to collect the initial dataset. These system included Au(111)-water and Au(110)-water interfaces with or without $CO_2$ containing 1 H$^+$ ion, 2 H$^+$ ion, 1 K$^+$ ion, and 2 K$^+$ ion, as well as the pure water interface. Additionally, Au(111)-water and Au(110)-water interfaces without $CO_2$ included 1 H$^+$ ion and 1 K$^+$ ion, 1 H$^+$ ion and 2 K$^+$ ion. For a comprehensive understanding of the methodology, refer to the section titled "Molecular Dynamics at a Constant Electrode Potential." Finally, a total of 48,000 structures were collected to form the initial dataset.

**Slow-growth method and free energy calculations**. The slow-growth approach[41] aims to explore the free energy profile along a geometric parameter $\zeta$ known as a CV. This variable can represent reaction coordinate of a system. It becomes possible to compute the free energy change between the initial state $\zeta_1$ to final equilibrium state $\zeta_2$ of the system by employing thermodynamic integration, as defined in

$$\Delta A_{1\rightarrow 2} = A(\zeta_2) - A(\zeta_1) = \int_{\zeta_1}^{\zeta_2} \frac{dA}{d\zeta} d\zeta = \int_{\zeta_1}^{\zeta_2} d\zeta \langle \frac{dH}{d\zeta} \rangle_\zeta \ , \tag{2}$$

in which, the derivative of the free energy with respect to the CV $\frac{dA}{d\zeta}$ is computed by averaging the derivative of Hamiltonian about the system with respect to the same variable over the ensemble $\langle \frac{dH}{d\zeta} \rangle_\zeta$, which is also referred to the potential of mean force. The operator $\langle ... \rangle$ is computed as the time-averaged result from a MD trajectory, where the system is constrained with the variable $\zeta$ held at a specified constant value. Then, the specific form of the PMF is derived from

$$\frac{dA}{d\zeta} = \frac{\langle Z^{-1/2}(-\lambda_\zeta + k_B TG)\rangle}{\langle Z^{-1/2}\rangle} \ , \tag{3}$$

where the $k_B$ and $\lambda_\zeta$ is the Boltzmann constant and the Lagrange multiplier associated with the variable $\zeta$, respectively. The Lagrange multiplier $\lambda_\zeta$ is introduced to impose on the reaction, thereby altering the Lagrangian to take the form:

$$L'(\vec{r},\vec{p},\zeta) = L(\vec{r},\vec{p}) + \lambda_\zeta[\zeta(\vec{r}) - \zeta] \ . \tag{4}$$

The factor Z is expressed through

$$Z = \sum_i \frac{1}{m_i}\left(\frac{\partial \zeta}{\partial \vec{r}_i}\right)^2 , \tag{5}$$

and the factor G is expressed through

$$G = \frac{1}{Z^2}\sum_{i,j}\frac{1}{m_i m_j}\frac{\partial \zeta}{\partial \vec{r}_i}\frac{\partial^2 \zeta}{\partial \vec{r}_i \partial \vec{r}_j}\frac{\partial \zeta}{\partial \vec{r}_j} \tag{6}$$

Here, the $m_i$ and $\vec{r}_i$ denote the atomic mass and position of the $i$th atom included in the CV, respectively. The value of $\lambda_\zeta$ can be computed employing the SHAKE algorithm[71]. The PMF calculation described above primarily utilizes the Blue-Moon method[40] due to its reliance on ensemble averaging. However, a key distinction arises with the Slow-Growth method, which introduces a constant rate of change $\dot{\zeta}$ for the $\zeta$ variable. This approach enables $\zeta$ to evolve gradually and automatically, circumventing the need for ensemble averaging. Consequently, the free energy calculation process can be expressed as follows:

$$\Delta A_{1\to 2} = \int_{\zeta_1}^{\zeta_2} d\zeta \, \langle\frac{dH}{d\zeta}\rangle_\zeta = \lim_{\dot{\zeta}\to 0}\int_{\zeta_1}^{\zeta_2}\frac{dH}{d\zeta}\dot{\zeta}dt . \tag{7}$$

We have implemented the corresponding function in Python, ensuring its adaptability for use with the MD code within the atomic simulation environment (ASE) package[72]. Furthermore, it is also compatible with CPMD, for which we have developed another code that will be presented at a later stage.

**Molecular dynamics at a constant electrode potential**. Following the Otani group's proposal, we implemented a scheme that performs MD simulations at a constant electrode potential[44]. This method connects the system to a fictitious potentiostat that facilitates electronic charge exchange with an external electron reservoir at a fixed potential. The modulate MD extends Nosé's Lagrangian to include the charge and its coupling to the potentiostate:

$$L = \sum_{i=1}^{N}\left(\frac{1}{2}m_i s^2 \dot{\vec{r}}_i^2 - \mathcal{E}(\vec{r}_i)\right) - \frac{1}{2}M_s \dot{s}^2 - g k_B T \ln s + \frac{1}{2}M_{n_e}\dot{n}_e^2 - \Phi_{pot}n_e , \tag{8}$$

in which, $s$ and $n_e$ are the thermal reservoir controlled by the thermostat coordinate and the system electron number, respectively. The particle mass and position are denoted by $m_i$ and $\vec{r}_i$ in the system, respectively. $T$, $\Phi$, and $g$ are the temperature, the electrode potential, and a constant equal to the system's degrees of freedom (including the charge degree of freedom), respectively. $M_s$ and $M_{n_e}$ are two fictitious masses. Then the equations of motion are derived as follow:

$$\dot{r} = \frac{p_i}{m_i} , \tag{9}$$

$$\dot{p}_i = F_i - p_i \frac{P_\xi}{M_\xi} , \tag{10}$$

$$\dot{n}_e = \frac{P_{n_e}}{M_{n_e}} , \tag{11}$$

$$\dot{P}_{n_e} = \Phi - \Phi_{pot} - P_{n_e}\frac{P_\xi}{M_\xi} , \tag{12}$$

$$\dot{\xi} = \frac{P_\xi}{M_\xi} , \tag{13}$$

$$\dot{P}_\xi = \sum_i \frac{p_i^2}{m_i} + \frac{P_{n_e}}{M_{n_e}} - g k_B T , \tag{14}$$

where, $\vec{p}_i$ denotes particle momenta in the system. The $\xi$ is a control variable and $P_\xi$ and

$M_\xi$ are the corresponding momenta and mass, respectively. The $P_{n_e}$ is the momenta of the system electron number. We developed a Python module for constant electrode potential MD simulations, based on the ASE interface[72]. It can work with various first-principles and NNP calculators, depending on the requirements.

**A versatile machine learning framework for neural network potentials, electrode potential and partial charge.** In order to perform MD simulations at a constant electrode potential, the NNP calculator needs to be able to capture the subtle perturbations of the PES caused by changes in electron number. At the same time, it also needs to be able to predict the electrode potential of the system. Therefore, the information of the system charge state must be encoded as machine learning features and be recognizable by the neural network. For example, the feature representations in the 4G-HDNNP[30,73] and SpookyNet[74] frameworks can meet this requirement. In this work, we used features of SpookeyNet as inputs for four dual-layer neural networks. Three of them predict the atomic local energy, the self-consistent atomic charge (for the long-range electrostatic potential), and the electrode potential. The fourth one predicts the atomic charge. We added this network because the second one may not capture the electrostatic potential well with Bader charges, causing errors in the PES. Thus, we had a separate network for atomic charge prediction.

In this work, we defined the total potential energy as

$$E_{pot} = \sum_{i=1}^{N} E_i \ , \tag{15}$$

where $E_i$ indicates the energy of the $i$th atom in the system and N is the number of the atoms in the system. The $E_i$ is defined as

$$E_i = E_i^{local} + E_i^{rep} + E_i^{ele} + E_i^{vdw} \ , \tag{16}$$

in which $E_i^{local}$, $E_i^{rep}$, $E_i^{ele}$ and $E_i^{vdw}$ indicate the local atomic energy, nuclear repulsion energy, electrostatic energy and D4 dispersion correction energy, respectively. Here, $E_i^{rep}$ and $E_i^{vdw}$ are defined to align with those reported in ref [74]. The details of $E_i^{local}$ and $E_i^{ele}$ will be elaborated in the following sections. The energy-conserving interatomic force $\vec{F}_i$ which are determined as the negative gradient of the potential energy relative to the atomic position $\vec{r}_i$ for the $i$th atom, as show in equation

$$\vec{F}_i = -\partial E_{pot}/\partial \vec{r}_i \ . \tag{17}$$

Upon evaluating all interaction modules of SpookyNet, the local atomic energy contributions $E_i^{local}$ are predicted utilizing atomic descriptors $\boldsymbol{f}_i$ of SpookeyNet, via the first dual-layer neural network

$$E_i^{local} = \boldsymbol{w}_{2_E}^T \varphi(\boldsymbol{w}_{1_E}^T \boldsymbol{f}_i + \tilde{b}_{i_E}) + \tilde{E}_{Z_i} \ . \tag{18}$$

Here, $\boldsymbol{w}_{1_E} \in \mathbb{R}^F$ and $\boldsymbol{w}_{2_E} \in \mathbb{R}^F$ represent the regression weight of the first and second layer neural networks, respectively. The function $\varphi(x)$ is the active function and $\tilde{b}_E \in \mathbb{R}$ represents the biases of the first layer neural network. $\tilde{E}_Z \in \mathbb{R}$ is element-dependent energy bias.

Aligned with previous assertions, the long-range electrostatic contribution to the potential energy of the $i$th atom ($E_i^{ele}$) is calculated using self-consistent atomic charges. The self-consistent atomic charge $q_i^s$ is predicted from atomic features $\boldsymbol{f}_i$ of SpookyNet in according with

$$q_i = \boldsymbol{w}_{2_q}^T \varphi\left(\boldsymbol{w}_{1_q}^T \boldsymbol{f}_i + \tilde{b}_{i_q}\right) + \tilde{q}_{Z_i} + \frac{1}{N}\left\{Q - \sum_{j=1}^{N}\left[\boldsymbol{w}_{2_q}^T \varphi\left(\boldsymbol{w}_{1_q}^T \boldsymbol{f}_i + \tilde{b}_{i_q}\right) + \tilde{q}_{Z_j}\right]\right\} \ . \tag{19}$$

Here, $\boldsymbol{w}_{1_q} \in \mathbb{R}^F$ and $\boldsymbol{w}_{2_q} \in \mathbb{R}^F$ denote the regression weights of the first and second layer neural networks within the second dual-layer neural network structure, respectively. The $\tilde{b}_q \in \mathbb{R}$ and $\tilde{q}_Z \in \mathbb{R}$ are the biases of the first layer neural network and element-dependent charge. Then, the Ewald summation can utilize the self-consistent atomic charges to calculate the complete long-range electrostatic energy. According to the Ewald summation method, the electrostatic potential $E_i^{ele}$ can be decomposed into the real space term $E_i^{real}$, the reciprocal space term $E_i^{recip}$, and the self-interaction correction term $E_i^{self}$, as show in equation

$$E_i^{ele} = E_i^{real} + E_i^{recip} + E_i^{self} . \tag{20}$$

The real space component is defined by

$$E_i^{real} = \frac{1}{2}\sum_{j \neq i}^{N_{neig}} q_i q_j \frac{erfc(\frac{r_{ij}}{\sqrt{2}\sigma})}{r_{ij}} , \tag{21}$$

where the $N_{neig}$ represents the number of neighboring atoms within the real space cutoff radius $r_{cut}^{real}$. The $\sigma$ represents the standard deviation of the Gaussian screening charges. The reciprocal space component is

$$E_i^{recip} = \frac{2\pi}{V} \sum_{|\vec{k}| \neq 0} \frac{\exp(\frac{-\sigma^2|\vec{k}|^2}{2})}{|\vec{k}|} \left| q_i \exp(i\vec{k}) \cdot r_i \right| . \tag{22}$$

Here, V denotes the volume of the unit cell and the summation is carried out over all reciprocal lattice vectors $|\vec{k}|$ that lie within the reciprocal lattice cutoff radius $r_{cut}^{recip}$. Finally, the self-interaction correction is given by

$$E_i^{self} = -\frac{q_i^2}{\sqrt{2\pi}\sigma} . \tag{23}$$

The width of the Gaussian screening charge distributions for Ewald summation, denoted by $\sigma$, is defined as

$$\sigma = \frac{1}{\sqrt{2\pi}} \times \left(\frac{V^2}{n_{atom}}\right)^{\frac{1}{6}} . \tag{24}$$

Here, $n_{atom}$ and $V$ are the number of atoms and the cell volume, respectively. The $r_{cut}^{real}$ is given by

$$r_{cut}^{real} = \sqrt{2 \times \log_{10} \alpha} \times \sigma , \tag{25}$$

in which $\alpha$ denotes the desired level of accuracy. The $r_{cut}^{recip}$ is given by

$$r_{cut}^{recip} = \frac{\sqrt{2 \times \log_{10} \alpha}}{\sigma} . \tag{26}$$

In this work, the third dual-layer neural network is employed to predict the reference-independent correction to the Fermi level $E_{Fermi}$ and $E_{Fermi}$ is modeled as

$$E_{Fermi} = \frac{\sum_i^N \left[ \boldsymbol{w}_{2_{Fermi}}^T \varphi(\boldsymbol{w}_{1_{Fermi}}^T f_i + \tilde{b}_{i_{Fermi}}) + \tilde{F}_{Z_i} \right]}{N}. \tag{27}$$

In this equation, $\boldsymbol{w}_{1_{Fermi}} \in \mathbb{R}^F$ and $\boldsymbol{w}_{2_{Fermi}} \in \mathbb{R}^F$ are the regression weights of the first and second layer neural networks. The $\tilde{b}_{Fermi} \in \mathbb{R}$ and $\tilde{F}_Z \in \mathbb{R}$ are the biases of the first layer neural network and element-dependent atomic Fermi level contribution, respectively.

Subsequently, the electrode potential can be determined using equation (1).

We have engineered a novel and independent dual-layer network specifically for the prediction of atomic charges. The breakthrough of this network lies in its approach which it trains directly with target charges, bypassing the need for fitting to long-range electrostatic potentials and thus sidestepping the issue of self-consistency. This method circumvents the potential decline in the accuracy of the PES and atomic charge predictions that can arise from long-range electrostatic interactions. While the charge calculation by this network aligns with the method for self-consistent charges based on equation (19), the parameters which it employs are uniquely determined by the novel network.

**Hyperparameters and Training.** All models based on SpookyNet[74] in this work utilize 3 interaction modules, 168 features, and a cutoff radius $r_{cut}$ of 5 Å, unless a different value is specified. In neural networks, weight initialization is done by assigning random (semi-)orthogonal matrices to the weights, and then scaling them according to the Glorot initialization scheme[75]. This scheme ensures that the weights are properly initialized for effective training and optimization. This step also breaks the symmetry among the network units, enabling each neuron to perform different computations during training. Moreover, the bias terms are all initialized with zeros.

The model parameters are optimized through an iterative training process that aims to minimize the loss function. This optimization is achieved using mini-batch gradient descent with the AMSGrad optimizer[76], which employs the generally recommended default momentum hyperparameters. In our study, we employed a cosine annealing schedule to dynamically adjust the learning rate during training. Specifically, the learning rate underwent smooth reductions at intervals of every 30 batches within an epoch. The entire training process spanned 130 epochs. Initially, we set the learning rate to $10^{-3}$, gradually decaying it to a minimum of $10^{-8}$. The period of the cosine annealing scheduler is determined by the total number of batches multiplied by 4.33 (130 / 30). The loss function is specified by

$$\mathcal{L} = a_E \mathcal{L}_E^{RMSE} + a_{\vec{F}} \mathcal{L}_{\vec{F}}^{RMSE} + a_{E_{Fermi}} \mathcal{L}_{E_{Fermi}}^{RMSE} + a_q \mathcal{L}_q^{RMSE} , \qquad (28)$$

in which, the $\mathcal{L}_E^{RMSE}$, $\mathcal{L}_{\vec{F}}^{RMSE}$, $\mathcal{L}_{E_{Fermi}}^{RMSE}$ and $a_q \mathcal{L}_q^{RMSE}$ represent the RMSE of energy, forces, corrected Fermi level, and atomic charges, respectively. The parameters $a_E$, $a_{\vec{F}}$, $a_{E_{Fermi}}$ and $a_q$ play a crucial role in determining the relative impact of each term on the overall loss. In this work, the $a_E$, $a_{\vec{F}}$, $a_{E_{Fermi}}$ and $a_q$ were set to 1, 50, 10, 1, respectively.

**Details on the active learning architecture.** We presented an active learning architecture for training reaction PES. The conventional approach of using AIMD to collect data is constrained by the data source and cannot efficiently explore the data space, even with enhanced sampling methods. This results in a huge computational cost. Moreover, our proposed variable-charge neural network model needs to calculate atomic charges and electrode potentials, which are computationally demanding. To overcome these challenges, we proposed an active learning framework that reduces the reliance on first-principles calculations. Our active learning framework employs a neural network committee, which consists of multiple neural networks with the same parameters but different random seeds. Each network can predict physical quantities of the system, such as average single-atom energy, single-atom energy, atomic forces, system stresses, system electrode potential, and atomic charges. The committee computes the mean and standard deviation of the predictions of each network for each selected physical quantity. The standard deviation indicates the model uncertainty at that point, analogous to the concept of Gaussian process regression. By integrating the active learning framework with MD simulations (with or without enhanced sampling), we can achieve structure evolution and filter out inaccurate or unseen structures from the structural evolution, thereby enhancing predictive ability of the model. The iterative process of the active learning framework is outlined as follows:

1. Training multiple models with the same parameters using different random seeds in software environments like Numpy[77] and Pytorch[78].
2. An integrative calculator that consists of multiple models is used to directly calculate the uncertainty (i.e. standard deviation) of the target physical quantities. Based on the calculator, constant potential MD simulations (enhanced sampling can be optionally included) are performed. Then, structures with unreliable or unstable predictions are screened out according to the magnitude of the uncertainty. The architecture adopts two screening rules: a) the structure is considered to have unreliable or unstable prediction when the uncertainties of all target physical quantities exceed the set threshold; b) the structure is considered to have unreliable or unstable prediction when the uncertainty of a certain target physical quantity exceeds the set threshold.
3. The invisibility of each structure generated through evolution from a specific initial structure but exhibiting inaccurate or unstable predictions is calculated according to predefined selection rules. Invisibility reflects the confidence of the model in predicting the structure, which is obtained by: weighting and summing the uncertainties of individual target physical quantities according to certain weight ratios, and defining the total uncertainty as invisibility. Then, the structures are sorted by invisibility from high to low, and a certain number of structures with the highest invisibility are filtered out. The prediction unreliability of these structures is the highest, which is helpful to further improve the model.
4. First-principles calculations are conducted to determine the target physical quantities of structures flagged as exhibiting potential inaccuracies or instabilities by the model.
5. Compare the proportion of uncertain or unstable structures in all evolved structures with the convergence threshold. If the proportion is lower, the target space is adequately sampled and the active learning process ends. If the proportion is higher, the target space requires more sampling, and the program repeats from step 1 until it converges.

In this work, we employ an active learning framework utilizing a committee of four neural networks. In step 3, the 100 most uncertain structures are identified and subject them to first-principles calculations. To establish the uncertainty threshold for each target physical quantity, we employed four initial models trained on the dataset. These models formed a neural network committee. Subsequently, we systematically explored all structures within the initial dataset, calculating the uncertainty associated with each target physical quantity in every structure. Finally, we identified the maximum uncertainty for each physical quantity and determined the threshold just below this value. The detailed results are presented in Supplementary Table 1. This approach enables us to quantitatively assess the predictions of the model for diverse targets, ensuring prediction credibility.

## Data availability

All the DFT datasets used to train veNNP and the molecular dynamics trajectory files generated using the trained veNNP are available from corresponding authors upon reasonable request.

## Code availability

The source code for the constant potential reactor is available from corresponding authors upon reasonable request.

## Acknowledgement

The authors thank the financial support from National Natural Science Foundation of China (21933006, 22203077, 223B2305), Natural Science Foundation of Henan Province (242300421129 and 232301420051) and the Ph.D. Candidate Research Innovation Fund of NKU School of Materials Science and Engineering.

## References

1. Choi C. *et al.* Highly active and stable stepped Cu surface for enhanced electrochemical $CO_2$ reduction to $C_2H_4$. *Nat. Catal.* **3**, 804-812 (2020).
2. Monteiro M. C. O., Dattila F., Hagedoorn B., García-Muelas R., López N. & Koper M. T. M. Absence of $CO_2$ electroreduction on copper, gold and silver electrodes without metal cations in solution. *Nat. Catal.* **4**, 654-662 (2021).
3. Jin Z. & Bard A. J. Surface Interrogation of Electrodeposited MnO(x) and CaMnO(3) Perovskites by Scanning Electrochemical Microscopy: Probing Active Sites and Kinetics for the Oxygen Evolution Reaction. *Angew Chem Int Ed Engl* **60**, 794-799 (2021).
4. of Electrochemistry C. S. The Top Ten Scientific Questions in Electrochemistry. *Journal of Electrochemistry* **30**, 2024121 (2024).
5. Hansen J. N. *et al.* Is There Anything Better than Pt for HER? *ACS Energy Lett.* **6**, 1175-1180 (2021).
6. Li P. *et al.* Hydrogen bond network connectivity in the electric double layer dominates the kinetic pH effect in hydrogen electrocatalysis on Pt. *Nat. Catal.* **5**, 900-911 (2022).
7. Durst J., Simon C., Hasché F. & Gasteiger H. A. Hydrogen Oxidation and Evolution Reaction Kinetics on Carbon Supported Pt, Ir, Rh, and Pd Electrocatalysts in Acidic Media. *J. Electrochem. Soc.* **162**, F190-F203 (2014).
8. Pan B., Wang Y. & Li Y. Understanding and leveraging the effect of cations in the electrical double layer for electrochemical $CO_2$ reduction. *Chem Catal.* **2**, 1267-1276 (2022).
9. Levell Z. *et al.* Emerging Atomistic Modeling Methods for Heterogeneous Electrocatalysis. *Chem. Rev.* **124**, 8620-8656 (2024).
10. Ovalle V. J., Hsu Y.-S., Agrawal N., Janik M. J. & Waegele M. M. Correlating hydration free energy and specific adsorption of alkali metal cations during $CO_2$ electroreduction on Au. *Nat. Catal.* **5**, 624-632 (2022).

11. Gu J., Liu S., Ni W., Ren W., Haussener S. & Hu X. Modulating electric field distribution by alkali cations for $CO_2$ electroreduction in strongly acidic medium. *Nat. Catal.* **5**, 268-276 (2022).
12. Li Z., Wang L., Sun L. & Yang W. Dynamic Cation Enrichment during Pulsed $CO_2$ Electrolysis and the Cation-Promoted Multicarbon Formation. *J. Am. Chem. Soc.*, (2024).
13. Shin S. J. *et al.* A unifying mechanism for cation effect modulating C1 and C2 productions from $CO_2$ electroreduction. *Nat. Commun.* **13**, 5482 (2022).
14. Wang Y. H. *et al.* In situ Raman spectroscopy reveals the structure and dissociation of interfacial water. *Nature* **600**, 81-85 (2021).
15. Hsu Y. S., Rathnayake S. T. & Waegele M. M. Cation effects in hydrogen evolution and $CO_2$-to-CO conversion: A critical perspective. *J. Chem. Phys.* **160**, (2024).
16. Zhang Z.-M. *et al.* Probing electrolyte effects on cation-enhanced $CO_2$ reduction on copper in acidic media. *Nat. Catal.* **7**, 807-817 (2024).
17. Li C. Y. *et al.* In situ probing electrified interfacial water structures at atomically flat surfaces. *Nat. Mater.* **18**, 697-701 (2019).
18. Zhang D., Wang R., Wang X. & Gogotsi Y. In situ monitoring redox processes in energy storage using UV–Vis spectroscopy. *Nat. Energy* **8**, 567-576 (2023).
19. Cao H., Waghray D., Knoppe S., Dehaen W., Verbiest T. & De Feyter S. Tailoring atomic layer growth at the liquid-metal interface. *Nat. Commun.* **9**, 4889 (2018).
20. Du X. *et al.* Machine-learning-accelerated simulations to enable automatic surface reconstruction. *Nat Comput. Sci.* **3**, 1034-1044 (2023).
21. Zhang Q. *et al.* Atomic dynamics of electrified solid-liquid interfaces in liquid-cell TEM. *Nature* **630**, 643-647 (2024).
22. Chan K. & Norskov J. K. Electrochemical Barriers Made Simple. *J. Phys. Chem. Lett.* **6**, 2663-2668 (2015).
23. Chan K. & Norskov J. K. Potential Dependence of Electrochemical Barriers from ab Initio Calculations. *J. Phys. Chem. Lett.* **7**, 1686-1690 (2016).
24. Le J., Iannuzzi M., Cuesta A. & Cheng J. Determining Potentials of Zero Charge of Metal Electrodes versus the Standard Hydrogen Electrode from Density-Functional-Theory-Based Molecular Dynamics. *Phys. Rev. Lett.* **119**, 016801 (2017).
25. Zhao X., Levell Z. H., Yu S. & Liu Y. Atomistic Understanding of Two-dimensional Electrocatalysts from First Principles. *Chem. Rev.* **122**, 10675-10709 (2022).
26. Le J.-B., Fan Q.-Y., Li J.-Q. & Cheng J. Molecular origin of negative component of Helmholtz capacitance at electrified Pt (111)/water interface. *Sci. Adv.* **6**, eabb1219 (2020).
27. Zhao X. & Liu Y. Origin of Selective Production of Hydrogen Peroxide by Electrochemical Oxygen Reduction. *J. Am. Chem. Soc.*, (2021).
28. Yu S., Levell Z., Jiang Z., Zhao X. & Liu Y. What Is the Rate-Limiting Step of Oxygen Reduction Reaction on Fe-N-C Catalysts? *J. Am. Chem. Soc.* **145**, 25352-25356 (2023).
29. Behler J. & Parrinello M. Generalized neural-network representation of high-dimensional potential-energy surfaces. *Phys. Rev. Lett.* **98**, 146401 (2007).
30. Behler J. Four Generations of High-Dimensional Neural Network Potentials. *Chem. Rev.* **121**, 10037-10072 (2021).
31. Lian Z., Dattila F. & López N. Stability and lifetime of diffusion-trapped oxygen in oxide-derived copper $CO_2$ reduction electrocatalysts. *Nat. Catal.* **7**, 401-411 (2024).
32. Freitas R. & Reed E. J. Uncovering the effects of interface-induced ordering of liquid on crystal

growth using machine learning. *Nat. Commun.* **11**, 3260 (2020).

33. Zhu J. *et al.* Superdiffusive Rotation of Interfacial Water on Noble Metal Surface. *J. Am. Chem. Soc.* **146**, 16281-16294 (2024).
34. Yang X., Bhowmik A., Vegge T. & Hansen H. A. Neural network potentials for accelerated metadynamics of oxygen reduction kinetics at Au-water interfaces. *Chem. Sci.* **14**, 3913-3922 (2023).
35. Zeng Z. *et al.* Mechanistic insight on water dissociation on pristine low-index $TiO_2$ surfaces from machine learning molecular dynamics simulations. *Nat. Commun.* **14**, 6131 (2023).
36. Rice P. S., Liu Z. P. & Hu P. Hydrogen Coupling on Platinum Using Artificial Neural Network Potentials and DFT. *J. Phys. Chem. Lett.* **12**, 10637-10645 (2021).
37. Hu Z.-Y., Luo L.-H., Shang C. & Liu Z.-P. Free Energy Pathway Exploration of Catalytic Formic Acid Decomposition on Pt-Group Metals in Aqueous Surroundings. *ACS Catal.* **14**, 7684-7695 (2024).
38. Amirbeigiarab R. *et al.* Atomic-scale surface restructuring of copper electrodes under $CO_2$ electroreduction conditions. *Nat. Catal.* **6**, 837-846 (2023).
39. Iannuzzi M., Laio A. & Parrinello M. Efficient exploration of reactive potential energy surfaces using Car-Parrinello molecular dynamics. *Phys. Rev. Lett.* **90**, 238302 (2003).
40. Carter E. A., Ciccotti G., Hynes J. T. & Kapral R. Constrained reaction coordinate dynamics for the simulation of rare events. *Chem. Phys. Lett.* **156**, 472-477 (1989).
41. Woo T. K., Margl P. M., Blöchl P. E. & Ziegler T. A combined Car−Parrinello QM/MM implementation for ab initio molecular dynamics simulations of extended systems: application to transition metal catalysis. *J. Phys. Chem. B* **101**, 7877-7880 (1997).
42. Hicks M. H., Nie W., Boehme A. E., Atwater H. A., Agapie T. & Peters J. C. Electrochemical $CO_2$ Reduction in Acidic Electrolytes: Spectroscopic Evidence for Local pH Gradients. *J. Am. Chem. Soc.* **146**, 25282-25289 (2024).
43. Fan M. *et al.* Cationic-group-functionalized electrocatalysts enable stable acidic CO2 electrolysis. *Nat. Catal.* **6**, 763-772 (2023).
44. Bonnet N., Morishita T., Sugino O. & Otani M. First-principles molecular dynamics at a constant electrode potential. *Phys. Rev. Lett.* **109**, 266101 (2012).
45. Yu M. & Trinkle D. R. Accurate and efficient algorithm for Bader charge integration. *J. Chem. Phys.* **134**, 064111 (2011).
46. Trasatti S. The absolute electrode potential: an explanatory note (Recommendations 1986). *Pure Appl. Chem.* **58**, 955-966 (1986).
47. Kolb D. & Schneider J. Surface reconstruction in electrochemistry: Au (100-(5× 20), Au (111)-(1× 23) and Au (110)-(1× 2). *Electrochim. Acta* **31**, 929-936 (1986).
48. Li L., Liu Y.-P., Le J.-B. & Cheng J. Unraveling molecular structures and ion effects of electric double layers at metal water interfaces. *Cell Reports Physical Science* **3**, (2022).
49. Hu X. *et al.* Understanding the role of axial O in $CO_2$ electroreduction on $NiN_4$ single-atom catalysts via simulations in realistic electrochemical environment. *J. Mater. Chem. A* **9**, 23515-23521 (2021).
50. Qin X., Vegge T. & Hansen H. A. Cation-Coordinated Inner-Sphere $CO_2$ Electroreduction at Au-Water Interfaces. *J. Am. Chem. Soc.* **145**, 1897-1905 (2023).
51. Qin X., Vegge T. & Hansen H. A. $CO_2$ activation at Au(110)-water interfaces: An ab initio molecular dynamics study. *J. Chem. Phys.* **155**, 134703 (2021).

52. Qin X., Hansen H. A., Honkala K. & Melander M. M. Cation-induced changes in the inner- and outer-sphere mechanisms of electrocatalytic $CO_2$ reduction. *Nat. Commun.* **14**, 7607 (2023).

53. Zhang Z., Zandkarimi B. & Alexandrova A. N. Ensembles of Metastable States Govern Heterogeneous Catalysis on Dynamic Interfaces. *Acc. Chem. Res.* **53**, 447-458 (2020).

54. Wang Y. Q. *et al.* Alkali Metal Cations Induce Structural Evolution on Au(111) During Cathodic Polarization. *J. Am. Chem. Soc.* **146**, 27713-27724 (2024).

55. Owen C. J., Xie Y., Johansson A., Sun L. & Kozinsky B. Low-index mesoscopic surface reconstructions of Au surfaces using Bayesian force fields. *Nat. Commun.* **15**, 3790 (2024).

56. Bonati L. *et al.* The role of dynamics in heterogeneous catalysis: Surface diffusivity and $N_2$ decomposition on Fe(111). *Proc. Natl. Acad. Sci.* **120**, e2313023120 (2023).

57. Bal K. M., Fukuhara S., Shibuta Y. & Neyts E. C. Free energy barriers from biased molecular dynamics simulations. *J. Chem. Phys.* **153**, 114118 (2020).

58. Fleurat-Lessard P. & Ziegler T. Tracing the minimum-energy path on the free-energy surface. *J. Chem. Phys.* **123**, 084101 (2005).

59. Bender J. T. *et al.* Understanding Cation Effects on the Hydrogen Evolution Reaction. *ACS Energy Lett.* **8**, 657-665 (2022).

60. Ringe S. *et al.* Double layer charging driven carbon dioxide adsorption limits the rate of electrochemical carbon dioxide reduction on Gold. *Nat. Commun.* **11**, 33 (2020).

61. Zhang Z. *et al.* Molecular understanding of the critical role of alkali metal cations in initiating $CO_2$ electroreduction on Cu(100) surface. *Nat. Commun.* **15**, 612 (2024).

62. Zhong W. *et al.* Electronic Spin Moment As a Catalytic Descriptor for Fe Single-Atom Catalysts Supported on $C_2N$. *J. Am. Chem. Soc.* **143**, 4405-4413 (2021).

63. Invernizzi M. & Parrinello M. Rethinking Metadynamics: From Bias Potentials to Probability Distributions. *J. Phys. Chem. Lett.* **11**, 2731-2736 (2020).

64. Blöchl P. E. Projector augmented-wave method. *Phys. Rev. B* **50**, 17953 (1994).

65. Kresse G. & Furthmüller J. Efficiency of ab-initio total energy calculations for metals and semiconductors using a plane-wave basis set. *Comp. Mater. Sci.* **6**, 15-50 (1996).

66. Kresse G. & Hafner J. Ab initio molecular dynamics for liquid metals. *Phys. Rev. B* **47**, 558-561 (1993).

67. Hammer B., Hansen L. B. & Nørskov J. K. Improved adsorption energetics within density-functional theory using revised Perdew-Burke-Ernzerhof functionals. *Phys. Rev. B* **59**, 7413 (1999).

68. Grimme S., Ehrlich S. & Goerigk L. Effect of the damping function in dispersion corrected density functional theory. *J. Comput. Chem.* **32**, 1456-1465 (2011).

69. Mathew K., Kolluru V. S. C., Mula S., Steinmann S. N. & Hennig R. G. Implicit self-consistent electrolyte model in plane-wave density-functional theory. *J. Chem. Phys.* **151**, 234101 (2019).

70. Mathew K., Sundararaman R., Letchworth-Weaver K., Arias T. A. & Hennig R. G. Implicit solvation model for density-functional study of nanocrystal surfaces and reaction pathways. *J. Chem. Phys.* **140**, 084106 (2014).

71. Ryckaert J.-P., Ciccotti G. & Berendsen H. J. Numerical integration of the cartesian equations of motion of a system with constraints: molecular dynamics of n-alkanes. *J. Comput. Phys.* **23**, 327-341 (1977).

72. Hjorth Larsen A. *et al.* The atomic simulation environment-a Python library for working with atoms. *J. Phys. Condens. Matter* **29**, 273002 (2017).

73. Ko T. W., Finkler J. A., Goedecker S. & Behler J. A fourth-generation high-dimensional neural network potential with accurate electrostatics including non-local charge transfer. *Nat. Commun.* **12**, 398 (2021).
74. Unke O. T., Chmiela S., Gastegger M., Schutt K. T., Sauceda H. E. & Muller K. R. SpookyNet: Learning force fields with electronic degrees of freedom and nonlocal effects. *Nat. Commun.* **12**, 7273 (2021).
75. Glorot X. & Bengio Y. Understanding the difficulty of training deep feedforward neural networks. In: *Proceedings of the thirteenth international conference on artificial intelligence and statistics*). JMLR Workshop and Conference Proceedings (2010).
76. Reddi S. J., Kale S. & Kumar S. On the convergence of Adam and beyond. *International Conference on Learning Representations* (2018).
77. Harris C. R. *et al.* Array programming with NumPy. *Nature* **585**, 357-362 (2020).
78. Paszke A. *et al.* Pytorch: An imperative style, high-performance deep learning library. *Adv. Neural Inf. Process. Syst.* **32**, 8024–8035 (2019).

# Supplementary information of "A constant-potential reactor framework for electrochemical reaction simulations"

Letian Chen[1], Yun Tian[2], Xu Hu[1], Suya Chen[1], Huijuan Wang[1], Xu Zhang[2]*, Zhen Zhou[1,2]*

[1] School of Materials Science and Engineering, Institute of New Energy Material Chemistry, Renewable Energy Conversion and Storage Center (ReCast), Key Laboratory of Advanced Energy Materials Chemistry (Ministry of Education), Nankai University, Tianjin 300350, China

[2] School of Chemical Engineering, Zhengzhou University, Zhengzhou 450001, Henan, China

* Corresponding author. E-mail: zzuzhangxu@zzu.edu.cn; zhouzhen@nankai.edu.cn

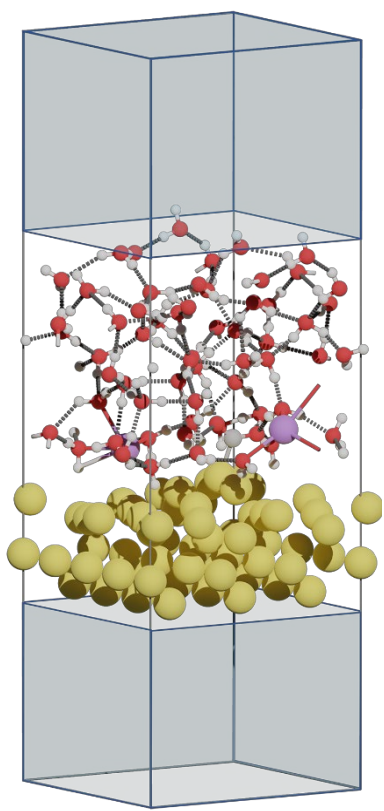

**Supplementary Figure 1. A schematic representation of a hybrid solvation model, incorporating both explicit and implicit solvent treatments within a supercell framework.** In the explicit region, atoms are depicted as spheres with colors corresponding to their element: white for H, red for O, light purple for K, light gray for C, and yellow for Au. The explicit region is enveloped by an implicit solvent representation, depicted in blue. Such a configuration enables the accurate modeling of the solution environment while preserving computational efficiency, facilitating the calculation of electrode potentials.

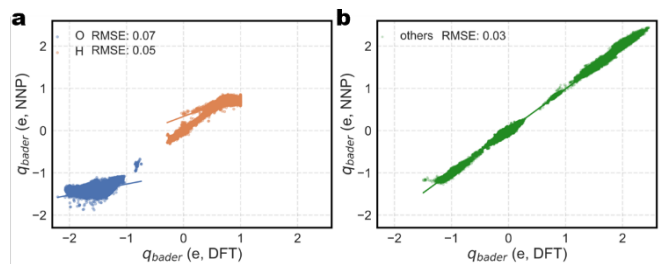

**Supplementary Figure 2. Predictive performance of the veNNP on atomic charges.** The plots compare the accuracy of veNNP and DFT for **a**, atomic charges of O and H, as well as **b**, those of other elements.

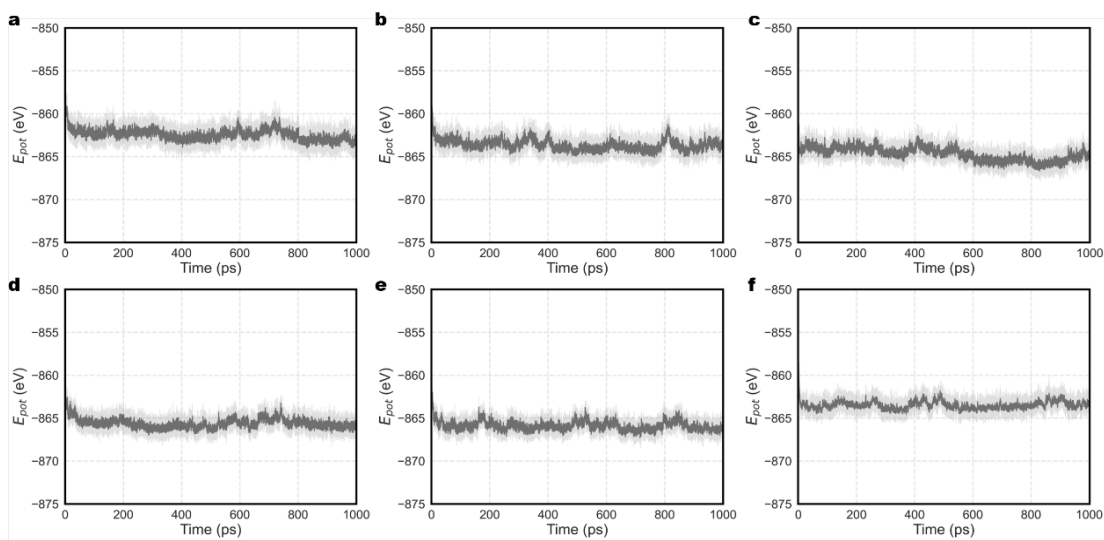

**Supplementary Figure 3. Total Energy of the Au(110)-water interface.** Total energies of the Au(110)-water interfaces were obtained from 1 ns molecular dynamics simulations under different conditions (**a**, 0.2 V vs SHE, **b**, 0 V vs SHE, **c**, -0.2 V vs SHE, **d**, -0.4 V vs SHE, **e**, -0.6 V vs SHE, **f**, constant charge). The gray line represents the results based on each frame, while the black line is the average obtained with a 200 fs moving window.

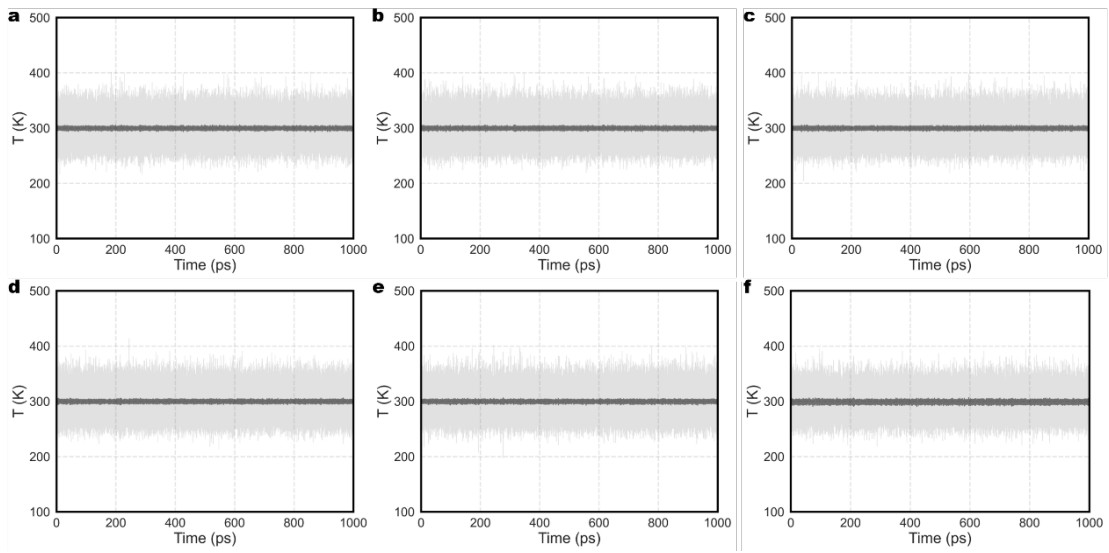

**Supplementary Figure 4. Temperature of the Au(110)-water interface.** Temperatures of the Au(110)-water interfaces were obtained from 1 ns molecular dynamics simulations under different conditions (**a**, 0.2 V vs SHE, **b**, 0 V vs SHE, **c**, -0.2 V vs SHE, **d**, -0.4 V vs SHE, **e**, -0.6 V vs SHE, **f**, constant charge). The gray line represents the results based on each frame, while the black line is the average obtained with a 200 fs moving window.

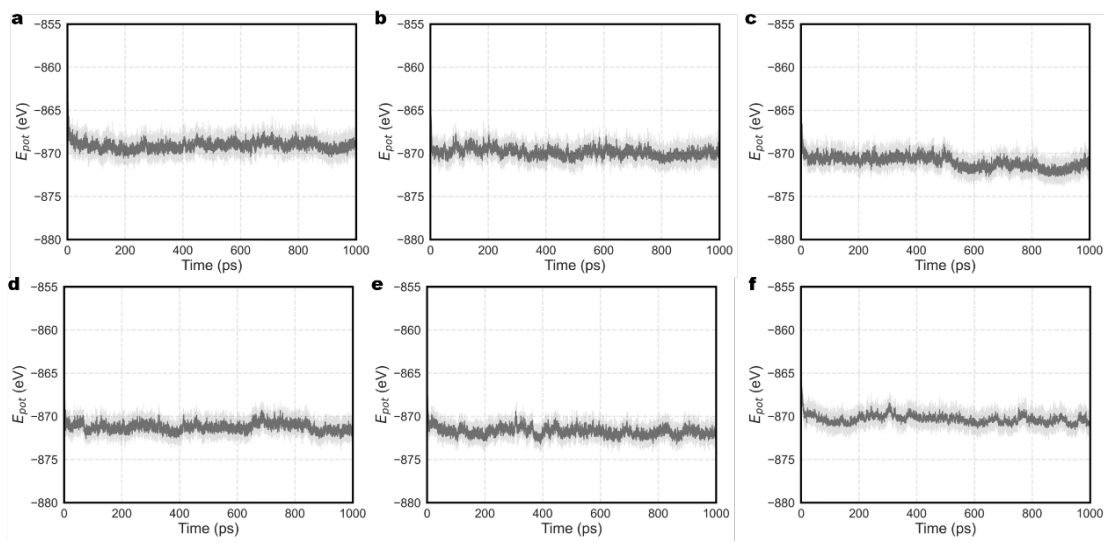

**Supplementary Figure 5. Total Energy of the Au(110)-water interface with 2 $K^+$ ions.** Total energies of the Au(110)-water interfaces with 2 $K^+$ ions were obtained from 1 ns molecular dynamics simulations under different conditions (**a**, 0.2 V vs SHE, **b**, 0 V vs SHE, **c**, -0.2 V vs SHE, **d**, -0.4 V vs SHE, **e**, -0.6 V vs SHE, **f**, constant charge). The gray line represents the results based on each frame, while the black line is the average obtained with a 200 fs moving window.

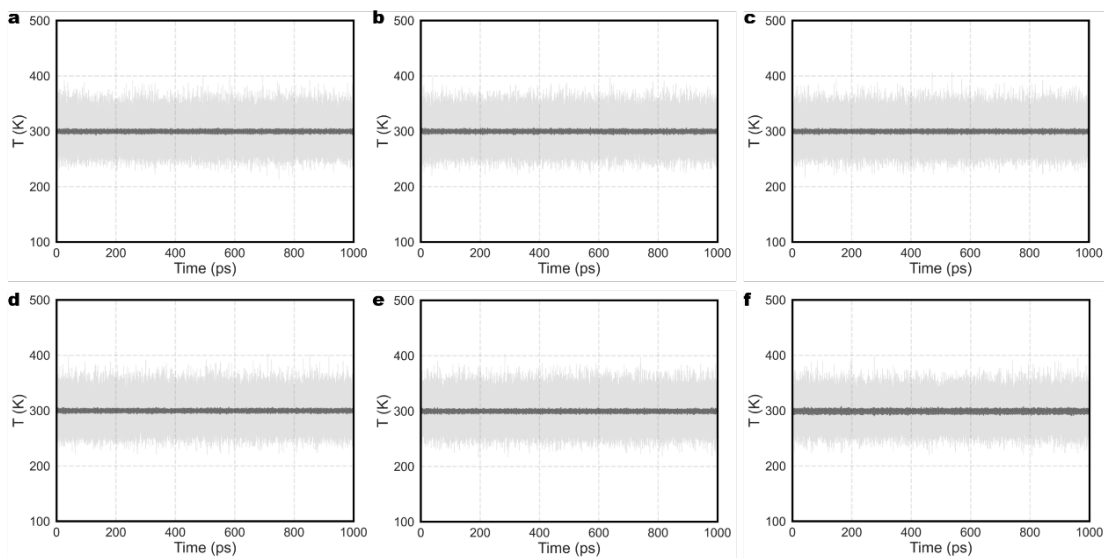

**Supplementary Figure 6. Temperature of the Au(110)-water interface with 2 K$^+$ ions.** Temperatures of the Au(110)-water interfaces with 2 K$^+$ ions were obtained from 1 ns molecular dynamics simulations under different conditions (**a**, 0.2 V vs SHE, **b**, 0 V vs SHE, **c**, -0.2 V vs SHE, **d**, -0.4 V vs SHE, **e**, -0.6 V vs SHE, **f**, constant charge). The gray line represents the results based on each frame, while the black line is the average obtained with a 200 fs moving window.

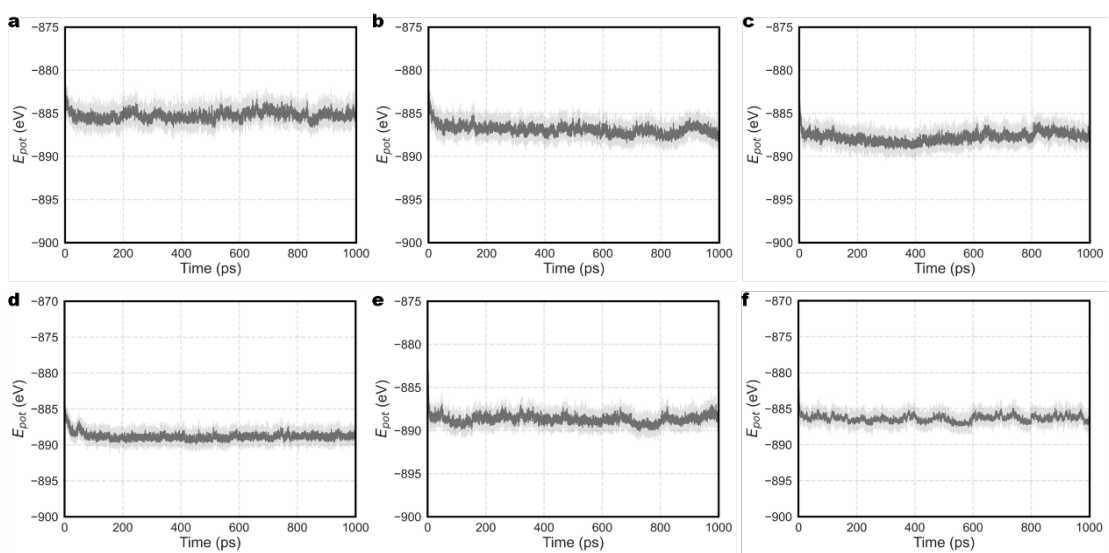

**Supplementary Figure 7. Total Energy of the Au(110)-water interface with a $CO_2$ molecule.** Total energies of the Au(110)-water interfaces with a $CO_2$ molecule were obtained from 1 ns molecular dynamics simulations under different conditions (**a**, 0.2 V vs SHE, **b**, 0 V vs SHE, **c**, -0.2 V vs SHE, **d**, -0.4 V vs SHE, **e**, -0.6 V vs SHE, **f**, constant charge). The gray line represents the results based on each frame, while the black line is the average obtained with a 200 fs moving window.

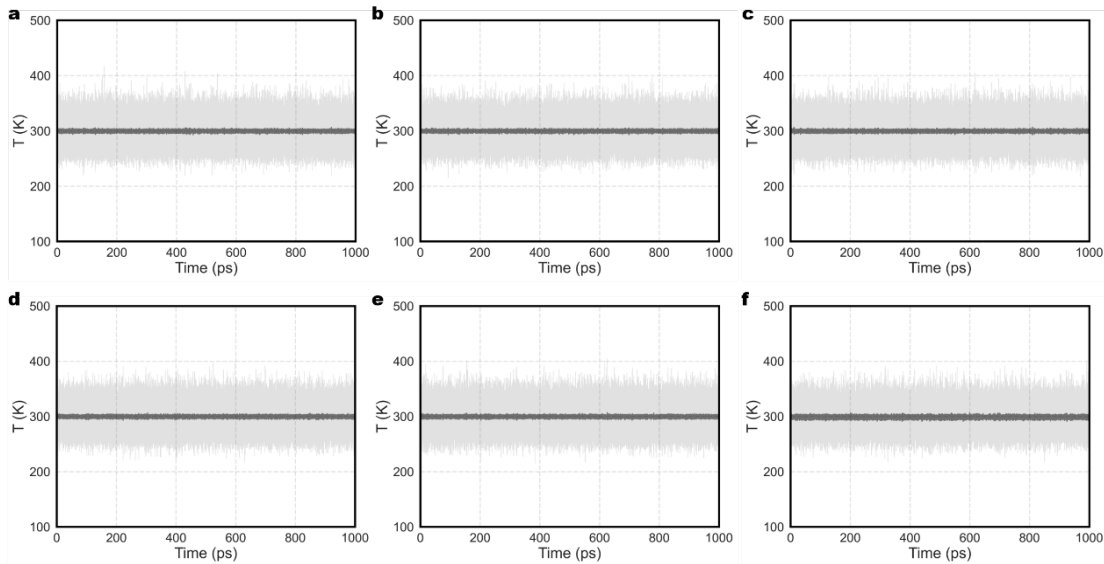

**Supplementary Figure 8. Temperature of the Au(110)-water interface with a $CO_2$ molecule.** Temperatures of the Au(110)-water interfaces with a $CO_2$ molecule were obtained from 1 molecular dynamics simulations under different conditions (**a**, 0.2 V vs SHE, **b**, 0 V vs SHE, **c**, -0.2 V vs SHE, **d**, -0.4 V vs SHE, **e**, -0.6 V vs SHE, **f**, constant charge). The gray line represents the results based on each frame, while the black line is the average obtained with a 200 fs moving window.

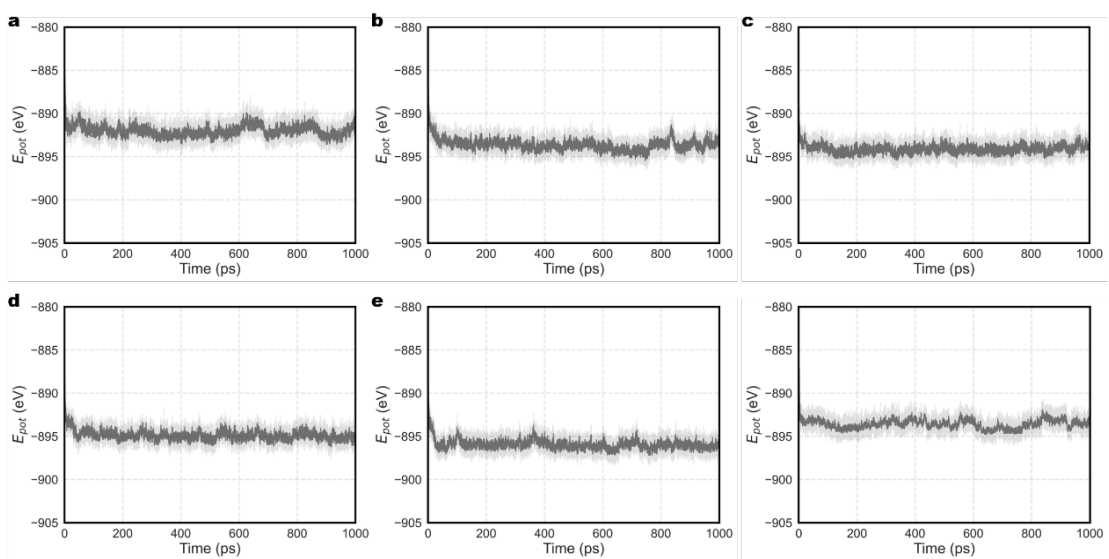

**Supplementary Figure 9. Total Energy of the Au(110)-water interface with 2 $K^+$ ions and a $CO_2$ molecule.** Total energies of the Au(110)-water interfaces with 2 $K^+$ ions and a $CO_2$ molecule were obtained from 1 ns molecular dynamics simulations under different conditions (**a**, 0.2 V vs SHE, **b**, 0 V vs SHE, **c**, -0.2 V vs SHE, **d**, -0.4 V vs SHE, **e**, -0.6 V vs SHE, **f**, constant charge). The gray line represents the results based on each frame, while the black line is the average obtained with a 200 fs moving window.

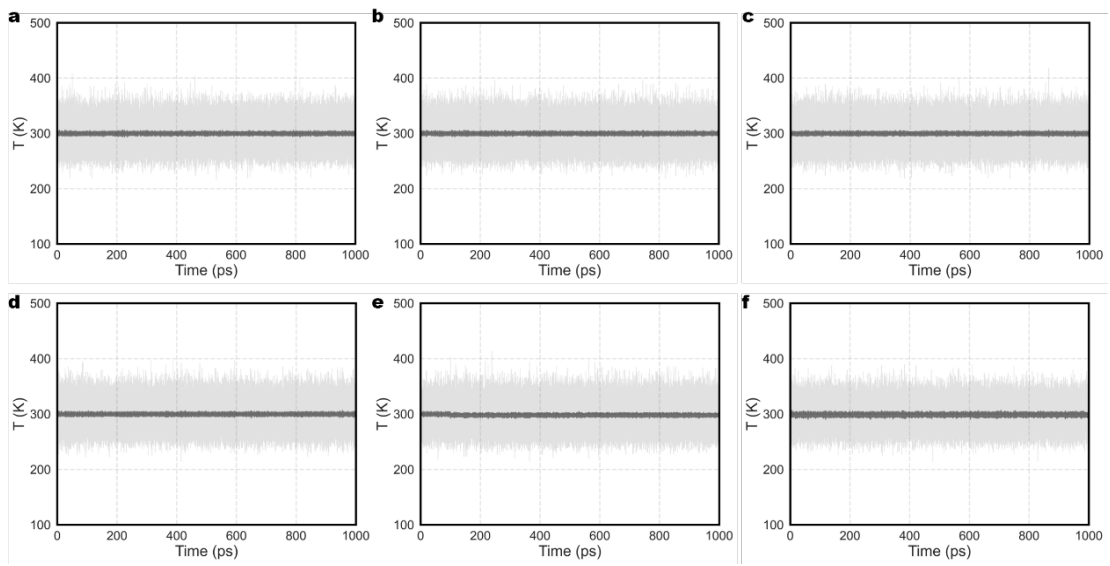

**Supplementary Figure 10. Temperature of the Au(110)-water interface with 2 K$^+$ ions and a CO$_2$ molecule.** Temperatures of the Au(110)-water interfaces with 2 K$^+$ ions and a CO$_2$ molecule were obtained from 1 ns molecular dynamics simulations under different conditions (**a**, 0.2 V vs SHE, **b**, 0 V vs SHE, **c**, -0.2 V vs SHE, **d**, -0.4 V vs SHE, **e**, -0.6 V vs SHE, **f**, constant charge). The gray line represents the results based on each frame, while the black line is the average obtained with a 200 fs moving window.

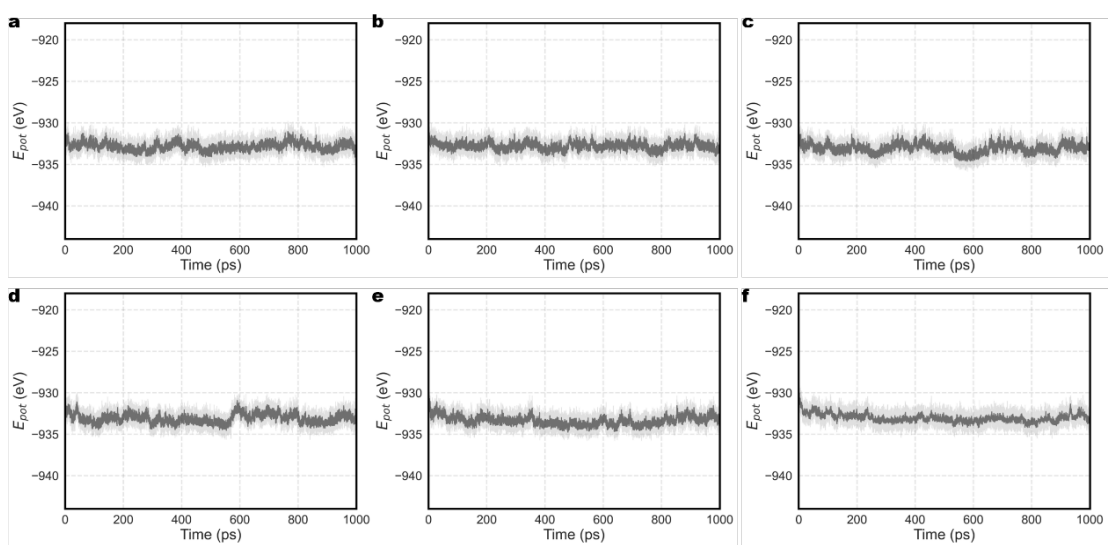

**Supplementary Figure 11. Total Energy of the Au(111)-water interface.** Total energies of the Au(111)-water interfaces were obtained from 1 ns molecular dynamics simulations under different conditions (**a**, 0.2 V vs SHE, **b**, 0 V vs SHE, **c**, -0.2 V vs SHE, **d**, -0.4 V vs SHE, **e**, -0.6 V vs SHE, **f**, constant charge). The gray line represents the results based on each frame, while the black line is the average obtained with a 200 fs moving window.

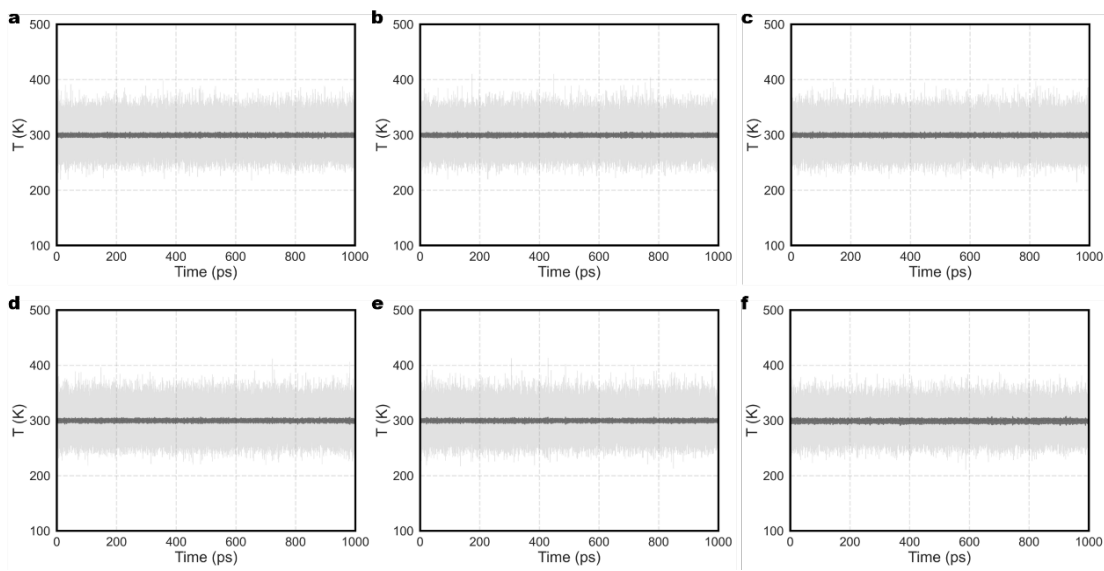

**Supplementary Figure 12. Temperature of the Au(111)-water.** Temperatures of the Au(111)-water interfaces were obtained from 1 ns molecular dynamics simulations under different conditions (**a**, 0.2 V vs SHE, **b**, 0 V vs SHE, **c**, -0.2 V vs SHE, **d**, -0.4 V vs SHE, **e**, -0.6 V vs SHE, **f**, constant charge). The gray line represents the results based on each frame, while the black line is the average obtained with a 200 fs moving window.

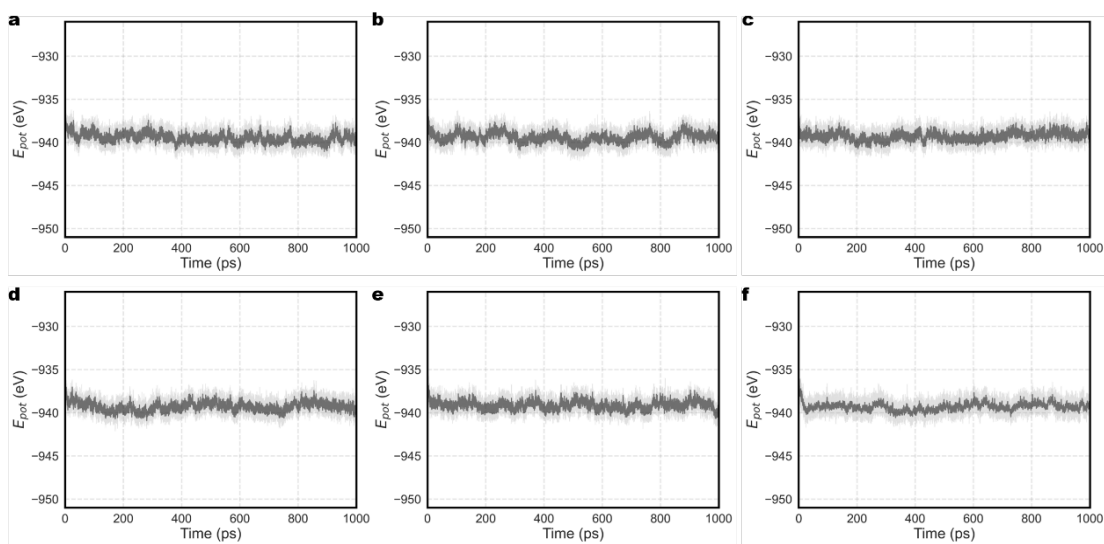

**Supplementary Figure 13. Total Energy of the Au(111)-water with 2 $K^+$ ions.** Total energies of the Au(111)-water interfaces with 2 $K^+$ ions were obtained from 1 ns molecular dynamics simulations under different conditions (**a**, 0.2 V vs SHE, **b**, 0 V vs SHE, **c**, -0.2 V vs SHE, **d**, -0.4 V vs SHE, **e**, -0.6 V vs SHE, **f**, constant charge). The gray line represents the results based on each frame, while the black line is the average obtained with a 200 fs moving window.

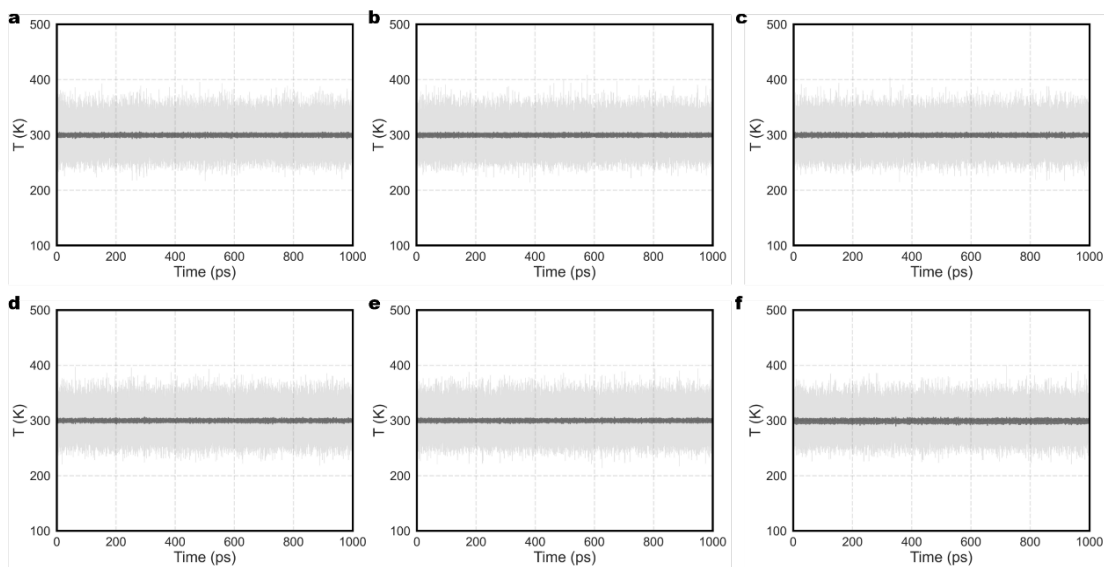

**Supplementary Figure 14. Temperature of the Au(111)-water with 2 K⁺ ions.** Temperatures of the Au(111)-water interfaces with 2 $K^+$ ions were obtained from 1 ns molecular dynamics simulations under different applied potentials (**a**, 0.2 V vs SHE, **b**, 0 V vs SHE, **c**, -0.2 V vs SHE, **d**, -0.4 V vs SHE, **e**, -0.6 V vs SHE, **f**, constant charge). The gray line represents the results based on each frame, while the black line is the average obtained with a 200 fs moving window.

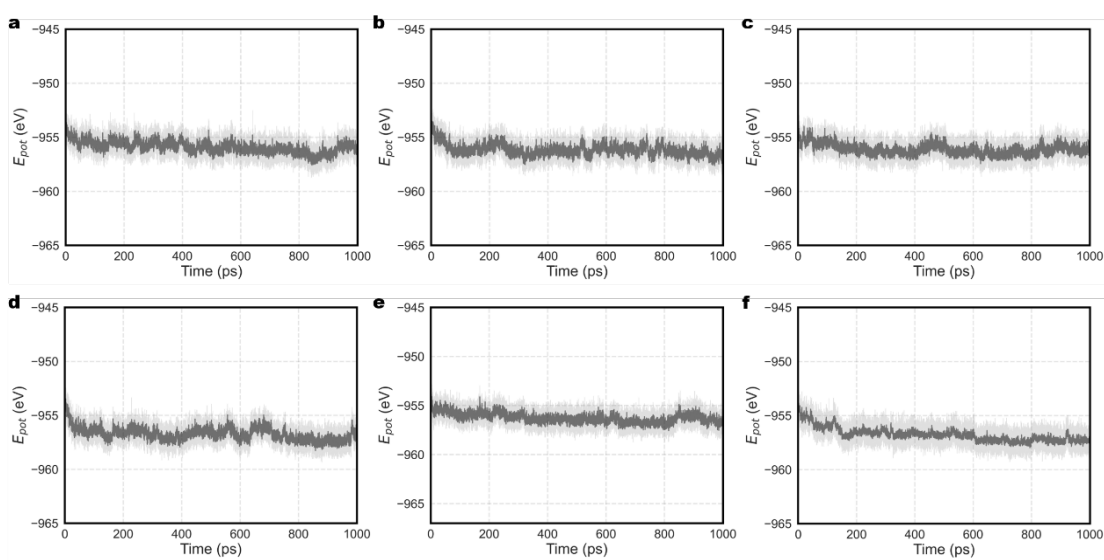

**Supplementary Figure 15. Total Energy of the Au(111)-water interface with a $CO_2$ molecule.** Total energies of the Au(111)-water interfaces with a $CO_2$ molecule were obtained from 1 ns molecular dynamics simulations under different conditions (**a**, 0.2 V vs SHE, **b**, 0 V vs SHE, **c**, -0.2 V vs SHE, **d**, -0.4 V vs SHE, **e**, -0.6 V vs SHE, **f**, constant charge). The gray line represents the results based on each frame, while the black line is the average obtained with a 200 fs moving window.

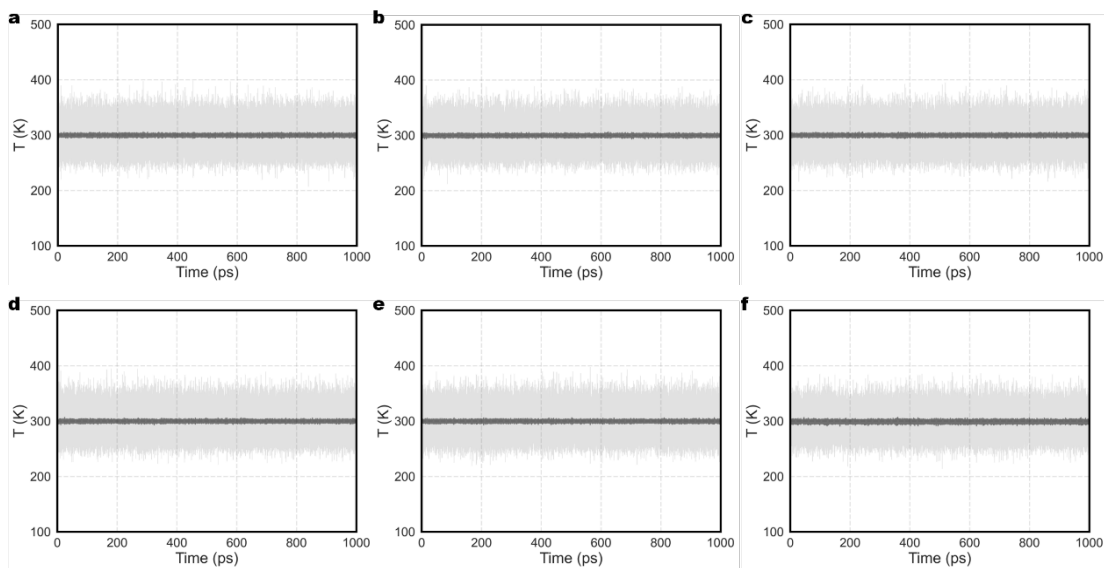

**Supplementary Figure 16. Temperature of the Au(111)-water interface with a CO2 molecule.** Temperatures of the Au(111)-water interfaces with a $CO_2$ molecule were obtained from 1 ns molecular dynamics simulations under different conditions (**a**, 0.2 V vs SHE, **b**, 0 V vs SHE, **c**, -0.2 V vs SHE, **d**, -0.4 V vs SHE, **e**, -0.6 V vs SHE, **f**, constant charge). The gray line represents the results based on each frame, while the black line is the average obtained with a 200 fs moving window.

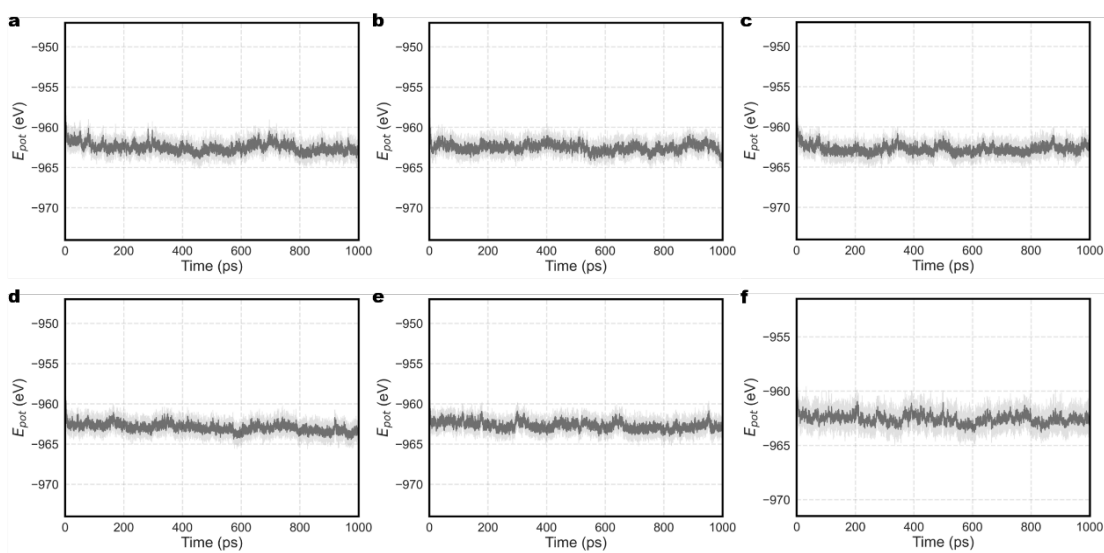

**Supplementary Figure 17. Total Energy of the Au(111)-water interface with 2 $K^+$ ions and a $CO_2$ molecule.** Total energies of the Au(111)-water interfaces with 2 $K^+$ ions and a $CO_2$ molecule were obtained from 1 ns molecular dynamics simulations under different conditions (**a**, 0.2 V vs SHE, **b**, 0 V vs SHE, **c**, -0.2 V vs SHE, **d**, -0.4 V vs SHE, **e**, -0.6 V vs SHE, **f**, constant charge). The gray line represents the results based on each frame, while the black line is the average obtained with a 200 fs moving window.

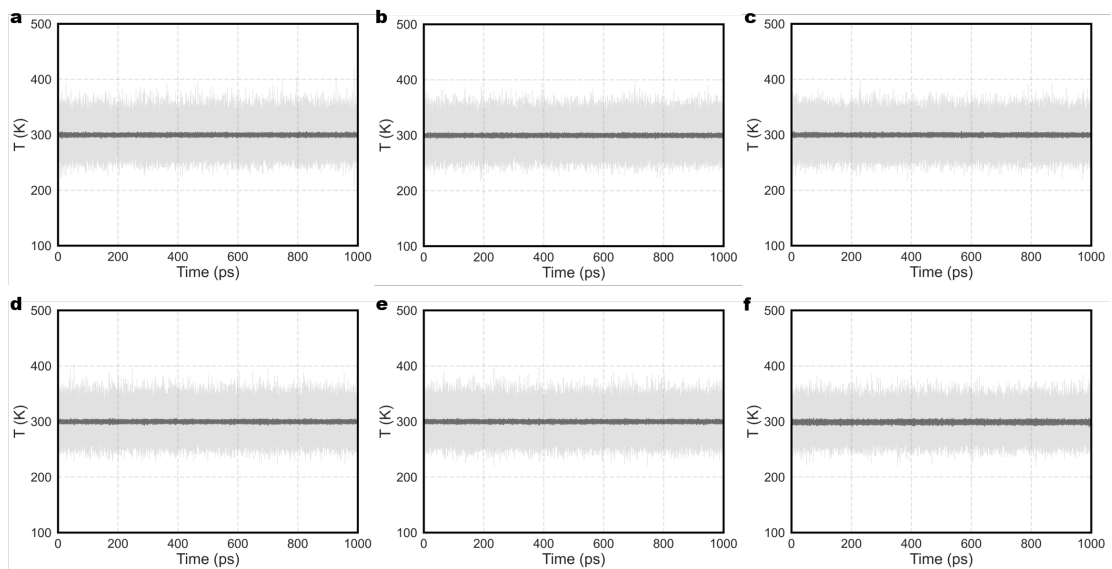

**Supplementary Figure 18. Temperature of the Au(111)-water interface with 2 K⁺ ions and a CO₂ molecule.** Temperatures of the Au(111)-water interfaces with 2 $K^+$ ions and a $CO_2$ molecule were obtained from 1 ns molecular dynamics simulations under different conditions (**a**, 0.2 V vs SHE, **b**, 0 V vs SHE, **c**, -0.2 V vs SHE, **d**, -0.4 V vs SHE, **e**, -0.6 V vs SHE, **f**, constant charge). The gray line represents the results based on each frame, while the black line is the average obtained with a 200 fs moving window.

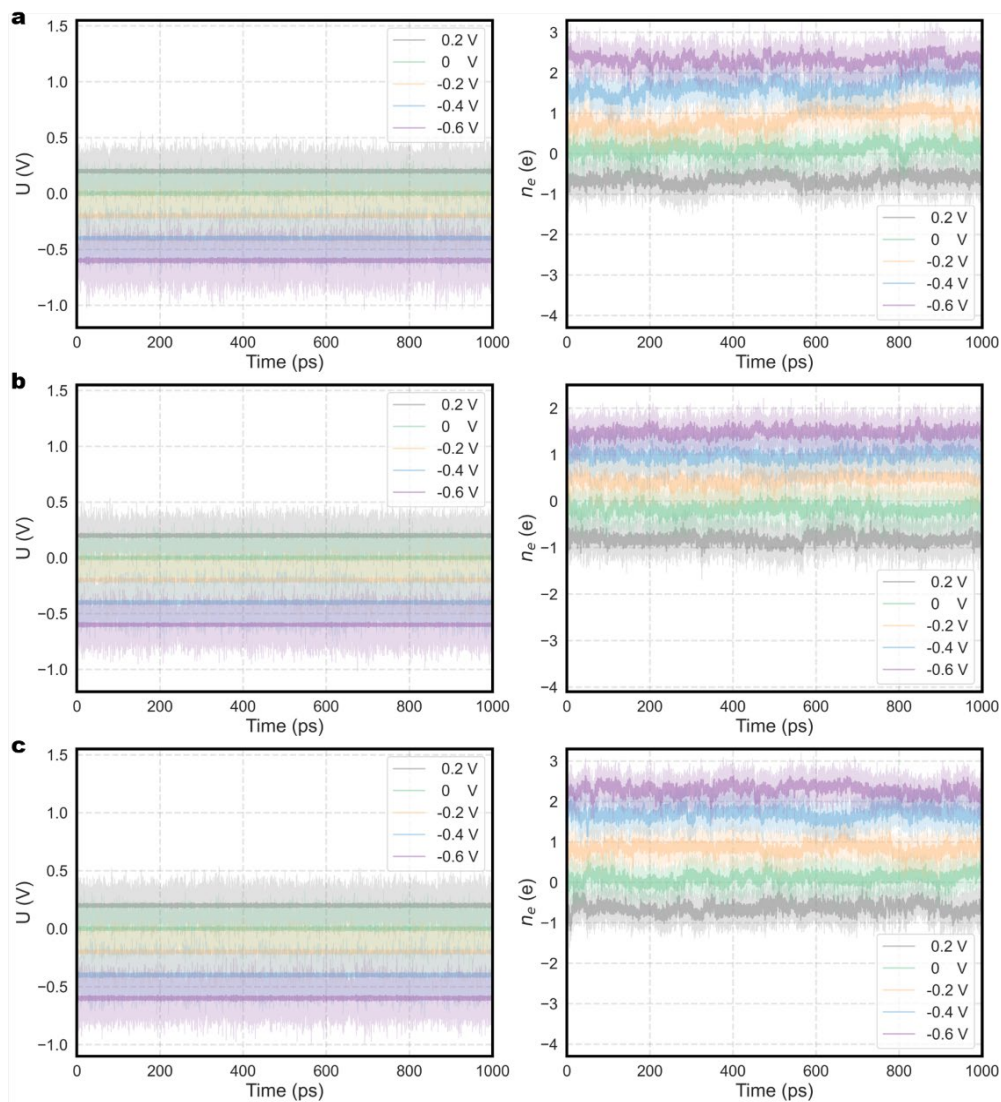

**Supplementary Figure 19. Applied potential and net electron count of Au-water interface.** Applied potentials and net electrons counts of **a**, the Au(110)-water interface, **b**, the Au(110)-water interface with 2 $K^+$ ions and **c**, the Au(110)-water interface with a $CO_2$ molecule. All results were obtained under a 1 ns constant potential (0.2, 0, -0.2, -0.4, -0.6 V vs SHE) molecular dynamics simulation. Applied potential is shown on the left, and net electron count is shown on the right. The light-colored line represents the results based on each frame, while the dark-colored line is the average obtained with a 200 fs moving window.

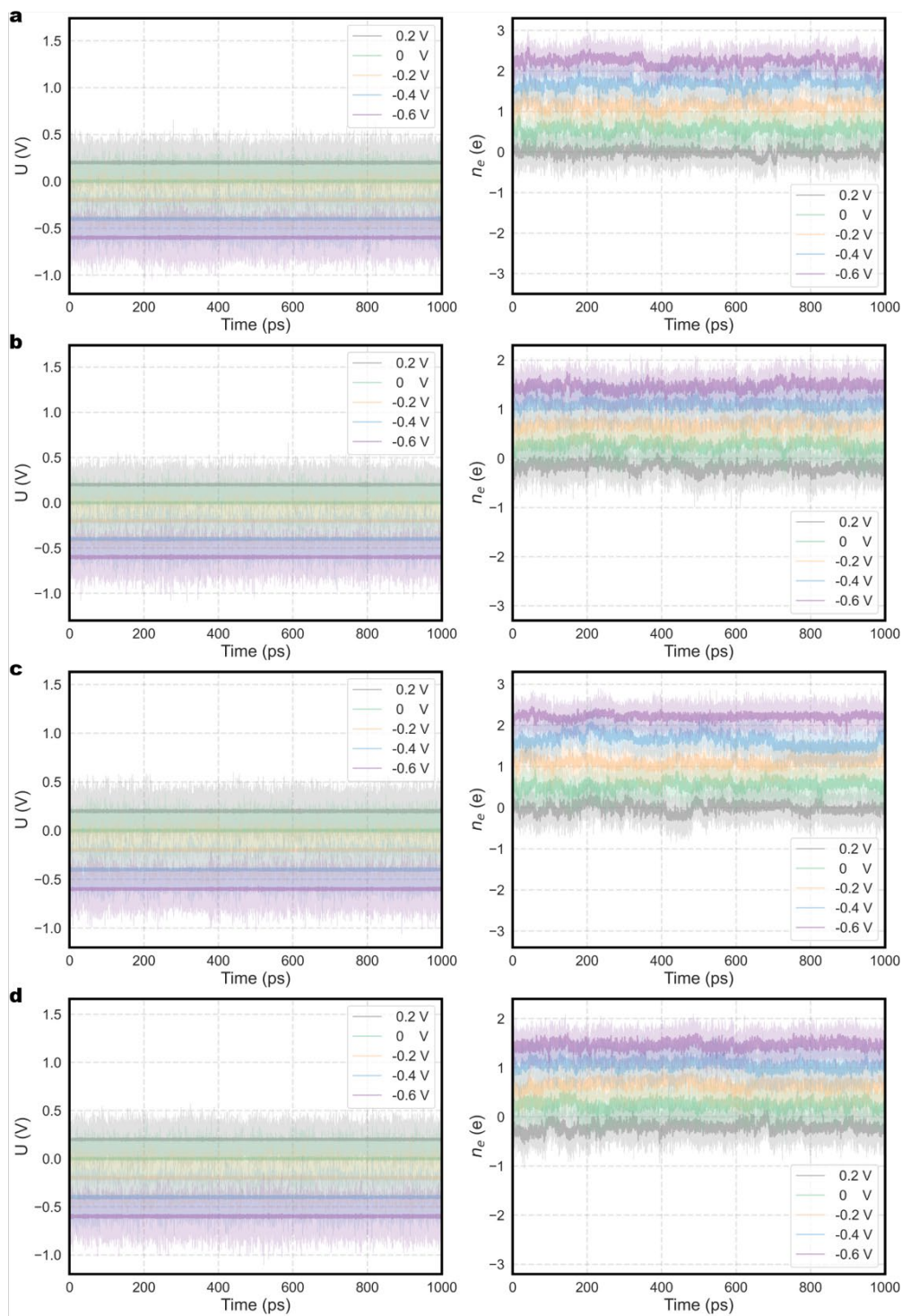

**Supplementary Figure 20. Applied potential and net electron count of Au-water interface.** Applied potentials and net electrons counts of **a**, the Au(111)-water interface, **b**, the Au(111)-water interface with 2 $K^+$ ions, **c**, the Au(111)-water interface with a $CO_2$ molecule, and **d**, the Au(111)-water interface with 2 $K^+$ ions and a $CO_2$ molecule. All results were obtained under a 1 ns constant potential (0.2, 0, -0.2, -0.4, -0.6 V vs SHE) molecular dynamics simulation. Applied potential is shown on the left, and net electron count is shown on the right. The light-colored line represents the results based on each frame, while the dark-colored line is the average obtained with a 200 fs moving window.

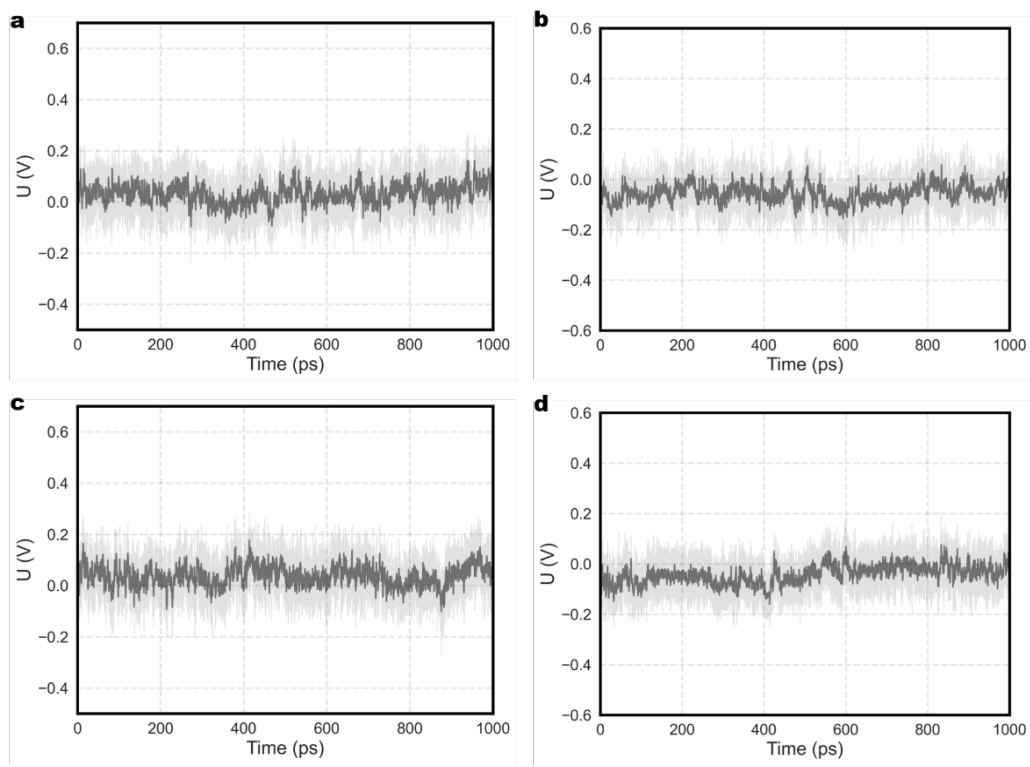

**Supplementary Figure 21. Applied potential of Au-water interface under a 1 ns constant charge molecular dynamics simulation.** Applied potentials of **a**, the Au(110)-water interface, **b**, the Au(110)-water interface with 2 $K^+$ ions, **c**, the Au(110)-water interface with a $CO_2$ molecule, and **d**, the Au(110)-water interface with 2 $K^+$ ions and a $CO_2$ molecule. The light-colored line represents the results based on each frame, while the dark-colored line is the average obtained with a 200 fs moving window.

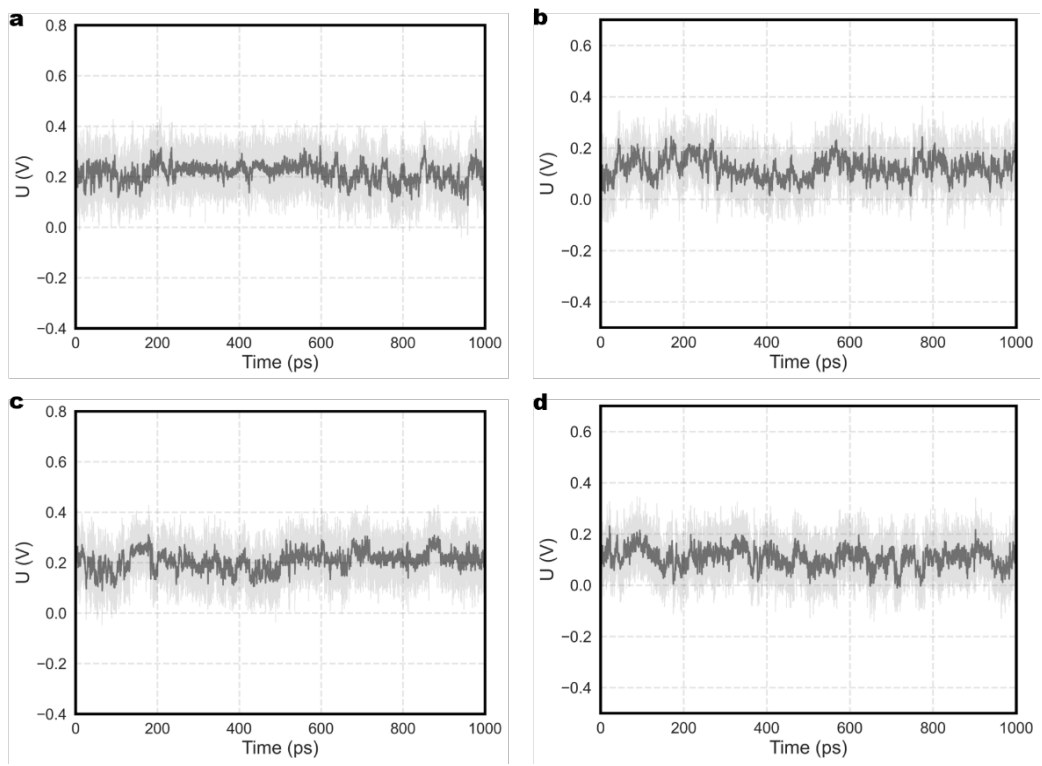

**Supplementary Figure 22. Applied potential of Au-water interface under a 1 ns constant charge molecular dynamics simulation.** Applied potentials of **a**, the Au(111)-water interface, **b**, the Au(111)-water interface with 2 $K^+$ ions, **c**, the Au(111)-water interface with a $CO_2$ molecule, and **d**, the Au(111)-water interface with 2 $K^+$ ions and a $CO_2$ molecule. The light-colored line represents the results based on each frame, while the dark-colored line is the average obtained with a 200 fs moving window.

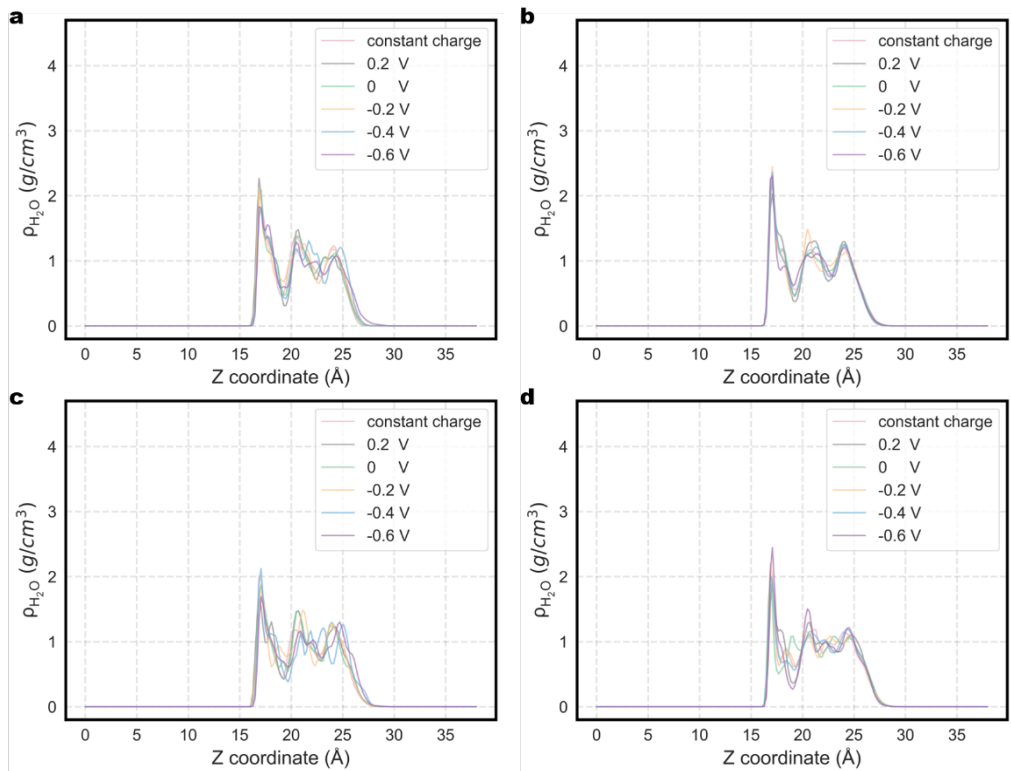

**Supplementary Figure 23. The distribution of water density along the z-axis normal to the Au(110)-water interface.** The profiles of water density along the z-axis normal to **a**, the Au(110)-water interface, **b**, the Au(110)-water interface with 2 $K^+$ ions, **c**, the Au(110)-water interface with a $CO_2$ molecule, and **d**, the Au(110)-water interface with 2 $K^+$ ions and a $CO_2$ molecule. The profiles of the water density were calculated based on the corresponding oxygen density. All results were obtained from 1 ns molecular dynamics simulations under different conditions (constant charge, 0.2 V vs SHE, 0 V vs SHE, -0.2 V vs SHE, -0.4 V vs SHE, -0.6 V vs SHE).

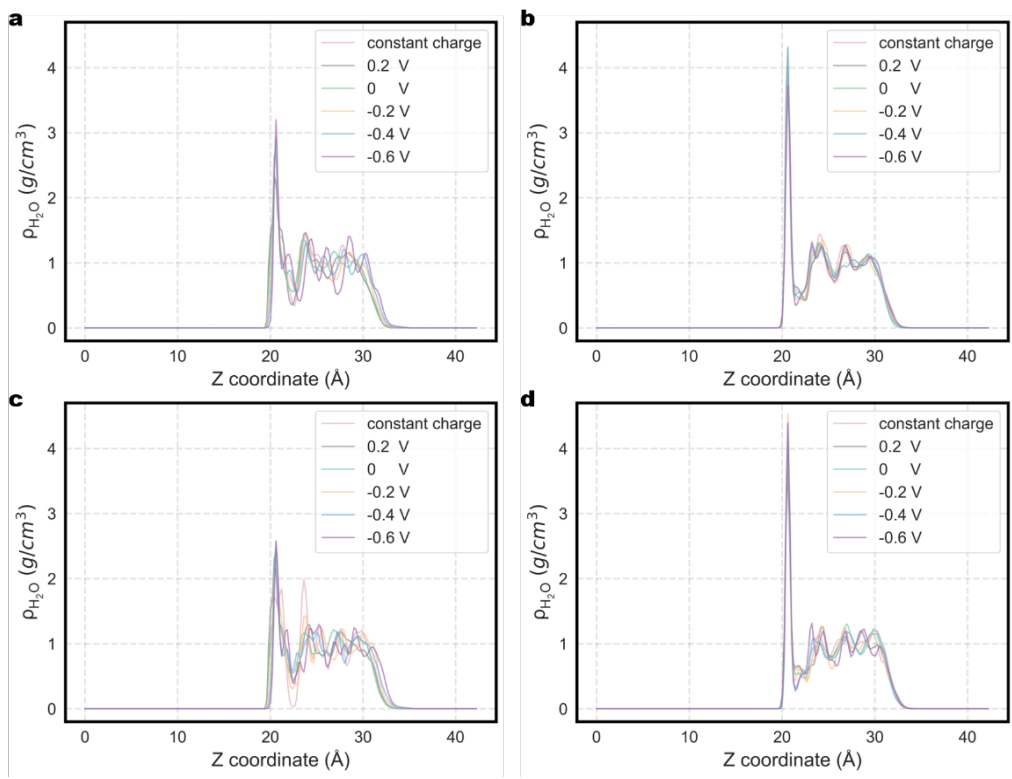

**Supplementary Figure 24. The distribution of water density along the z-axis normal to the Au(111)-water interface.** The profiles of water density along the z-axis normal to **a**, the Au(111)-water interface, **b**, the Au(111)-water interface with 2 $K^+$ ions, **c**, the Au(111)-water interface with a $CO_2$ molecule, and **d**, the Au(111)-water interface with 2 $K^+$ ions and a $CO_2$ molecule. The profiles of the water density were calculated based on the corresponding oxygen density. All results were obtained from 1 ns molecular dynamics simulations under different conditions (constant charge, 0.2 V vs SHE, 0 V vs SHE, -0.2 V vs SHE, -0.4 V vs SHE, -0.6 V vs SHE).

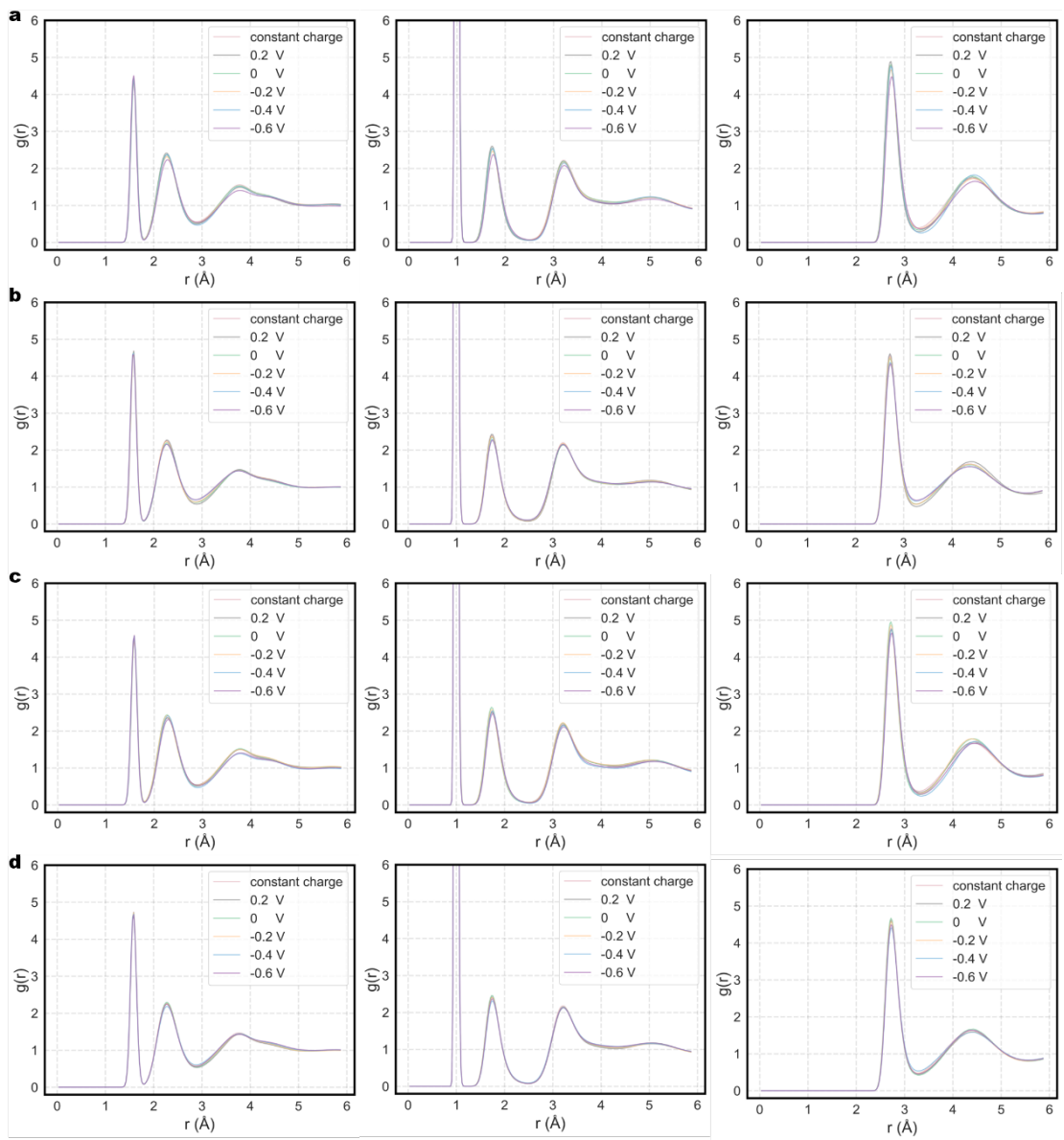

**Supplementary Figure 25. Comparison of radial distribution functions for the Au(110)-water interface.** Radial distribution functions of **a**, the Au(110)-water interface, **b**, the Au(110)-water interface with 2 $K^+$ ions, **c**, the Au(110)-water interface with a $CO_2$ molecule, and **d**, the Au(110)-water interface with 2 $K^+$ ions and a $CO_2$ molecule. Radial distribution functions are shown for H-H pairs on the left, O-H pairs in the middle, and O-O pairs on the right. All results were obtained from 1 ns molecular dynamics simulations under different conditions (constant charge, 0.2 V vs SHE, 0 V vs SHE, -0.2 V vs SHE, -0.4 V vs SHE, -0.6 V vs SHE).

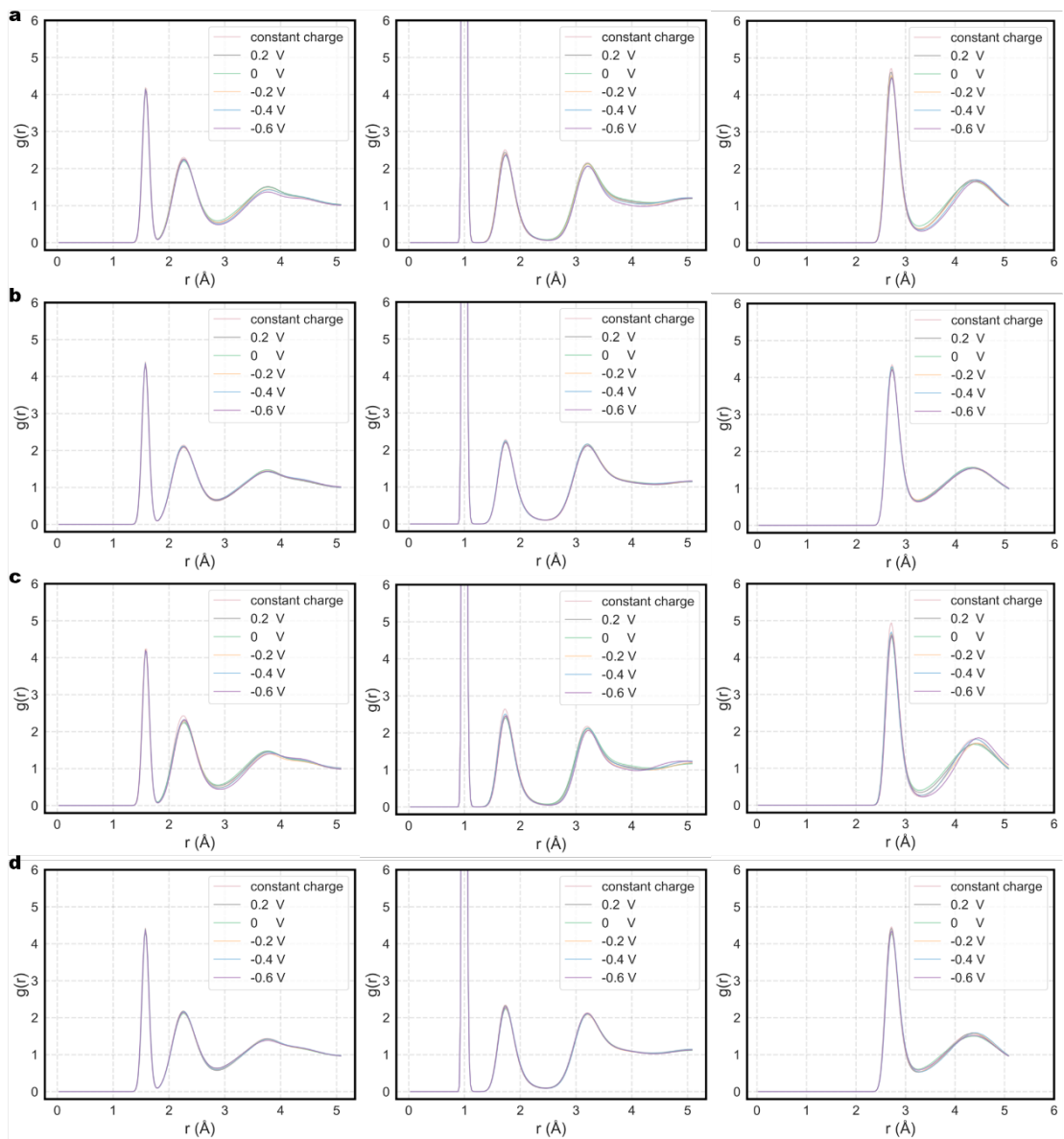

**Supplementary Figure 26. Comparison of radial distribution functions for the Au(111)-water interface.** Radial distribution functions of **a**, the Au(111)-water interface, **b**, the Au(111)-water interface with 2 $K^+$ ions, **c**, the Au(111)-water interface with a $CO_2$ molecule, and **d**, the Au(111)-water interface with 2 $K^+$ ions and a $CO_2$ molecule. Radial distribution functions are shown for H-H pairs on the left, O-H pairs in the middle, and O-O pairs on the right. All results were obtained from 1 ns molecular dynamics simulations under different conditions (constant charge, 0.2 V vs SHE, 0 V vs SHE, -0.2 V vs SHE, -0.4 V vs SHE, -0.6 V vs SHE).

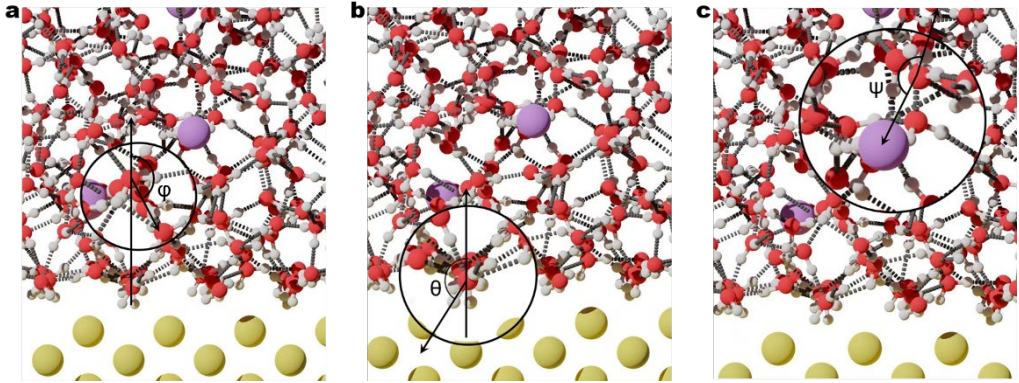

**Supplementary Figure 27. The schematic of the defined angle in interfacial water. a**, The angle φ between the bisector of the water molecule and the surface normal, **b**, the angle θ between the O-H bond of the water molecule and the surface normal, and **c**, the angle ψ between the bisector of the water molecule and the vector from the oxygen atom of the water molecule to the potassium ion.

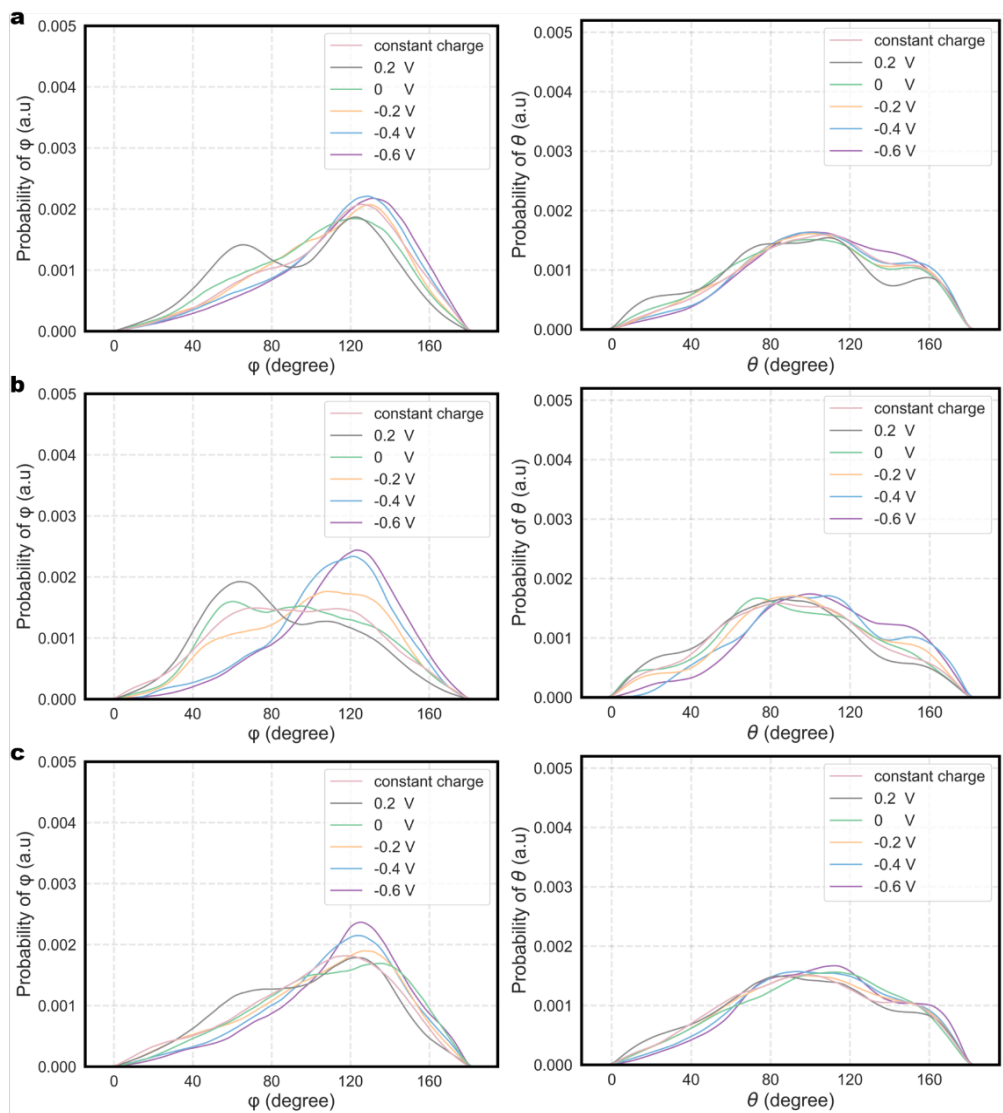

**Supplementary Figure 28. Comparison of probability distribution profiles for angles φ and θ at the Au(110)-water interface.** Profiles for angles φ (left) and θ (right) of the interfacial water at **a**, the Au(110)-water interface with 2 K$^+$ ions, **b**, the Au(110)-water interface with a CO$_2$ molecule, and **c**, the Au(110)-water interface with 2 K$^+$ ions and a CO$_2$ molecule. The specific definitions of these angles are provided in Supplementary Figure 27 a and b. All results were obtained from 1 ns molecular dynamics simulations under different conditions (constant charge, 0.2 V vs SHE, 0 V vs SHE, -0.2 V vs SHE, -0.4 V vs SHE, -0.6 V vs SHE).

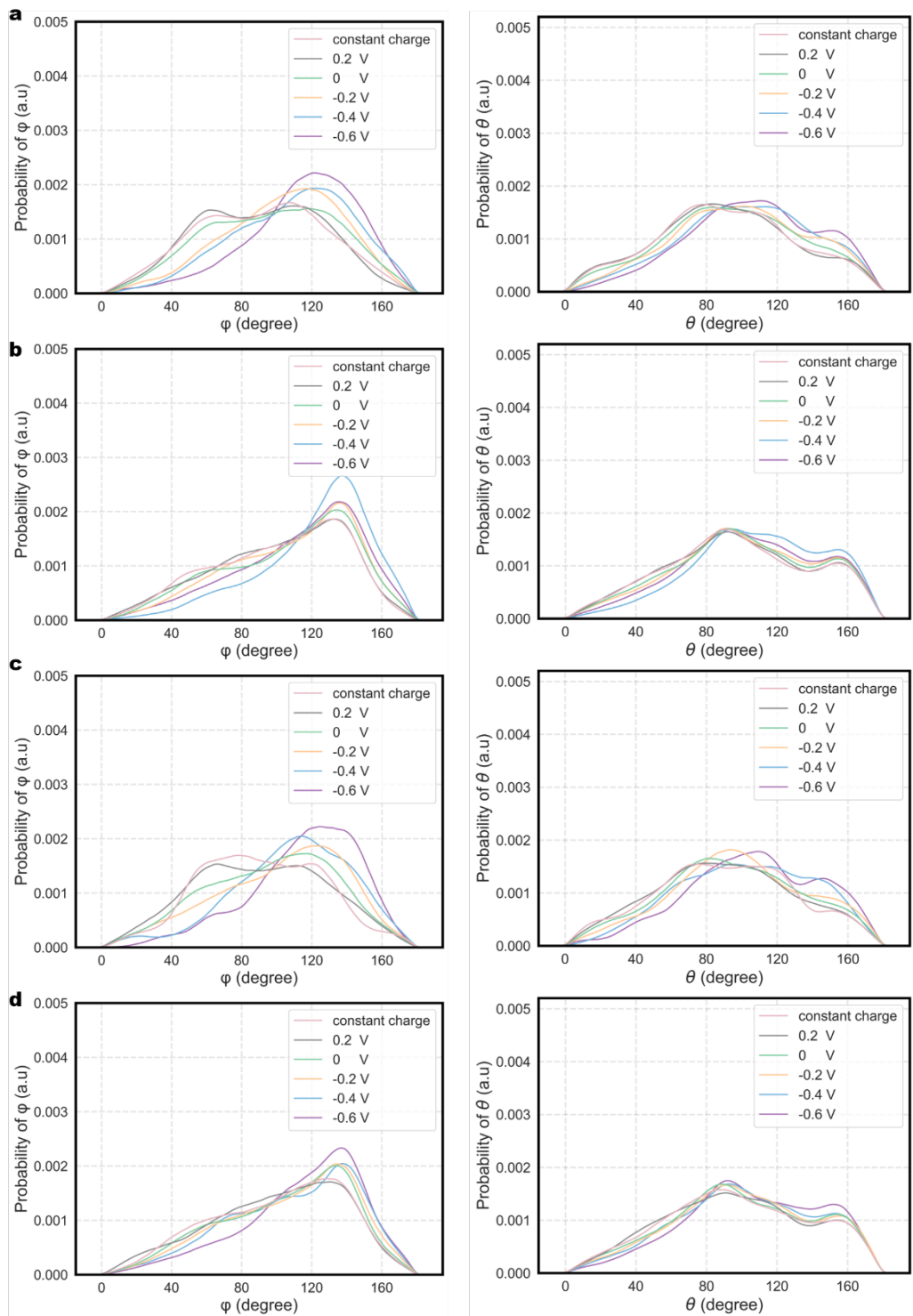

**Supplementary Figure 29. Comparison of probability distribution profiles for angles φ and θ at the Au(111)-water interface.** Profiles for angles φ (left) and θ (right) of the interfacial water at **a**, the Au(111)-water interface, **b**, the Au(111)-water interface with 2 $K^+$ ions, **c**, the Au(111)-water interface with a CO2 molecule, and **d**, the Au(111)-water interface with 2 $K^+$ ions and a CO2 molecule. The specific definitions of these angles are provided in Supplementary Figure 27 a and b. All results were obtained from 1 ns molecular dynamics simulations under different conditions (constant charge, 0.2 V vs SHE, 0 V vs SHE, -0.2 V vs SHE, -0.4 V vs SHE, -0.6 V vs SHE).

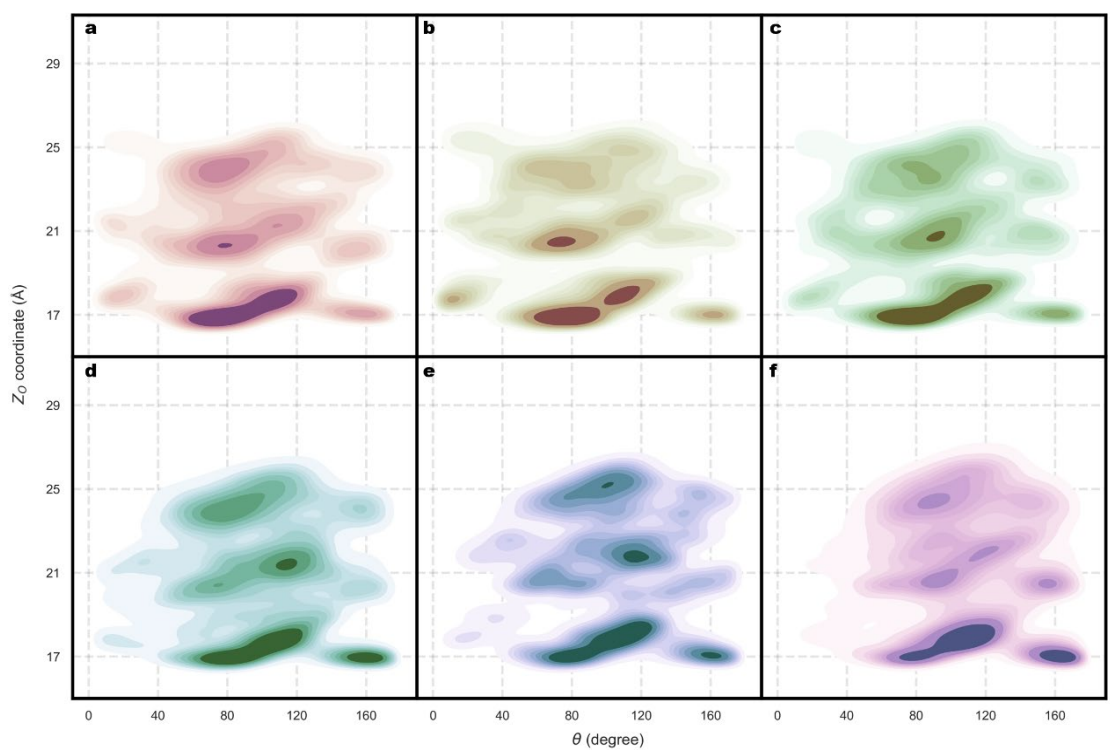

**Supplementary Figure 30. Probability distribution profile for angle θ and the Z-coordinate of the corresponding oxygen atom in water molecule at the Au(110)-water interface.** Probability distribution profiles for angle θ and the Z-coordinate of the corresponding oxygen atom in water molecule at the Au(110)-water interface were obtained from 1 ns molecular dynamics simulations under different conditions (**a**, constant charge, **b**, 0.2 V vs SHE, **c**, 0 V vs SHE, **d**, -0.2 V vs SHE, **e**, -0.4 V vs SHE, **f**, -0.6 V vs SHE). The specific definition of the angle θ are provided in Supplementary Figure 27 b.

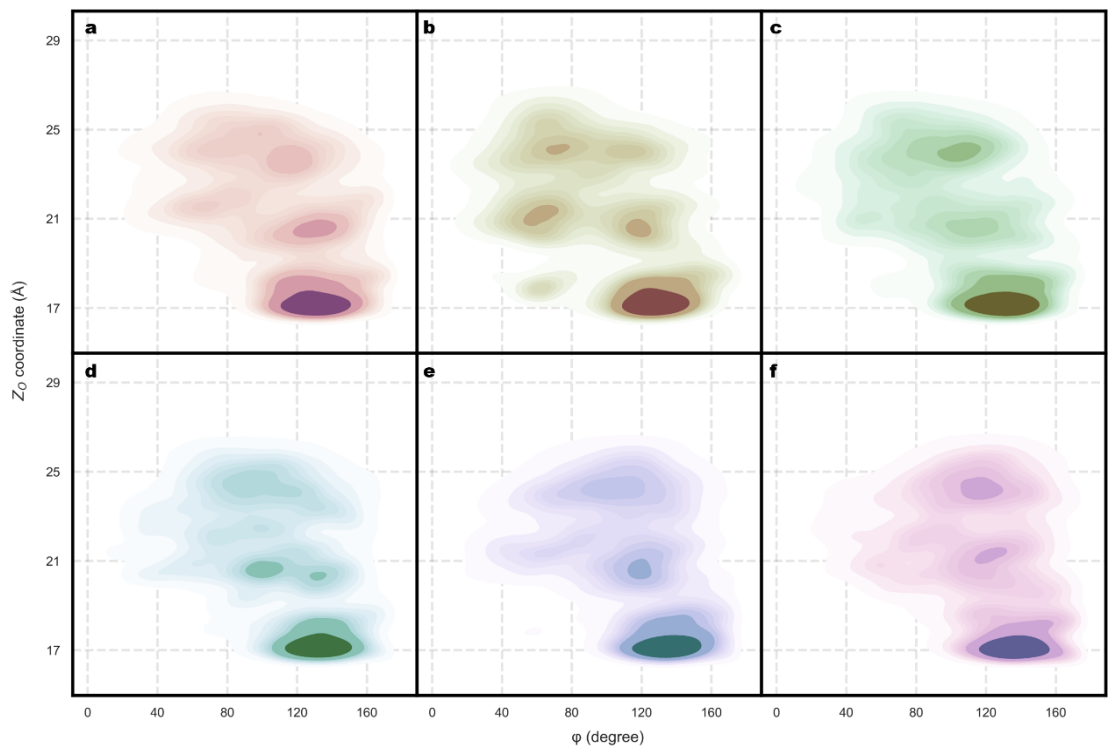

**Supplementary Figure 31. Probability distribution profile for angle φ and the Z-coordinate of the corresponding oxygen atom in water molecule at the Au(110)-water interface with 2 $K^+$ ions.** Probability distribution profiles for angle φ and the Z-coordinate of the corresponding oxygen atom in water molecule at the Au(110)-water interface with 2 $K^+$ ions were obtained from 1 ns molecular dynamics simulations under different conditions (**a**, constant charge, **b**, 0.2 V vs SHE, **c**, 0 V vs SHE, **d**, -0.2 V vs SHE, **e**, -0.4 V vs SHE, **f**, -0.6 V vs SHE). The specific definition of the angle φ are provided in Supplementary Figure 27 a.

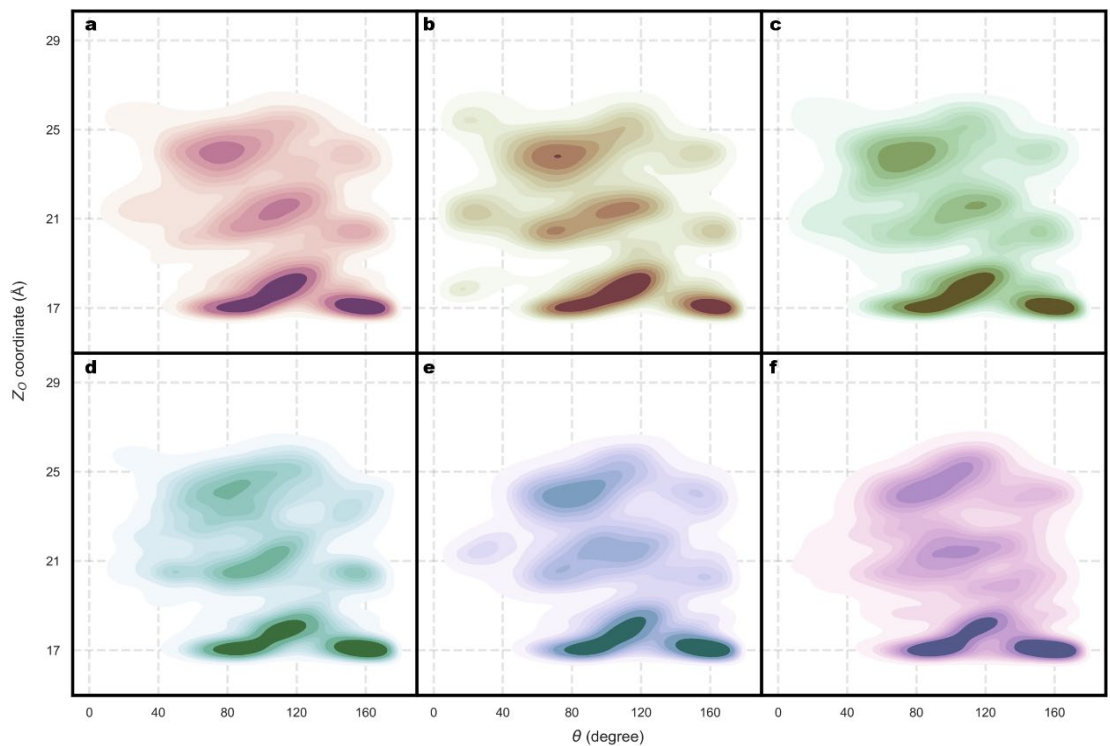

**Supplementary Figure 32. Probability distribution profile for angle θ and the Z-coordinate of the corresponding oxygen atom in water molecule at the Au(110)-water interface with 2 K$^+$ ions.** Probability distribution profiles for angle θ and the Z-coordinate of the corresponding oxygen atom in water molecule at the Au(110)-water interface with 2 K$^+$ were obtained from 1 ns molecular dynamics simulations under different conditions (**a**, constant charge, **b**, 0.2 V vs SHE, **c**, 0 V vs SHE, **d**, -0.2 V vs SHE, **e**, -0.4 V vs SHE, **f**, -0.6 V vs SHE). The specific definition of the angle θ are provided in Supplementary Figure 27 b.

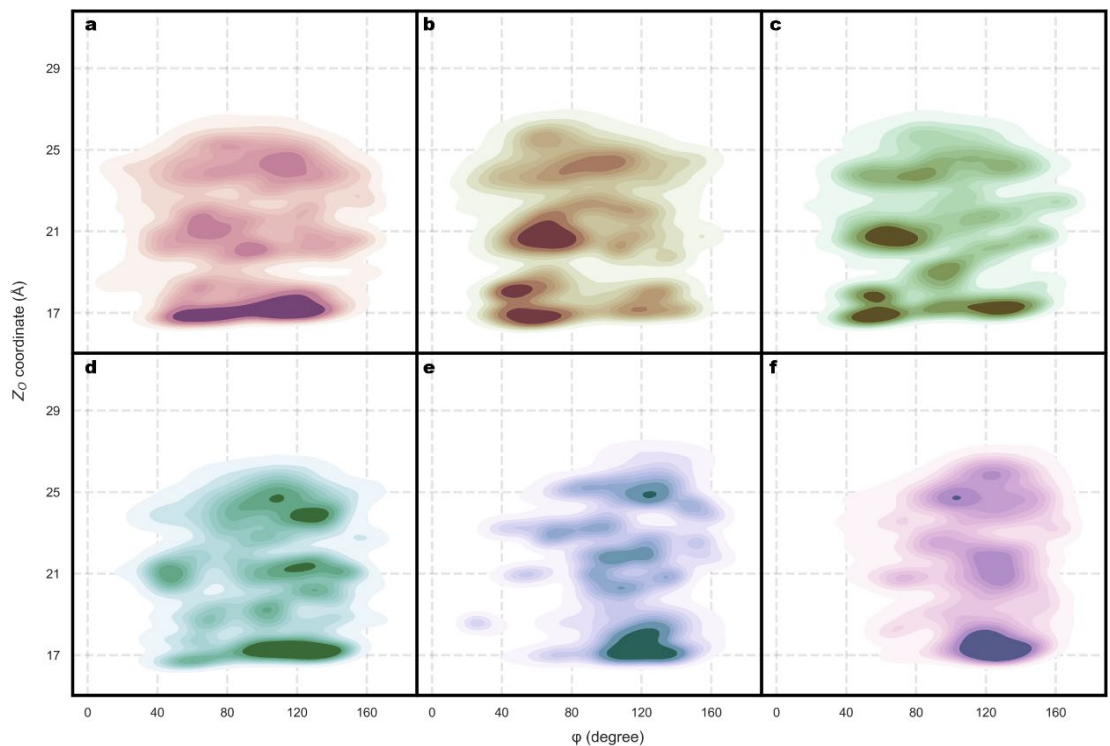

**Supplementary Figure 33. Probability distribution profile for angle φ and the Z-coordinate of the corresponding oxygen atom in water molecule at the Au(110)-water interface with a $CO_2$ molecule.** Probability distribution profiles for angle φ and the Z-coordinate of the corresponding oxygen atom in water molecule at the Au(110)-water interface with a $CO_2$ molecule were obtained from 1 ns molecular dynamics simulations under different conditions (**a**, constant charge, **b**, 0.2 V vs SHE, **c**, 0 V vs SHE, **d**, -0.2 V vs SHE, **e**, -0.4 V vs SHE, **f**, -0.6 V vs SHE). The specific definition of the angle φ are provided in Supplementary Figure 27 a.

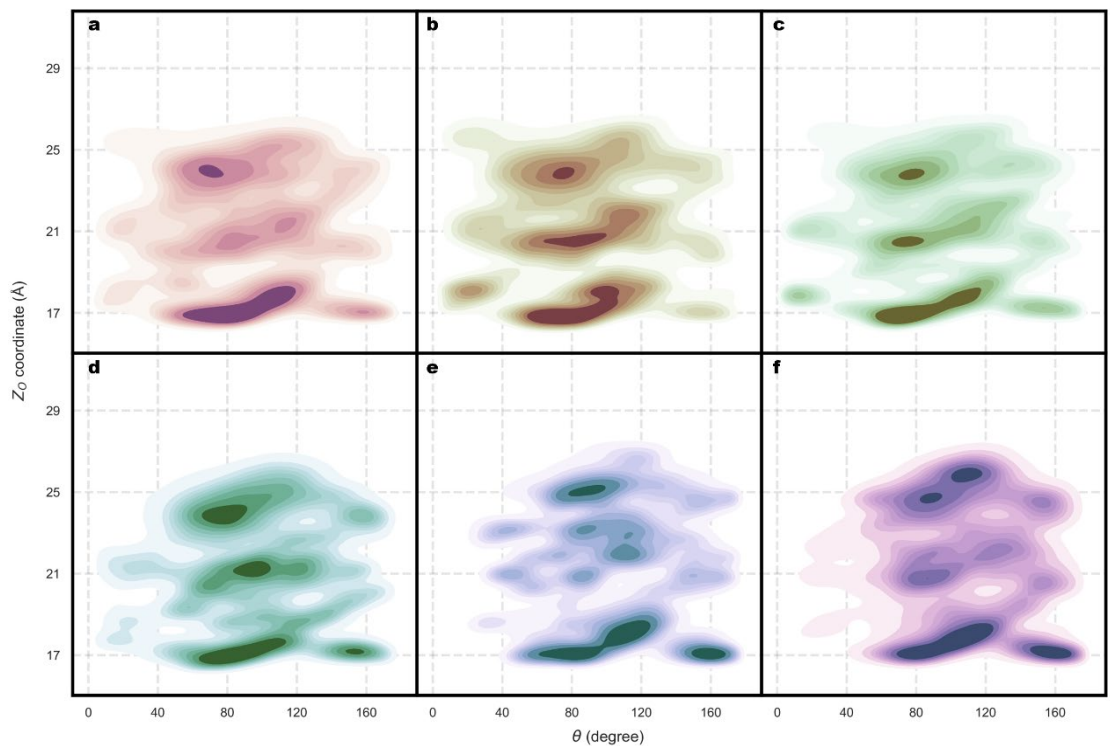

**Supplementary Figure 34. Probability distribution profile for angle θ and the Z-coordinate of the corresponding oxygen atom in water molecule at the Au(110)-water interface with a $CO_2$ molecule.** Probability distribution profiles for angle θ and the Z-coordinate of the corresponding oxygen atom in water molecule at the Au(110)-water interface with a $CO_2$ molecule were obtained from 1 ns molecular dynamics simulations under different conditions (**a**, constant charge, **b**, 0.2 V vs SHE, **c**, 0 V vs SHE, **d**, -0.2 V vs SHE, **e**, -0.4 V vs SHE, **f**, -0.6 V vs SHE). The specific definition of these angle θ are provided in Supplementary Figure 27 b.

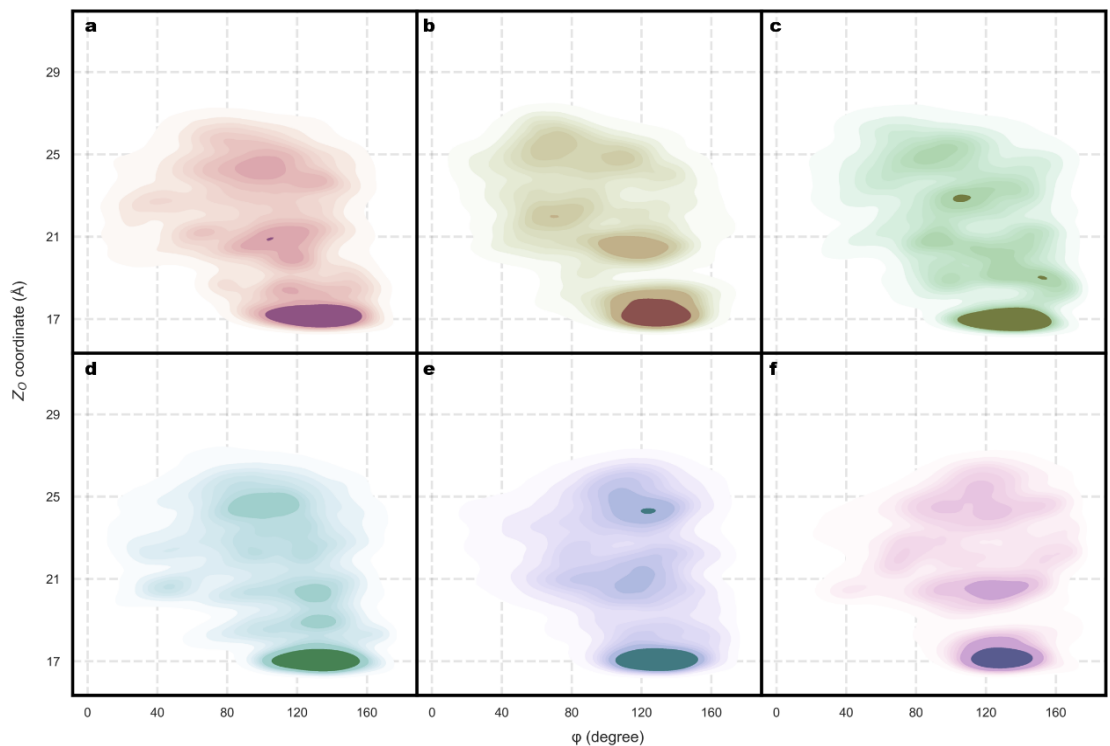

**Supplementary Figure 35 Probability distribution profile for angle φ and the Z-coordinate of the corresponding oxygen atom in water molecule at the Au(110)-water interface with 2 $K^+$ ions and a $CO_2$ molecule.** Probability distribution profiles for angle φ and the Z-coordinate of the corresponding oxygen atom in water molecule at the Au(110)-water interface with 2 $K^+$ ions and a $CO_2$ molecule were obtained from 1 ns molecular dynamics simulations under different conditions (**a**, constant charge, **b**, 0.2 V vs SHE, **c**, 0 V vs SHE, **d**, -0.2 V vs SHE, **e**, -0.4 V vs SHE, **f**, -0.6 V vs SHE). The specific definition of these angle φ are provided in Supplementary Figure 27 a.

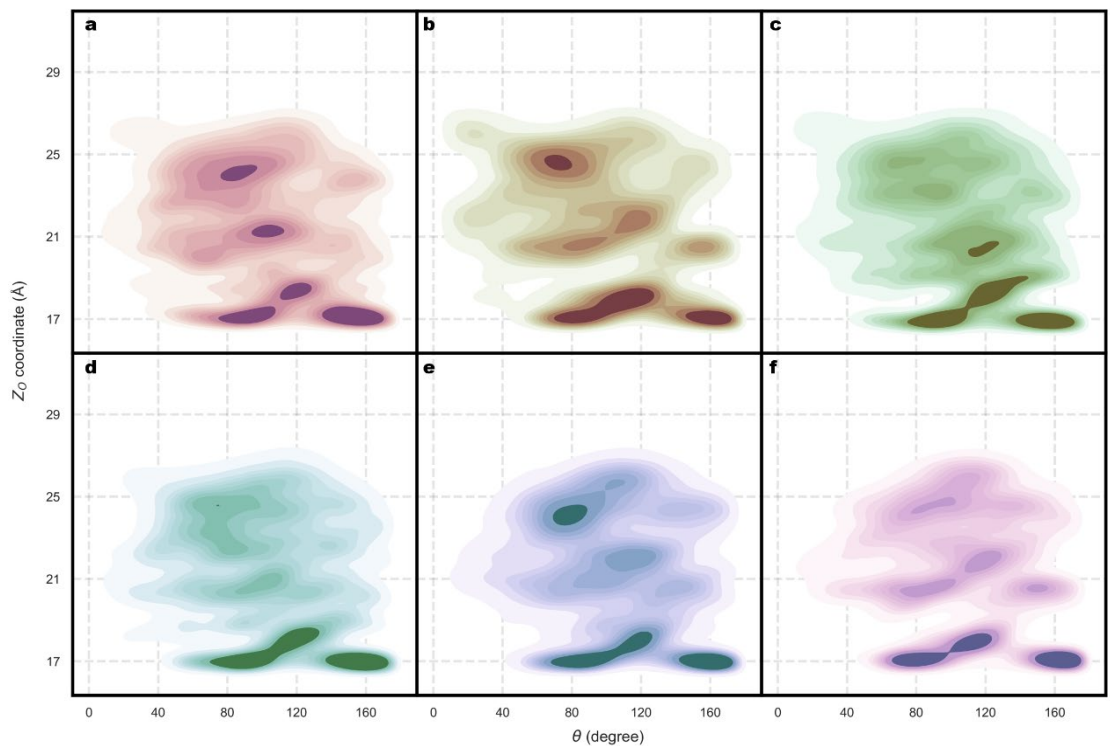

**Supplementary Figure 36. Probability distribution profile for angle θ and the Z-coordinate of the corresponding oxygen atom in water molecule at the Au(110)-water interface with 2 $K^+$ ions and a $CO_2$ molecule.** Probability distribution profiles for angle θ and the Z-coordinate of the corresponding oxygen atom in water molecule at the Au(110)-water interface with 2 $K^+$ ions and a $CO_2$ molecule were obtained from 1 ns molecular dynamics simulations under different conditions (**a**, constant charge, **b**, 0.2 V vs SHE, **c**, 0 V vs SHE, **d**, -0.2 V vs SHE, **e**, -0.4 V vs SHE, **f**, -0.6 V vs SHE). The specific definition of the angle θ are provided in Supplementary Figure 27 b.

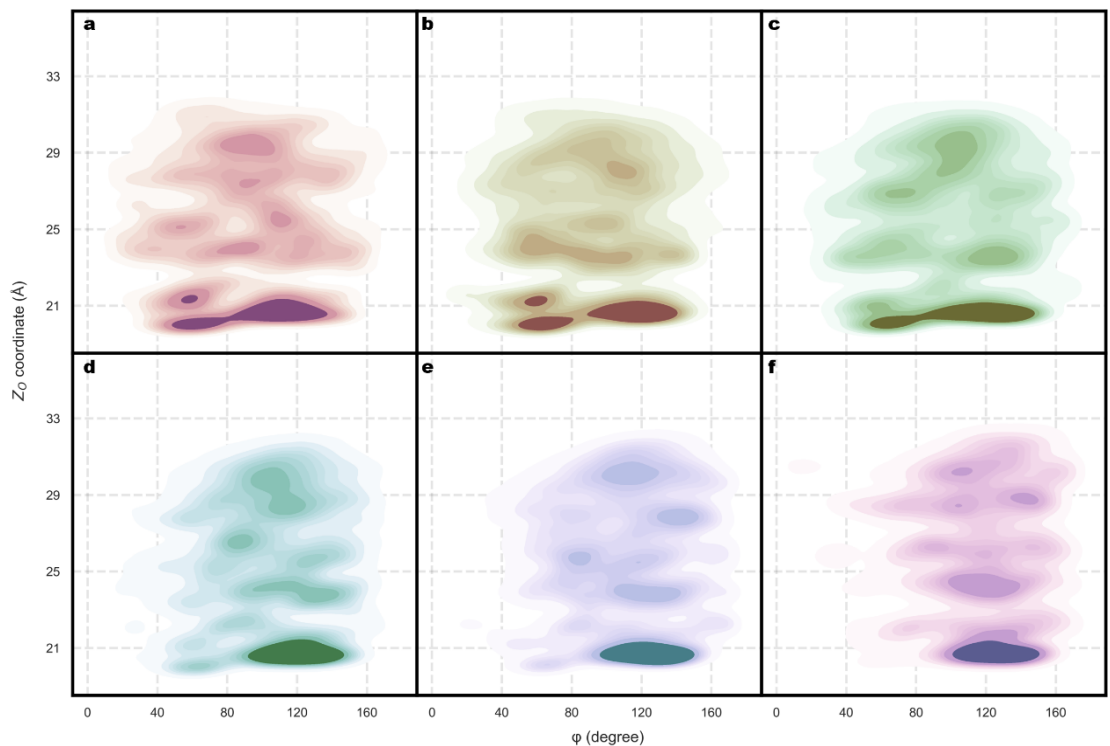

**Supplementary Figure 37. Probability distribution profile for angle φ and the Z-coordinate of the corresponding oxygen atom in water molecule at the Au(111)-water interface.** Probability distribution profiles for angle φ and the Z-coordinate of the corresponding oxygen atom in water molecule at the Au(111)-water interface were obtained from 1 ns molecular dynamics simulations under different conditions (**a**, constant charge, **b**, 0.2 V vs SHE, **c**, 0 V vs SHE, **d**, -0.2 V vs SHE, **e**, -0.4 V vs SHE, **f**, -0.6 V vs SHE). The specific definition of the angle φ are provided in Supplementary Figure 27 a.

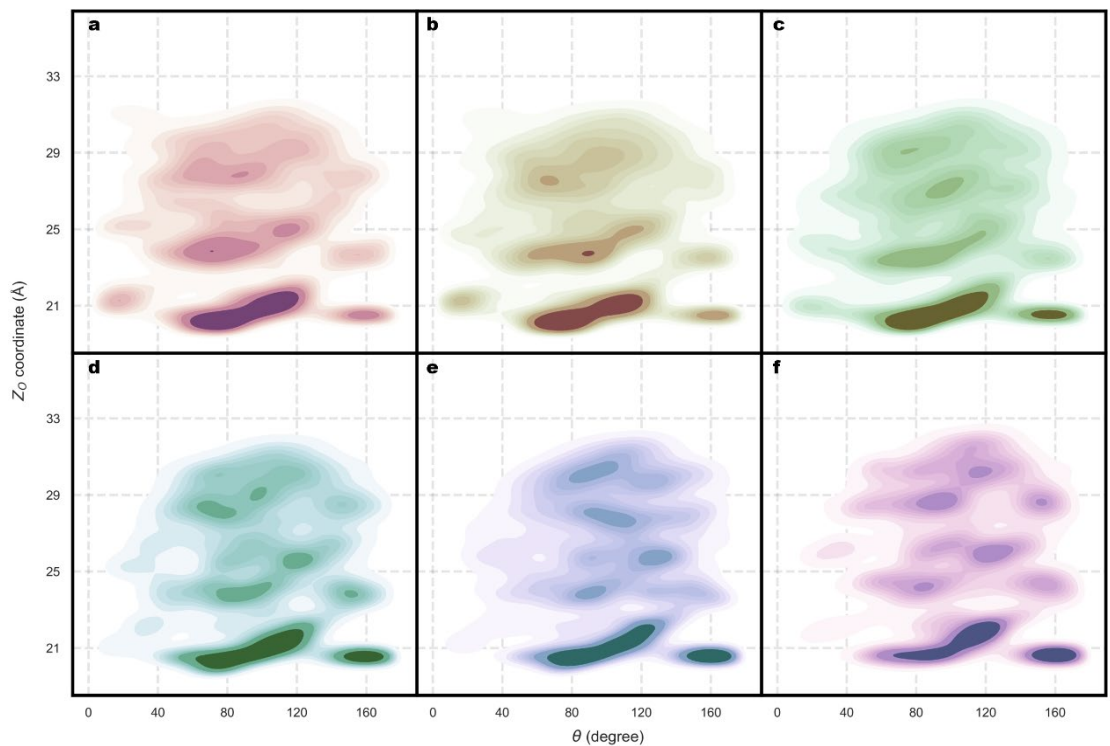

**Supplementary Figure 38. Probability distribution profiles for angles θ and the Z-coordinate of the corresponding oxygen atom in water molecule at the Au(111)-water interface.** Probability distribution profiles for angle θ and the Z-coordinate of the corresponding oxygen atom in water molecule at the Au(111)-water interface were obtained from 1 ns molecular dynamics simulations under different conditions (**a**, constant charge, **b**, 0.2 V vs SHE, **c**, 0 V vs SHE, **d**, -0.2 V vs SHE, **e**, -0.4 V vs SHE, **f**, -0.6 V vs SHE). The specific definition of the angle are θ provided in Supplementary Figure 27 b.

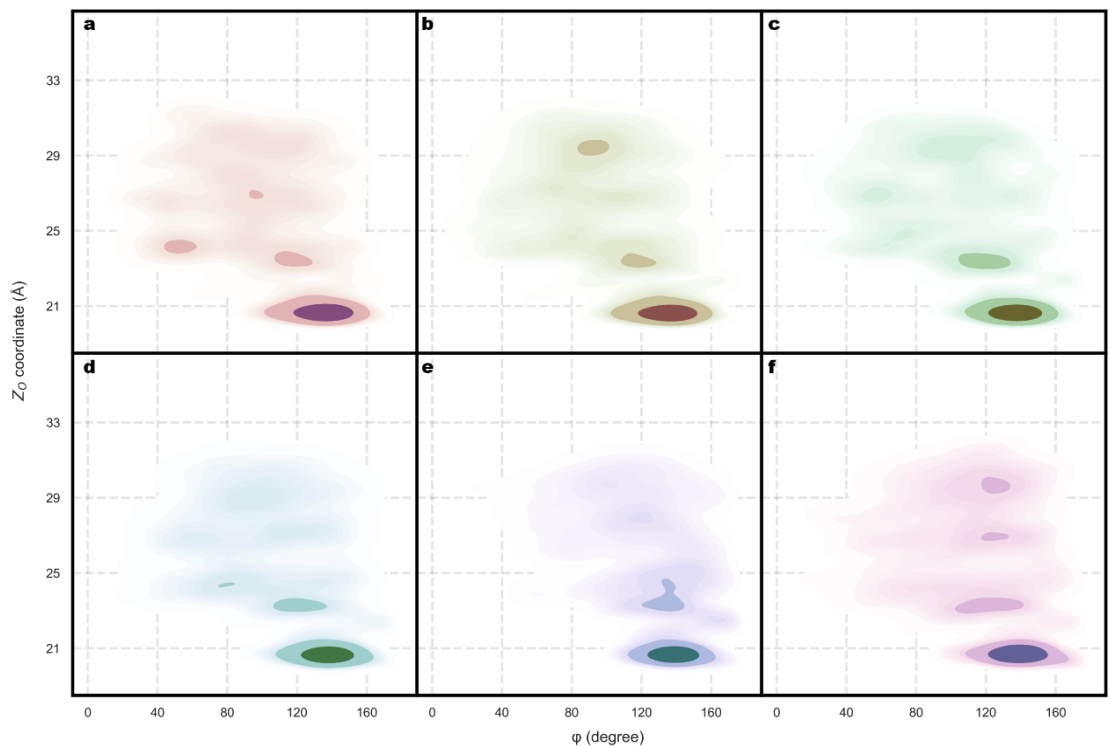

**Supplementary Figure 39. Probability distribution profile for angle φ and the Z-coordinate of the corresponding oxygen atom in water molecule at the Au(111)-water interface with 2 K$^+$ ions.** Probability distribution profiles for angle φ and the Z-coordinate of the corresponding oxygen atom in water molecule at the Au(111)-water interface with 2 K$^+$ ions were obtained from 1 ns molecular dynamics simulations under different conditions (**a**, constant charge, **b**, 0.2 V vs SHE, **c**, 0 V vs SHE, **d**, -0.2 V vs SHE, **e**, -0.4 V vs SHE, **f**, -0.6 V vs SHE). The specific definition of the angle φ are provided in Supplementary Figure 27 a.

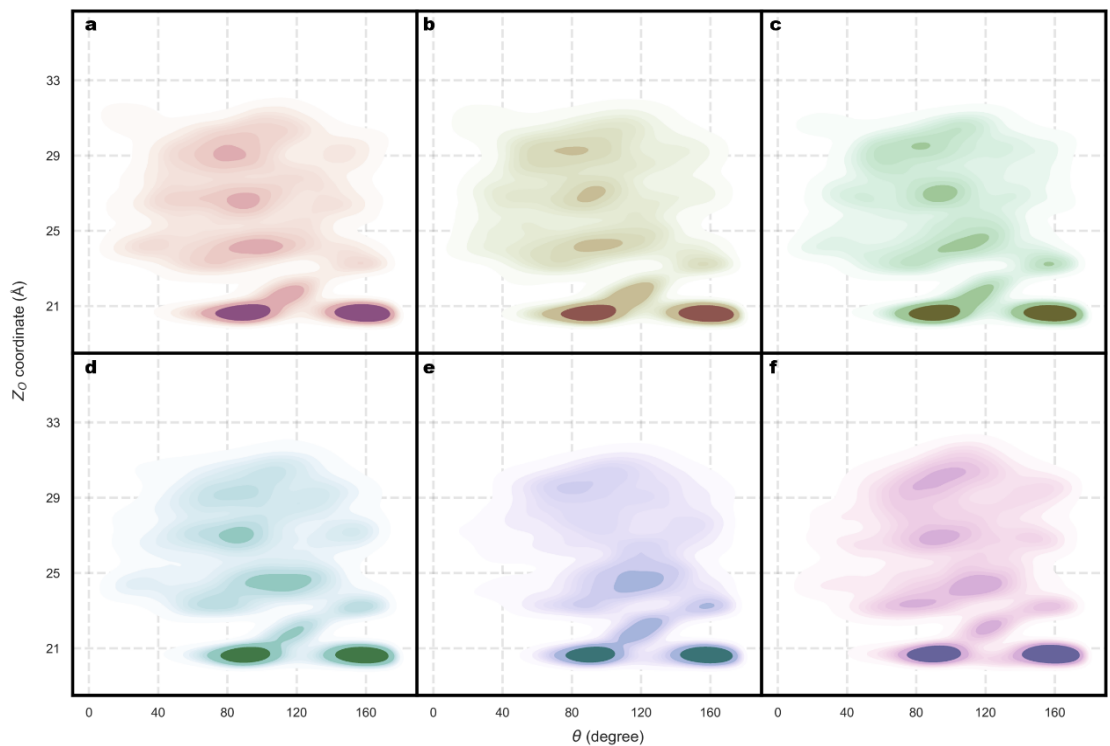

**Supplementary Figure 40. Probability distribution profile for angle θ and the Z-coordinate of the corresponding oxygen atom in water molecule at the Au(111)-water interface with 2 $K^+$ ions.** Probability distribution profiles for angle θ and the Z-coordinate of the corresponding oxygen atom in water molecule at the Au(111)-water interface with 2 $K^+$ ions were obtained from 1 ns molecular dynamics simulations under different conditions (**a**, constant charge, **b**, 0.2 V vs SHE, **c**, 0 V vs SHE, **d**, -0.2 V vs SHE, **e**, -0.4 V vs SHE, **f**, -0.6 V vs SHE). The specific definition of the angle θ are provided in Supplementary Figure 27 b.

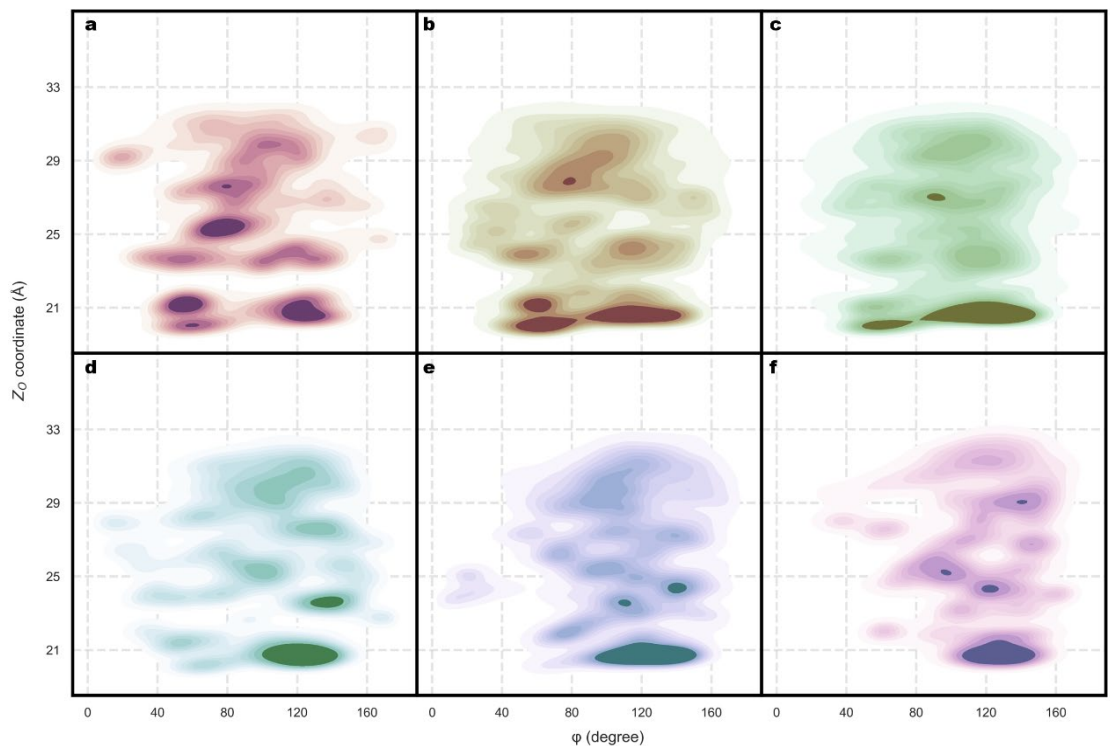

**Supplementary Figure 41. Probability distribution profile for angle φ and the z-coordinate of the corresponding oxygen atom in water molecule at the Au(111)-water interface with a $CO_2$ molecule.** Probability distribution profiles for angle φ and the z-coordinate of the corresponding oxygen atom in water molecule at the Au(111)-water interface with a $CO_2$ molecule were obtained from 1 ns molecular dynamics simulations under different conditions (**a**, constant charge, **b**, 0.2 V vs SHE, **c**, 0 V vs SHE, **d**, -0.2 V vs SHE, **e**, -0.4 V vs SHE, **f**, -0.6 V vs SHE). The specific definition of the angle φ are provided in Supplementary Figure 27 a.

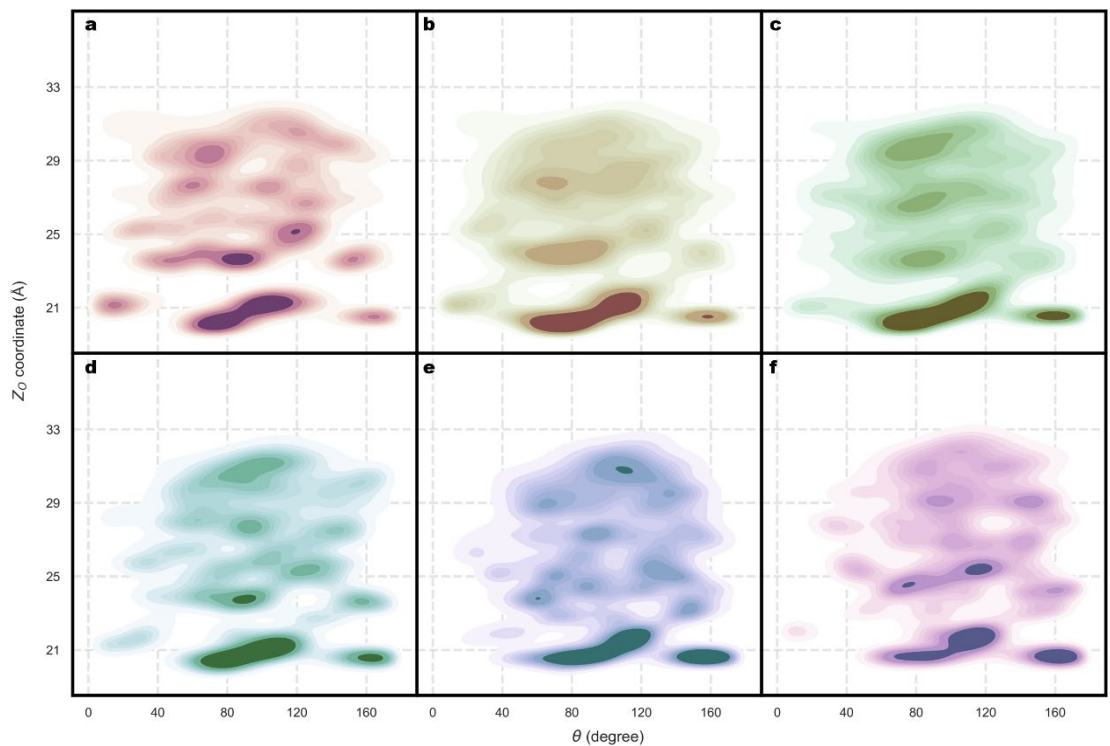

**Supplementary Figure 42. Probability distribution profile for angle θ and the Z-coordinate of the corresponding oxygen atom in water molecule at the Au(110)-water interface with a $CO_2$ molecule.** Probability distribution profiles for angle θ and the Z-coordinate of the corresponding oxygen atom in water molecule at the Au(110)-water interface with a $CO_2$ molecule were obtained from 1 ns molecular dynamics simulations under different conditions (**a**, constant charge, **b**, 0.2 V vs SHE, **c**, 0 V vs SHE, **d**, -0.2 V vs SHE, **e**, -0.4 V vs SHE, **f**, -0.6 V vs SHE). The specific definition of the angle θ are provided in Supplementary Figure 27 b.

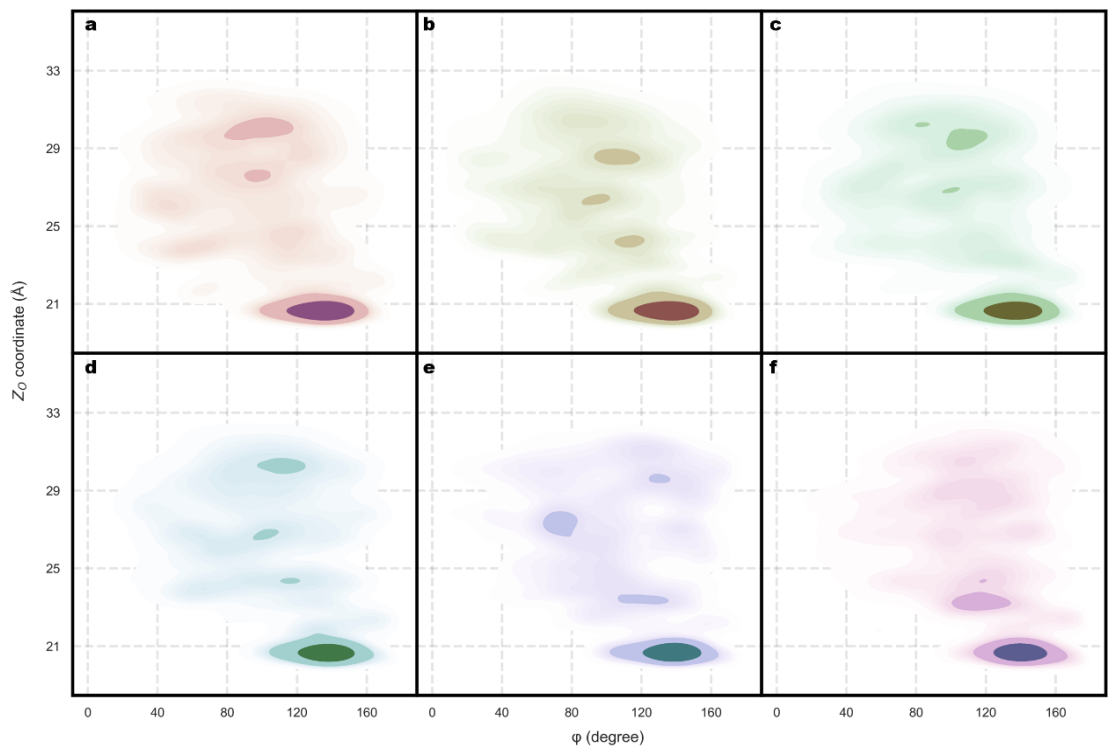

**Supplementary Figure 43. Probability distribution profile for angle φ and the Z-coordinate of the corresponding oxygen atom in water molecule at the Au(110)-water interface with 2 $K^+$ ions and a $CO_2$ molecule.** Probability distribution profiles for angle φ and the Z-coordinate of the corresponding oxygen atom in water molecule at the Au(110)-water interface with 2 $K^+$ ions and a $CO_2$ molecule were obtained from 1 ns molecular dynamics simulations under different conditions (**a**, constant charge, **b**, 0.2 V vs SHE, **c**, 0 V vs SHE, **d**, -0.2 V vs SHE, **e**, -0.4 V vs SHE, **f**, -0.6 V vs SHE). The specific definition of the angle φ are provided in Supplementary Figure 27 a.

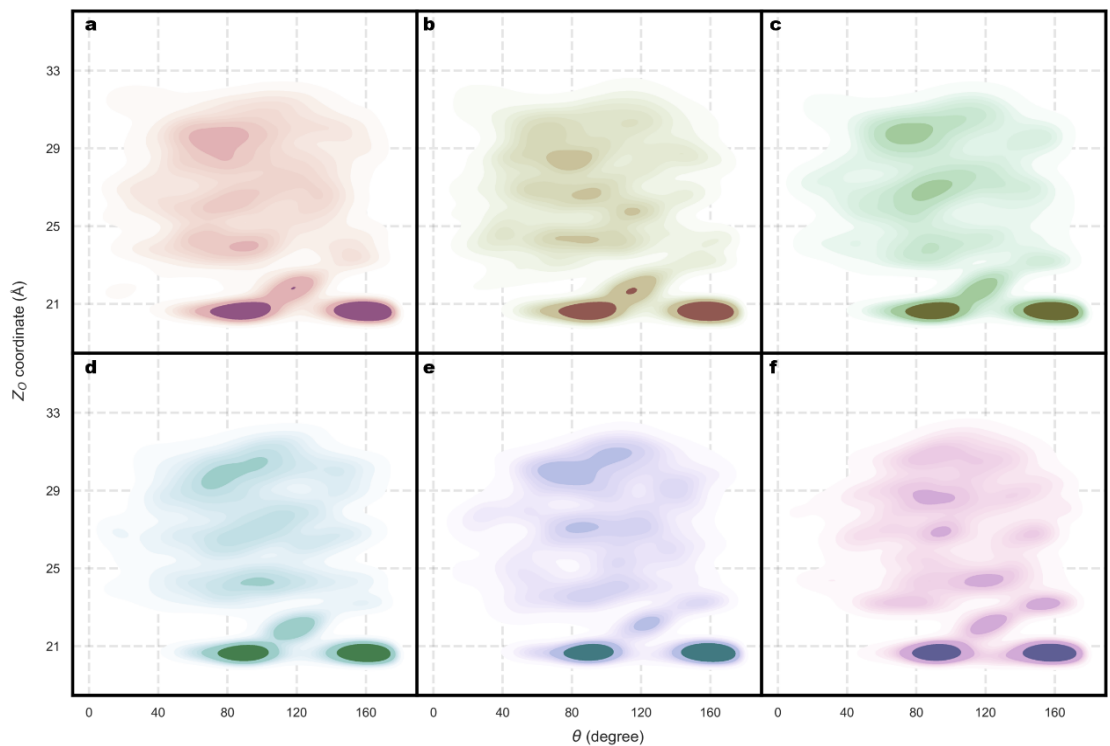

**Supplementary Figure 44. Probability distribution profile for angle φ and the Z-coordinate of the corresponding oxygen atom in water molecule at the Au(110)-water interface with 2 K$^+$ ions and a CO$_2$ molecule.** Probability distribution profiles for angle φ and the Z-coordinate of the corresponding oxygen atom in water molecule at the Au(110)-water interface with 2 K$^+$ ions and a CO$_2$ molecule were obtained from 1 ns molecular dynamics simulations under different conditions (**a**, constant charge, **b**, 0.2 V vs SHE, **c**, 0 V vs SHE, **d**, -0.2 V vs SHE, **e**, -0.4 V vs SHE, **f**, -0.6 V vs SHE). The specific definition of the angle φ are provided in Supplementary Figure 27 a.

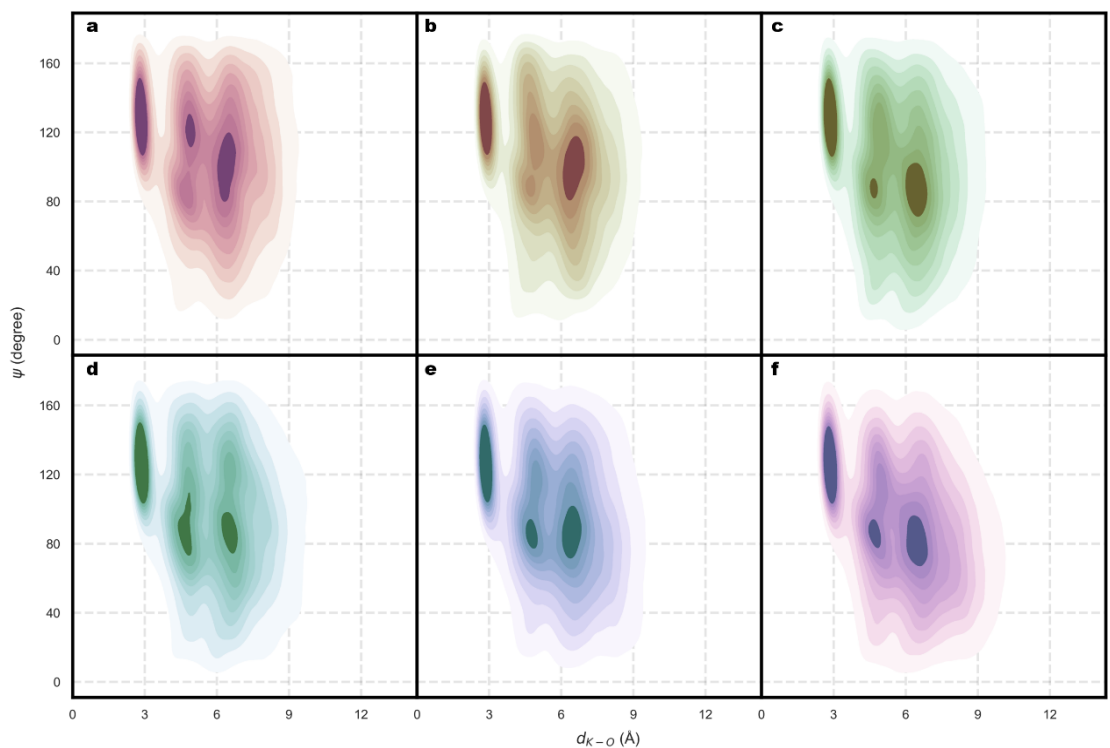

**Supplementary Figure 45. Probability distribution profile for the angle ψ and the distance between K$^+$ ions and corresponding oxygen atom in water molecule at the Au(110)-water interface with 2 K$^+$ ions.** Probability distribution profiles for the angle ψ and the distances between K$^+$ ions and corresponding oxygen atom in water molecule at the Au(110)-water interface with 2 K$^+$ ions were obtained from 1 ns molecular dynamics simulations under different conditions (**a**, constant charge, **b**, 0.2 V vs SHE, **c**, 0 V vs SHE, **d**, -0.2 V vs SHE, **e**, -0.4 V vs SHE, **f**, -0.6 V vs SHE). The specific definition of the angle ψ are provided in Supplementary Figure 27 c.

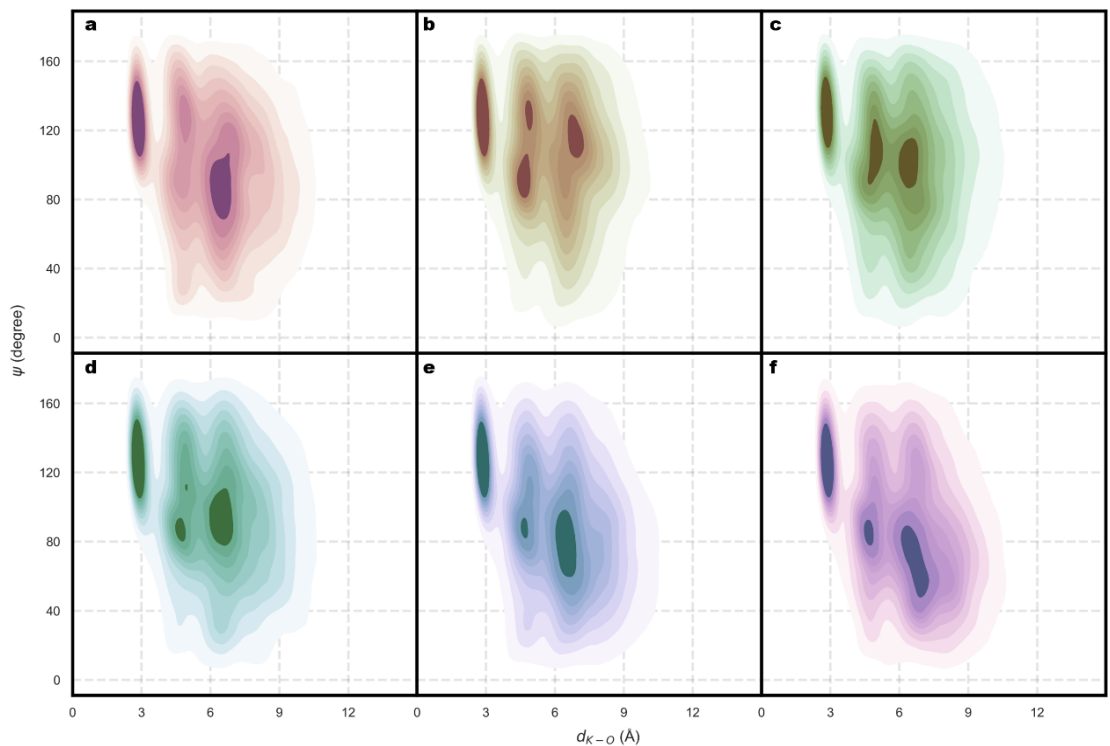

**Supplementary Figure 46. Probability distribution profile for the angle ψ and the distance between K$^+$ ions and corresponding oxygen atom in water molecule at the Au(110)-water interface with 2 K$^+$ ions and a CO$_2$ molecule.** Probability distribution profiles for the angle ψ and the distance between K$^+$ ion and corresponding oxygen atom in water molecule at the Au(110)-water interface with 2 K$^+$ ions and a CO$_2$ molecule were obtained from 1 ns molecular dynamics simulations under different conditions (**a**, constant charge, **b**, 0.2 V vs SHE, **c**, 0 V vs SHE, **d**, -0.2 V vs SHE, **e**, -0.4 V vs SHE, **f**, -0.6 V vs SHE). The specific definition of the angle ψ are provided in Supplementary Figure 27 c.

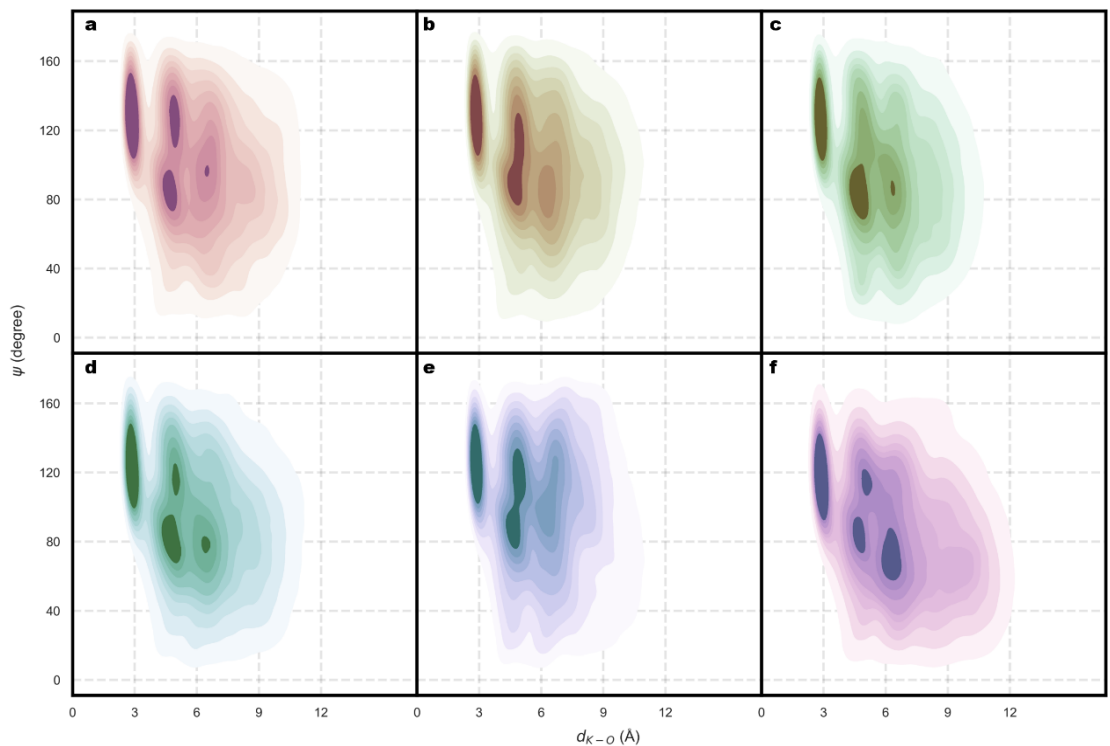

**Supplementary Figure 47. Probability distribution profile for the angle ψ and the distance between K⁺ ions and corresponding oxygen atom in water molecule at the Au(111)-water interface with 2 K⁺ ions.** Probability distribution profiles for the angle ψ and the distances between K⁺ ions and corresponding oxygen atom in water molecule at the Au(111)-water interface with 2 K⁺ ions were obtained from 1 ns molecular dynamics simulations under different conditions (**a**, constant charge, **b**, 0.2 V vs SHE, **c**, 0 V vs SHE, **d**, -0.2 V vs SHE, **e**, -0.4 V vs SHE, **f**, -0.6 V vs SHE). The specific definition of the angle ψ are provided in Supplementary Figure 27 c.

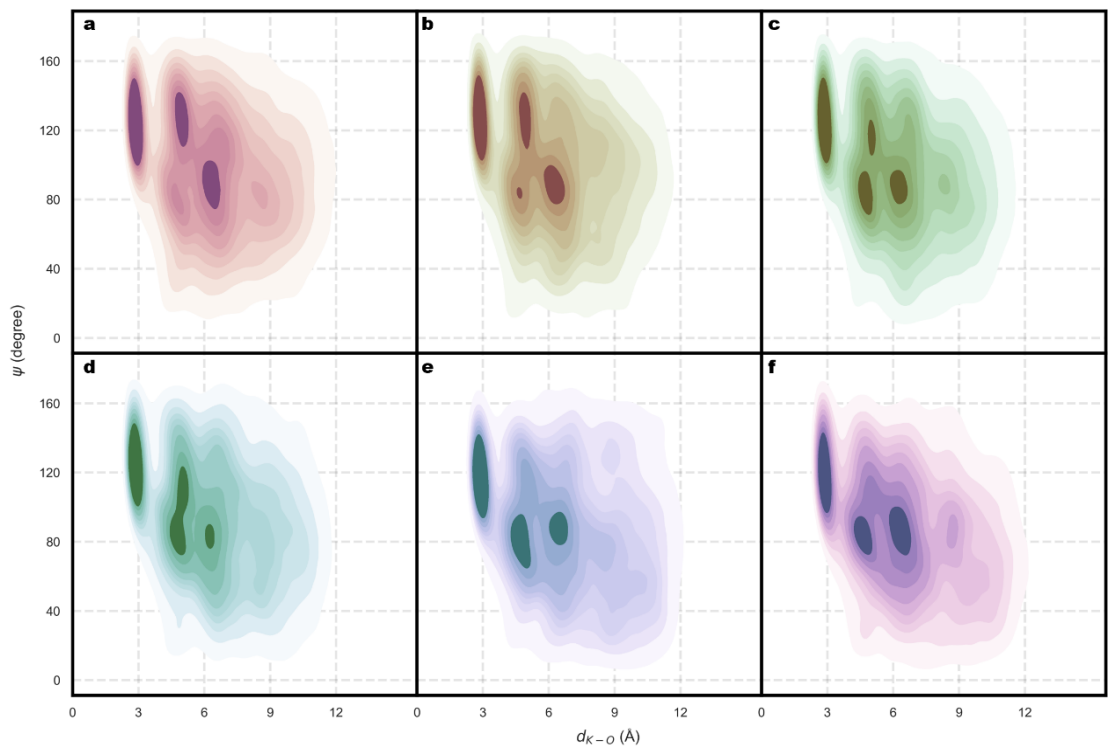

**Supplementary Figure 48. Probability distribution profile for the angle ψ and the distance between K$^+$ ions and corresponding oxygen atom in water molecules at the Au(111)-water interface with 2 K$^+$ ions and a CO$_2$ molecule.** Probability distribution profiles for the angle ψ and the distances between K$^+$ ions and corresponding oxygen atom in water molecule at the Au(111)-water interface with 2 K$^+$ ions and a CO$_2$ molecule were obtained from 1 ns molecular dynamics simulations under different conditions (**a**, constant charge, **b**, 0.2 V vs SHE, **c**, 0 V vs SHE, **d**, -0.2 V vs SHE, **e**, -0.4 V vs SHE, **f**, -0.6 V vs SHE). The specific definition of the angle ψ are provided in Supplementary Figure 27 c.

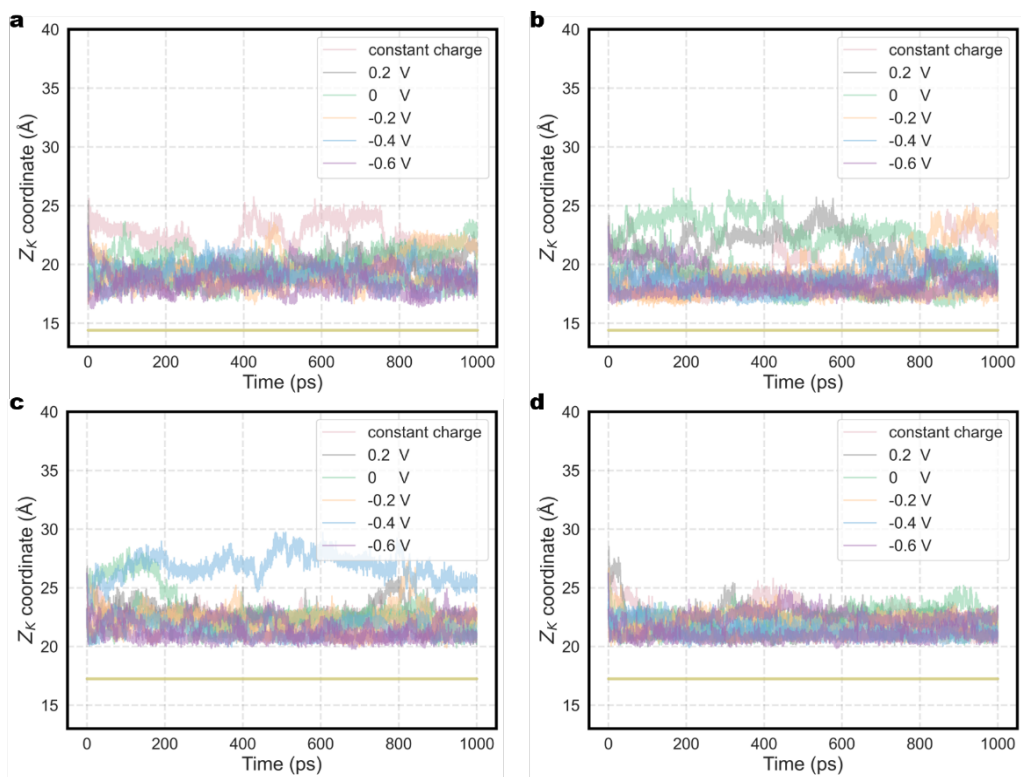

**Supplementary Figure 49. Variation of the Z-coordinate of K ions at the Au-water interface.** The Z-coordinate of K ions at **a**, the Au(110)-water interface with 2 $K^+$ ions, **b**, the Au(110)-water interface with 2 $K^+$ ions and a $CO_2$ molecule. **c**, the Au(111)-water interface with 2 $K^+$ ions, **d**, the Au(111)-water interface with 2 $K^+$ ions and a $CO_2$ molecule. All results were obtained from 1 ns molecular dynamics simulations under different conditions (constant charge, 0.2 V vs SHE, 0 V vs SHE, -0.2 V vs SHE, -0.4 V vs SHE, -0.6 V vs SHE). The yellow line at the bottom represents the initial position of the topmost surface of Au.

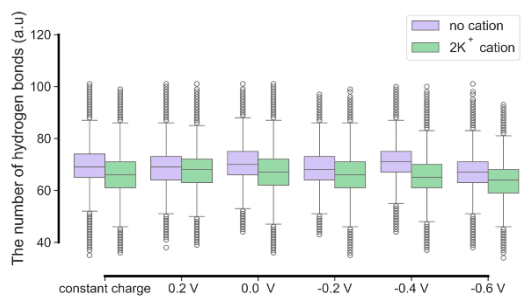

**Supplementary Figure 50. Statistic of hydrogen bonds at the Au(110)-water interface with a $CO_2$ molecule.** Statistics of hydrogen bonds at the Au(110)-water interface with a $CO_2$ molecule were obtained from 1 ns molecular dynamics simulations under different conditions (constant charge, 0.2 V vs SHE, 0 V vs SHE, -0.2 V vs SHE, -0.4 V vs SHE, -0.6 V vs SHE). In the box plot, the green and purple colors represent the differences between solutions with and without $K^+$ ions, respectively.

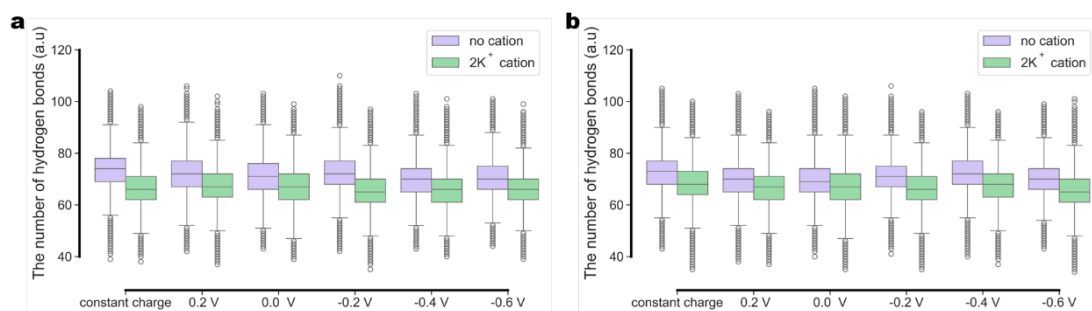

**Supplementary Figure 51. Statistic of hydrogen bonds at the Au(111)-water interface.** Statistics of hydrogen bonds at **a**, the Au(111)-water interface, and **b**, the Au(111)-water interface with a $CO_2$ molecule. All results were obtained from 1 ns molecular dynamics simulations under different conditions (constant charge, 0.2 V vs SHE, 0 V vs SHE, -0.2 V vs SHE, -0.4 V vs SHE, -0.6 V vs SHE). In the box plot, the purple and green colors represent the differences between solutions with and without $K^+$ ions, respectively.

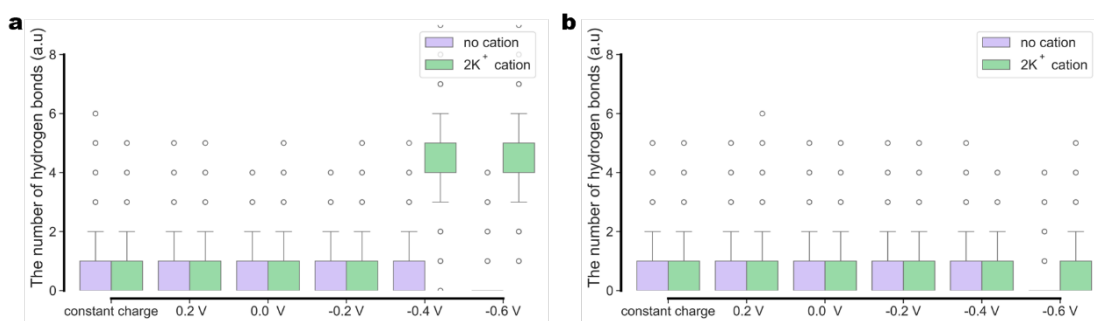

**Supplementary Figure 52. Statistic of hydrogen bonds formed by the oxygen atoms in the $CO_2$ molecule at the Au-water interface.** Statistics of hydrogen bonds formed by the oxygen atoms in the $CO_2$ molecule **a**, the Au(110)-water interface, and **b**, the Au(111)-water interface. All results were obtained from 1 ns molecular dynamics simulations under different conditions (constant charge, 0.2 V vs SHE, 0 V vs SHE, -0.2 V vs SHE, -0.4 V vs SHE, -0.6 V vs SHE). In the box plot, the green and purple colors represent the differences between solutions with and without $K^+$ ions, respectively.

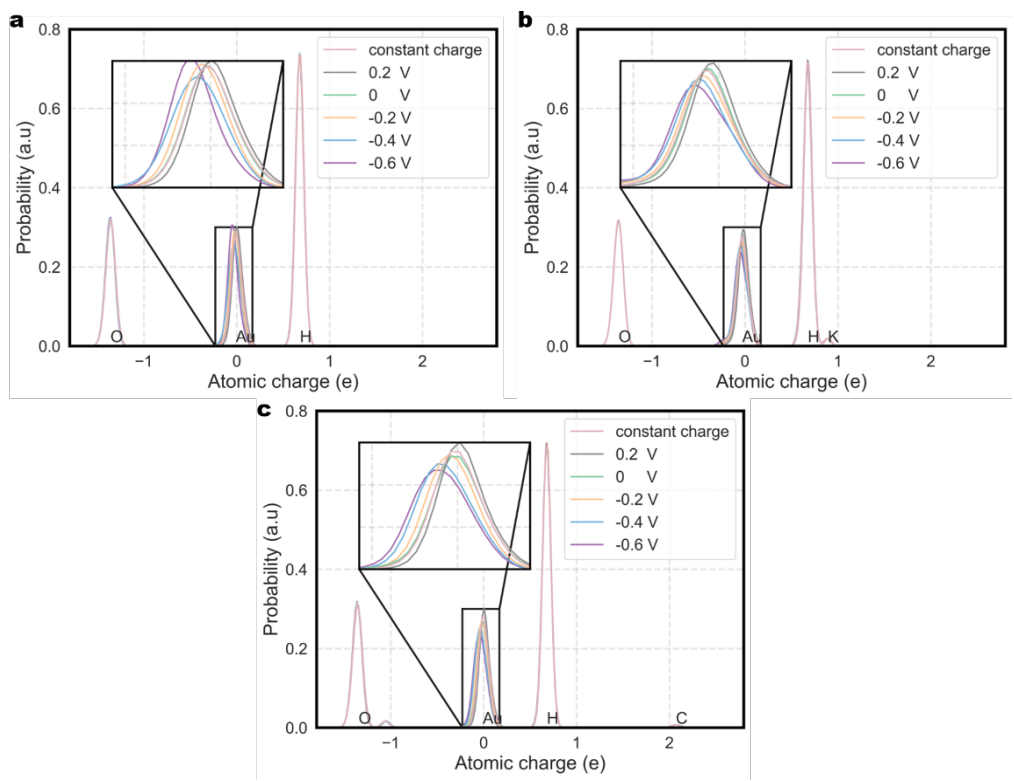

**Supplementary Figure 53. Comparison of probability distribution profile for the Bader charges of various elements at the Au(110)-water interface.** Probability distribution profiles for the Bader charges of various elements at **a**, the Au(110)-water interface, **b**, the Au(110)-water interface with 2 $K^+$ ions, and **c**, the Au(110)-water interface with a $CO_2$ molecule. All results were obtained from 1 ns molecular dynamics simulations under different conditions (constant charge, 0.2 V vs SHE, 0 V vs SHE, -0.2 V vs SHE, -0.4 V vs SHE, -0.6 V vs SHE).

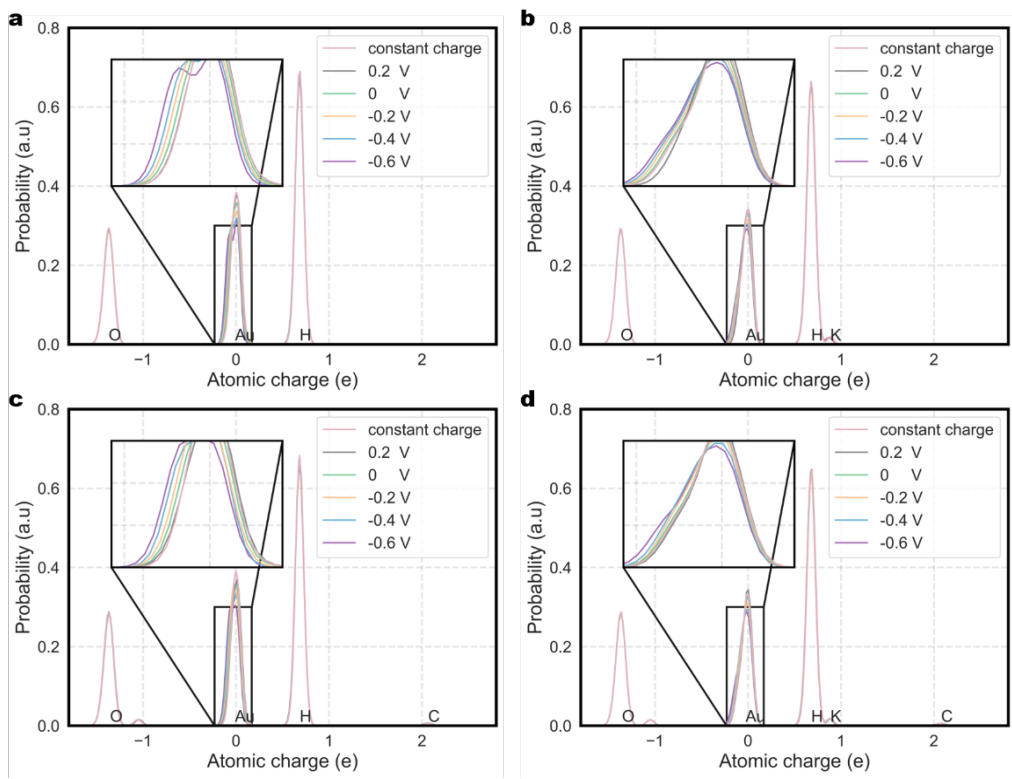

**Supplementary Figure 54. Comparison of probability distribution profile for the Bader charges of various elements at the Au(111)-water interface.** Probability distribution profile for the Bader charges of various elements at **a**, the Au(111)-water interface, **b**, the Au(111)-water interface with 2 $K^+$ ions, **c**, the Au(111)-water interface with a $CO_2$ molecule, and d, the Au(111)-water interface with 2 $K^+$ ions and a $CO_2$ molecule. All results were obtained from 1 ns molecular dynamics simulations under different conditions (constant charge, 0.2 V vs SHE, 0 V vs SHE, -0.2 V vs SHE, -0.4 V vs SHE, -0.6 V vs SHE).

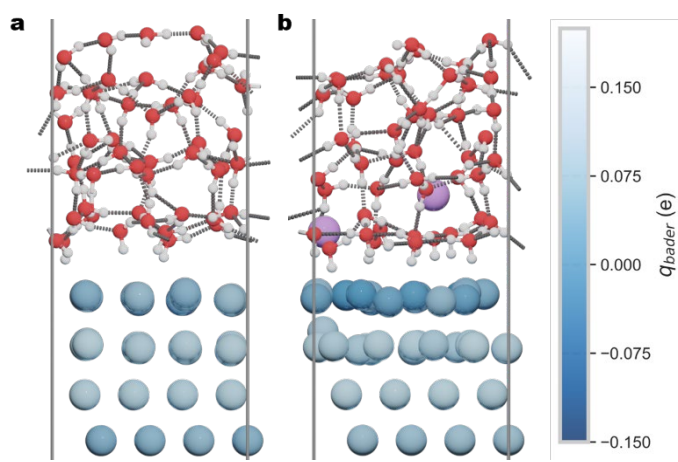

**Supplementary Figure 55. Snapshots of the Au(111)-water interface at -0.6 V vs SHE. a**, Pure water interface. **b**, Interface with 2 $K^+$ ions. The Au atoms are colored according to their Bader charges (see color bar). Light purple, white, and red spheres represent K, H, and O atoms, respectively.

**Supplementary Note 1. Determining the initial collective variable for $CO_2$ adsorption.**

In our nanosecond (ns) scale molecular dynamics (MD) simulations, the distance between $CO_2$ molecules and the Au surface was not constrained. Consequently, $CO_2$ could drift more than 7 Å away from the surface. The slow-growth method involves incrementally changing a defined collective variable (CV) until a target state is reached. For $CO_2$ adsorption, this CV is typically the distance between the carbon atom in $CO_2$ and a specific exposed Au site. When $CO_2$ is far from the surface, the slow-growth method requires more steps to reach the adsorbed state, increasing the computational cost, especially when multiple calculations are needed for reliable free energy barriers[1]. Previous studies indicate that $CO_2$ adsorption initiates within 3-4 Å of the surface[2-4]. Moreover, including the approach of $CO_2$ from larger distances introduces free energy changes associated with mass transfer, which can confound our analysis of the chemical adsorption process. Therefore, to focus solely on chemical adsorption, we first identify the endpoint of the mass transfer process for each system at different applied potentials, which we refer to as the $CO_2$ physical adsorption point. This physisorbed state serves as the starting point for our slow-growth calculations.

To ensure consistency in our subsequent free energy calculations, we first performed $CO_2$ desorption simulations starting from the two spontaneously adsorbed $CO_2$ configurations observed in our initial MD simulations (discussed in the main text). Then, these two desorbed structures were used as the starting points for determining the physisorbed $CO_2$ location in each system. This procedure guarantees that all subsequent free energy calculations consistently begin with the relevant CV for $CO_2$ adsorption. Supplementary Figure 56 shows the applied potential, net charge, and free energy changes during the two $CO_2$ desorption simulations. Experimental conditions typically maintain a constant potential. It requires allowing the system charge to fluctuate, as demonstrated by our constant potential approach which maintains a fixed applied potential while the system net charge changes significantly. Therefore, accurate electrocatalysis simulations must break from the traditional constant charge paradigm. Notably, $CO_2$ desorption is more difficult at -0.6 V than at -0.4 V. To ensure sufficient separation between $CO_2$ and the surface, the final value of the CV (C-Au distance) is set to 4.5 Å.

Next, we explored the physisorption sites of $CO_2$ on both Au(110) and Au(111). As we are not concerned with the precise free energy of the mass transfer process, only a single slow-growth simulation was performed for each system. The collective variable (CV) was defined as the distance between the carbon atom of $CO_2$ and a specific Au site, with a target distance of 2.8 Å. As discussed in the main text, the Au(110) surface undergoes significant reconstruction during the simulation. We considered two types of sites on this surface, protruding sites (locations with significant elevation) and flat sites (relatively planar locations). For Au(111), which remains largely planar within our simulation timescale, we randomly selected a single Au site for analysis. This random selection accounts for the lateral diffusion of Au atoms within the periodic simulation cell, as identical atom indices do not guarantee structural equivalence over time. Thus, we examined physisorption at a representative flat site on the Au(111) surface. Supplementary Figures 57-64 show the potential of mean force (PMF) and free energy profiles for $CO_2$ physisorption on Au(110) surface, both with and without cations, at the two site types described above. The physisorption point is defined as the metastable position closest to the surface within the physisorption region. In the free energy profiles, this corresponds to the minimum C-Au distance at a free energy minimum, indicated by the pink dots in Supplementary

Figures 61-64. These points are taken as the physisorption locations for each system. Supplementary Figures 65 and 71 demonstrate that the constant potential reactor maintains a fixed applied potentials throughout these processes. Furthermore, compared to chemisorption, the relatively small changes in system charge confirm that these processes are indeed consistent with physisorption. The $CO_2$ physisorption process on Au(111) is shown in Supplementary Figures 66-70 and 72. These figures depict physisorption at a representative flat surface site, both with and without cations.

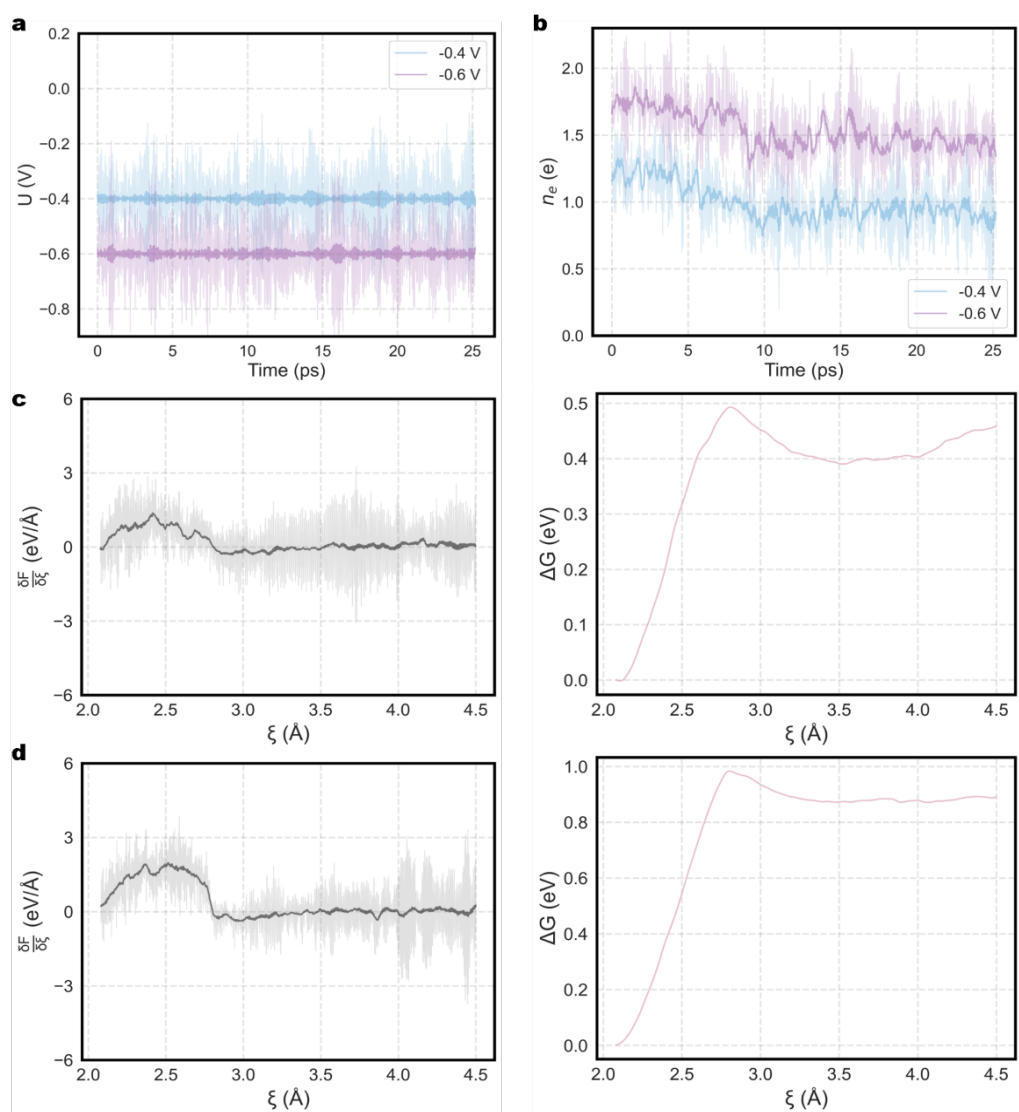

**Supplementary Figure 56. The desorption process of adsorbed $CO_2$.** $CO_2$ is adsorbed at a protruding surface site at the Au(110)-water interface with 2 $K^+$ ions. The protruding surface site refers to the Au site on the Au-water interface that become raised compared to the original flat surface after surface reconstruction. **a**, Applied potential and **b**, net electron counts of the Au(110)-water interface with 2 $K^+$ ions. The PMF (left) and $\Delta G$ (right) obtained using the slow-growth method under **c**, -0.4 V vs SHE and **d**, -0.6 V vs SHE. The light-colored line represents the results based on each frame, while the dark-colored line is the average obtained with a 200 fs moving window.

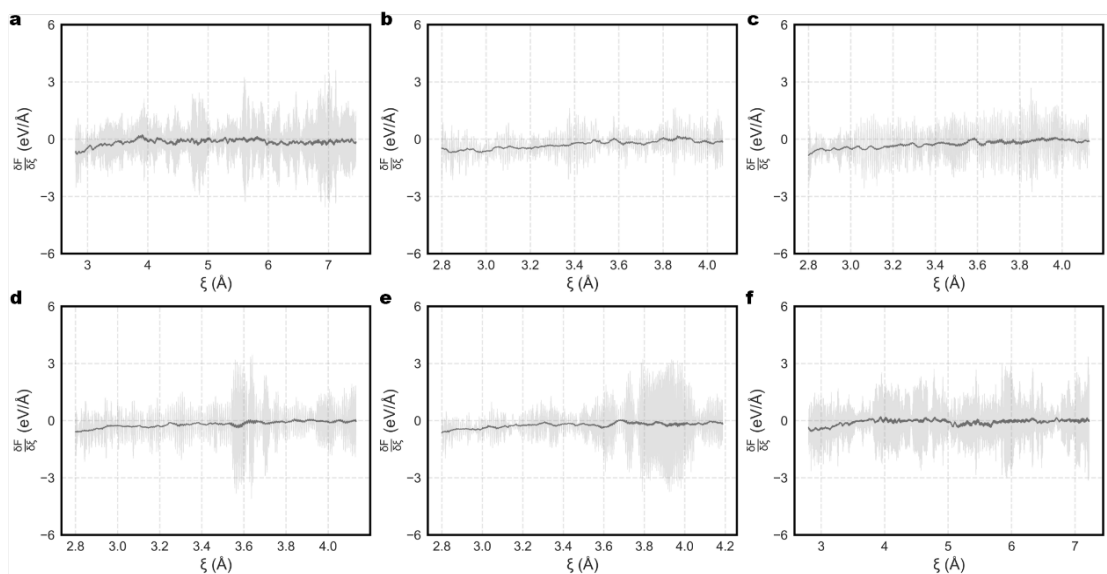

**Supplementary Figure 57. PMF for physical adsorption of CO$_2$ on a flat surface site at the Au(110)-water interface.** The flat surface site refers to the Au site on the Au-water interface that remain relatively smooth after surface reconstruction, compared to the protruding surface. The PMFs for physical adsorption of CO2 on a flat surface site at the Au(110)-water interfaces obtained using the slow-growth method under different conditions (a, constant charge, b, 0.2 V vs SHE, c, 0 V vs SHE, d, -0.2 V vs SHE, e, -0.4 V vs SHE, f, -0.6 V vs SHE). The light-colored line represents the results based on each frame, while the dark-colored line is the average obtained with a 200 fs moving window.

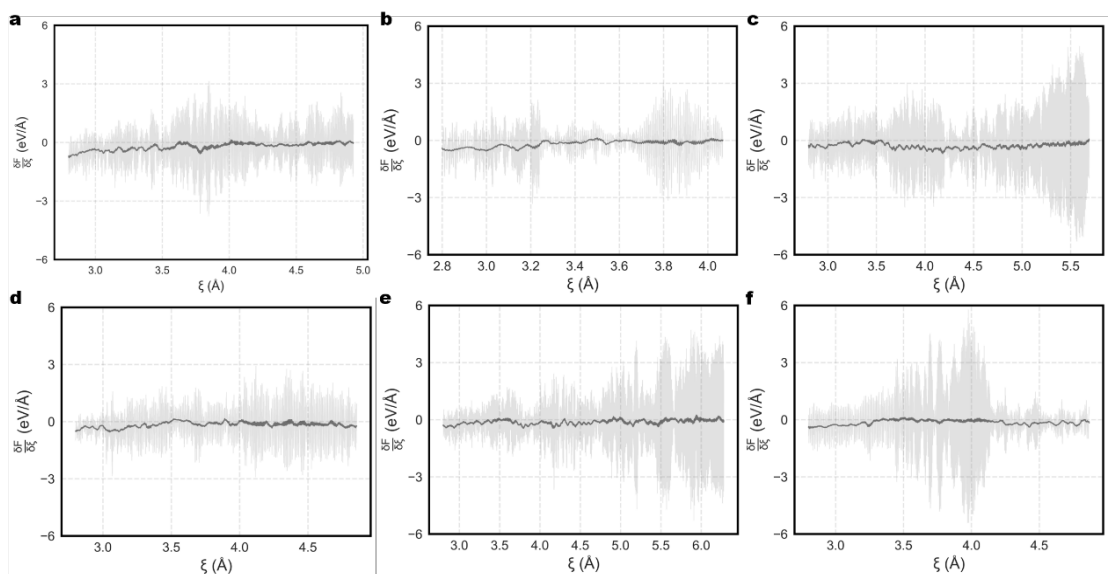

**Supplementary Figure 58. PMF for physical adsorption of $CO_2$ on a protruding surface site at the Au(110)-water interface.** The protruding surface site refers to the Au site on the Au-water interface that become raised compared to the original flat surface after surface reconstruction. The PMFs for physical adsorption of $CO_2$ on a protruding surface site at the Au(110)-water interfaces obtained using the slow-growth method under different conditions (**a**, constant charge, **b**, 0.2 V vs SHE, **c**, 0 V vs SHE, **d**, -0.2 V vs SHE, **e**, -0.4 V vs SHE, **f**, -0.6 V vs SHE). The light-colored line represents the results based on each frame, while the dark-colored line is the average obtained with a 200 fs moving window.

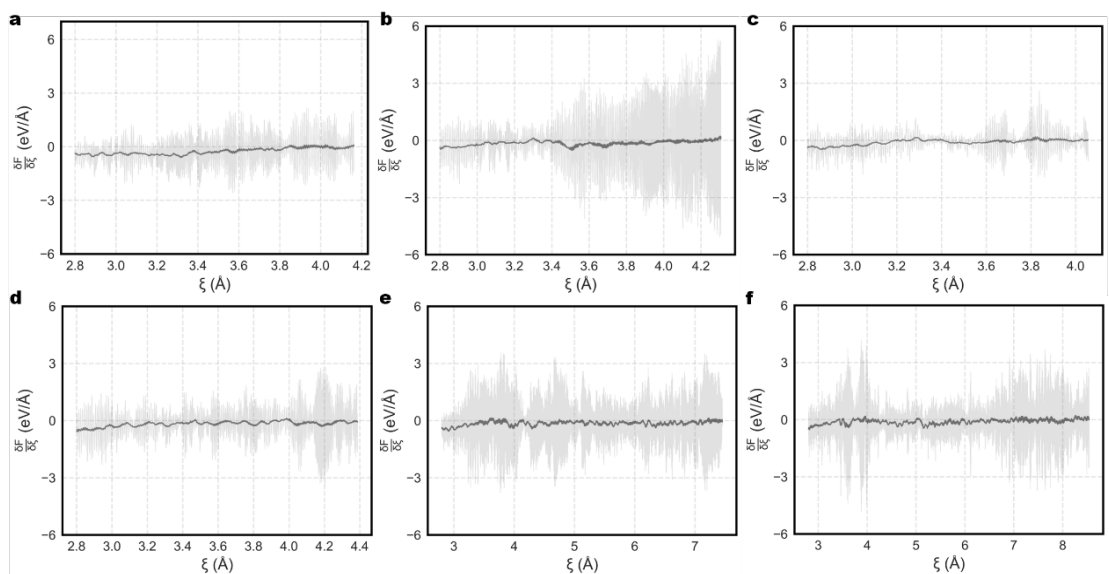

**Supplementary Figure 59. PMF for physical adsorption of $CO_2$ on a flat surface site at the Au(110)-water interface with 2 $K^+$ ions.** The flat surface site refers to the Au site on the Au-water interface that remain relatively smooth after surface reconstruction, compared to the protruding surface. The PMFs for physical adsorption of $CO_2$ on a flat surface site at the Au(110)-water interfaces with 2 $K^+$ ions obtained using the slow-growth method under different conditions (**a**, constant charge, **b**, 0.2 V vs SHE, **c**, 0 V vs SHE, **d**, -0.2 V vs SHE, **e**, -0.4 V vs SHE, **f**, -0.6 V vs SHE). The light-colored line represents the results based on each frame, while the dark-colored line is the average obtained with a 200 fs moving window.

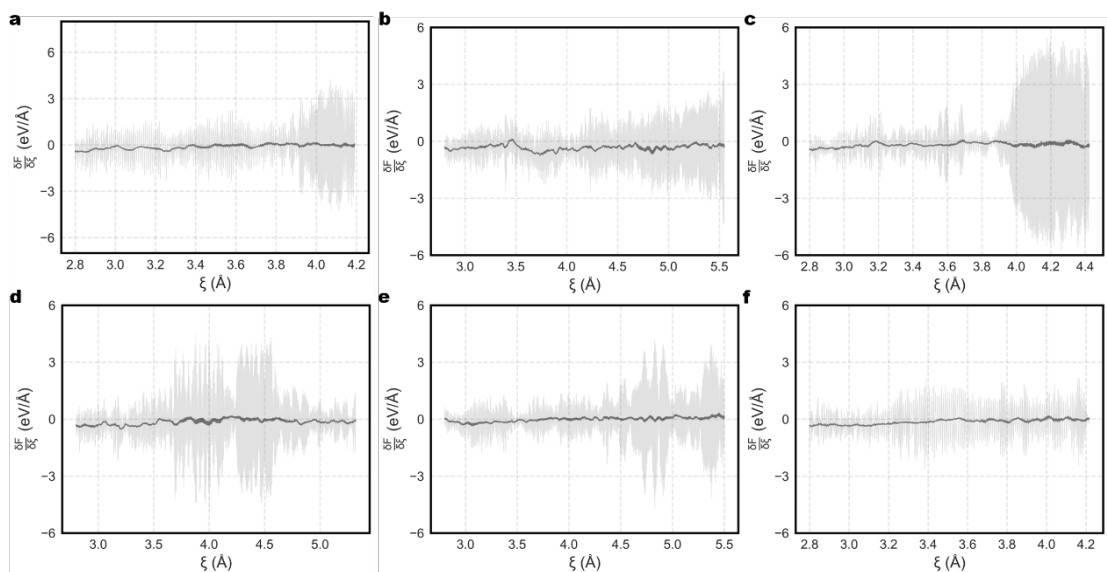

**Supplementary Figure 60. PMF for physical adsorption of $CO_2$ on a protruding surface site at the Au(110)-water interface with 2 $K^+$ ions.** The protruding surface site refers to the Au site on the Au-water interface that become raised compared to the original flat surface after surface reconstruction. The PMFs for physical adsorption of $CO_2$ on a protruding surface site at the Au(110)-water interfaces with 2 $K^+$ ions obtained using the slow-growth method under different conditions (**a**, constant charge, **b**, 0.2 V vs SHE, **c**, 0 V vs SHE, **d**, -0.2 V vs SHE, **e**, -0.4 V vs SHE, **f**, -0.6 V vs SHE). The light-colored line represents the results based on each frame, while the dark-colored line is the average obtained with a 200 fs moving window.

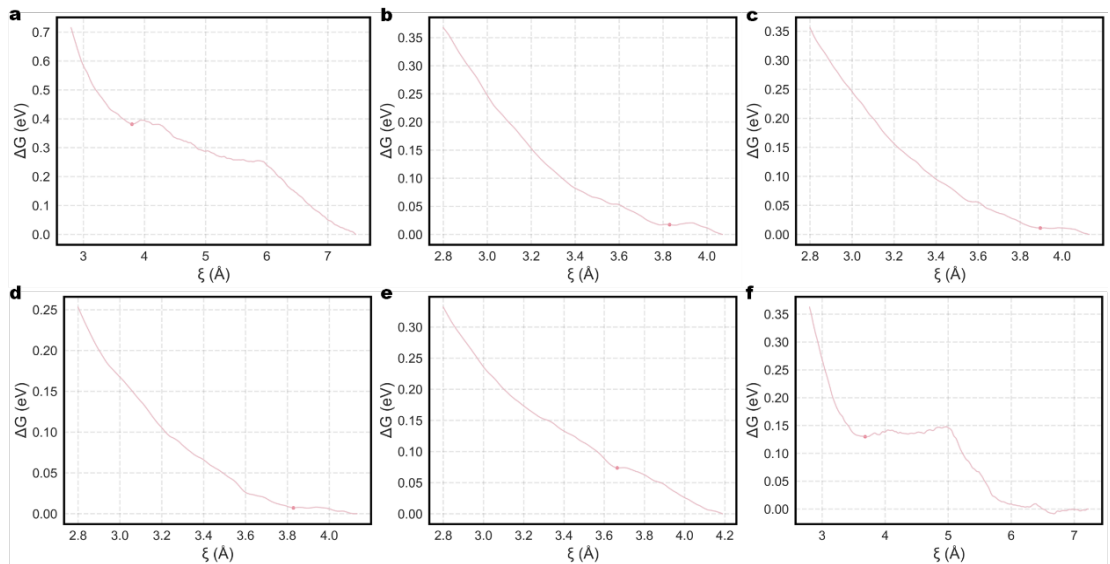

**Supplementary Figure 61. ΔG for physical adsorption of $CO_2$ on a flat surface site at the Au(110)-water interface.** The flat surface site refers to the Au site on the Au-water interface that remain relatively smooth after surface reconstruction, compared to the protruding surface. The ΔGs for physical adsorption of $CO_2$ on a flat surface site at the Au(110)-water interfaces obtained using the slow-growth method under different conditions (**a**, constant charge, **b**, 0.2 V vs SHE, **c**, 0 V vs SHE, **d**, -0.2 V vs SHE, **e**, -0.4 V vs SHE, **f**, -0.6 V vs SHE). The pink dots indicate the identified stable physical adsorption positions of $CO_2$.

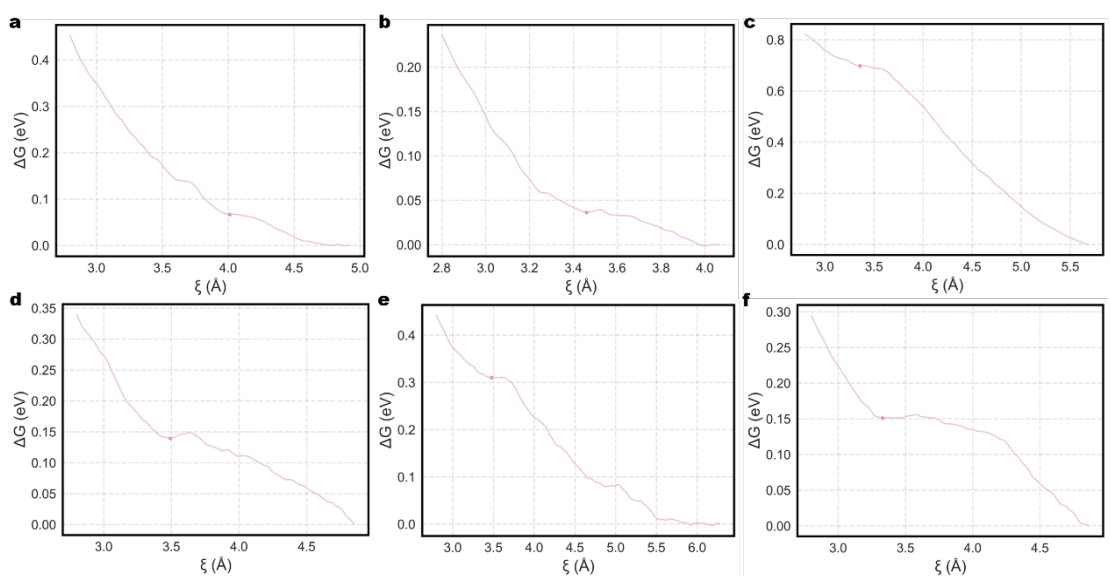

**Supplementary Figure 62. ΔG for physical adsorption of $CO_2$ on a protruding surface site at the Au(110)-water interface.** The protruding surface site refers to the Au site on the Au-water interface that become raised compared to the original flat surface after surface reconstruction. The ΔGs for physical adsorption of $CO_2$ on a protruding surface site at the Au(110)-water interfaces obtained using the slow-growth method under different conditions (**a**, constant charge, **b**, 0.2 V vs SHE, **c**, 0 V vs SHE, **d**, -0.2 V vs SHE, **e**, -0.4 V vs SHE, **f**, -0.6 V vs SHE). The pink dots indicate the identified stable physical adsorption positions of $CO_2$.

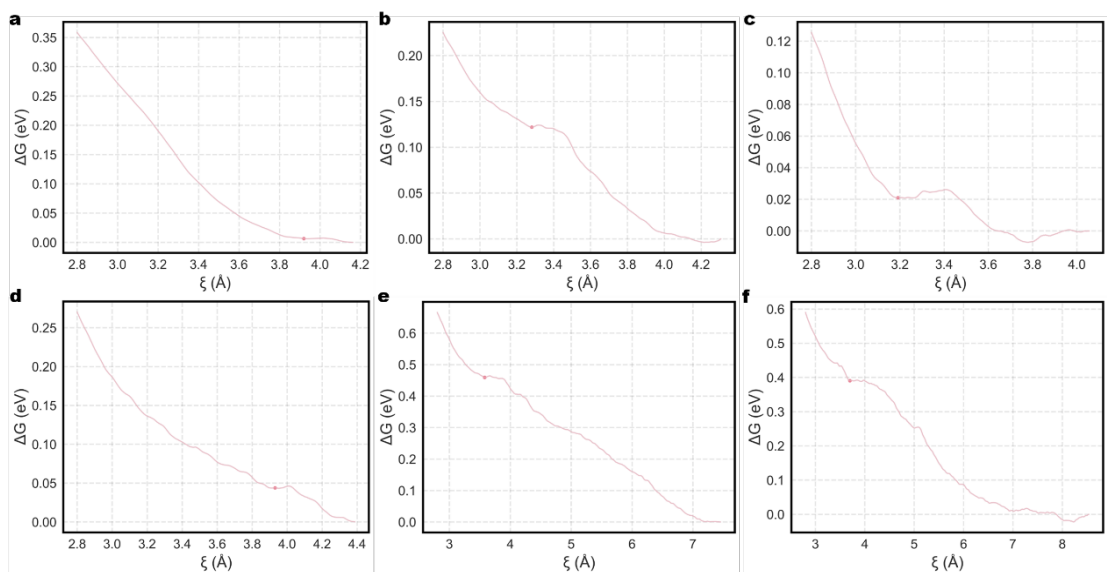

**Supplementary Figure 63. ΔG for physical adsorption of $CO_2$ on a flat surface site at the Au(110)-water interface with 2 $K^+$ ions.** The flat surface site refers to the Au site on the Au-water interface that remain relatively smooth after surface reconstruction, compared to the protruding surface. The ΔGs for physical adsorption of $CO_2$ on a flat surface site at the Au(110)-water interface with 2 $K^+$ ions obtained using the slow-growth method under different conditions (**a**, constant charge, **b**, 0.2 V vs SHE, **c**, 0 V vs SHE, **d**, -0.2 V vs SHE, **e**, -0.4 V vs SHE, **f**, -0.6 V vs SHE). The pink dots indicate the identified stable physical adsorption positions of $CO_2$.

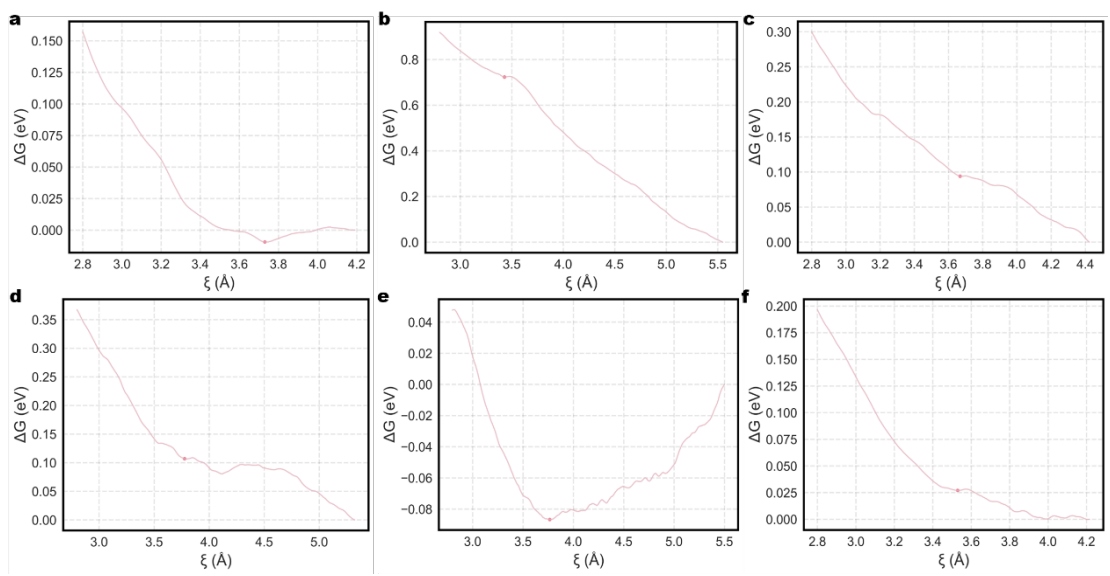

**Supplementary Figure 64. ΔG for physical adsorption of $CO_2$ on a protruding surface site at the Au(110)-water interface with 2 $K^+$ ions.** The protruding surface site refers to the Au site on the Au-water interface that become raised compared to the original flat surface after surface reconstruction. The ΔGs for physical adsorption of $CO_2$ on a protruding surface site at the Au(110)-water interfaces with 2 $K^+$ ions obtained using the slow-growth method under different conditions (**a**, constant charge, **b**, 0.2 V vs SHE, **c**, 0 V vs SHE, **d**, -0.2 V vs SHE, **e**, -0.4 V vs SHE, **f**, -0.6 V vs SHE). The pink dots indicate the identified stable physical adsorption positions of $CO_2$.

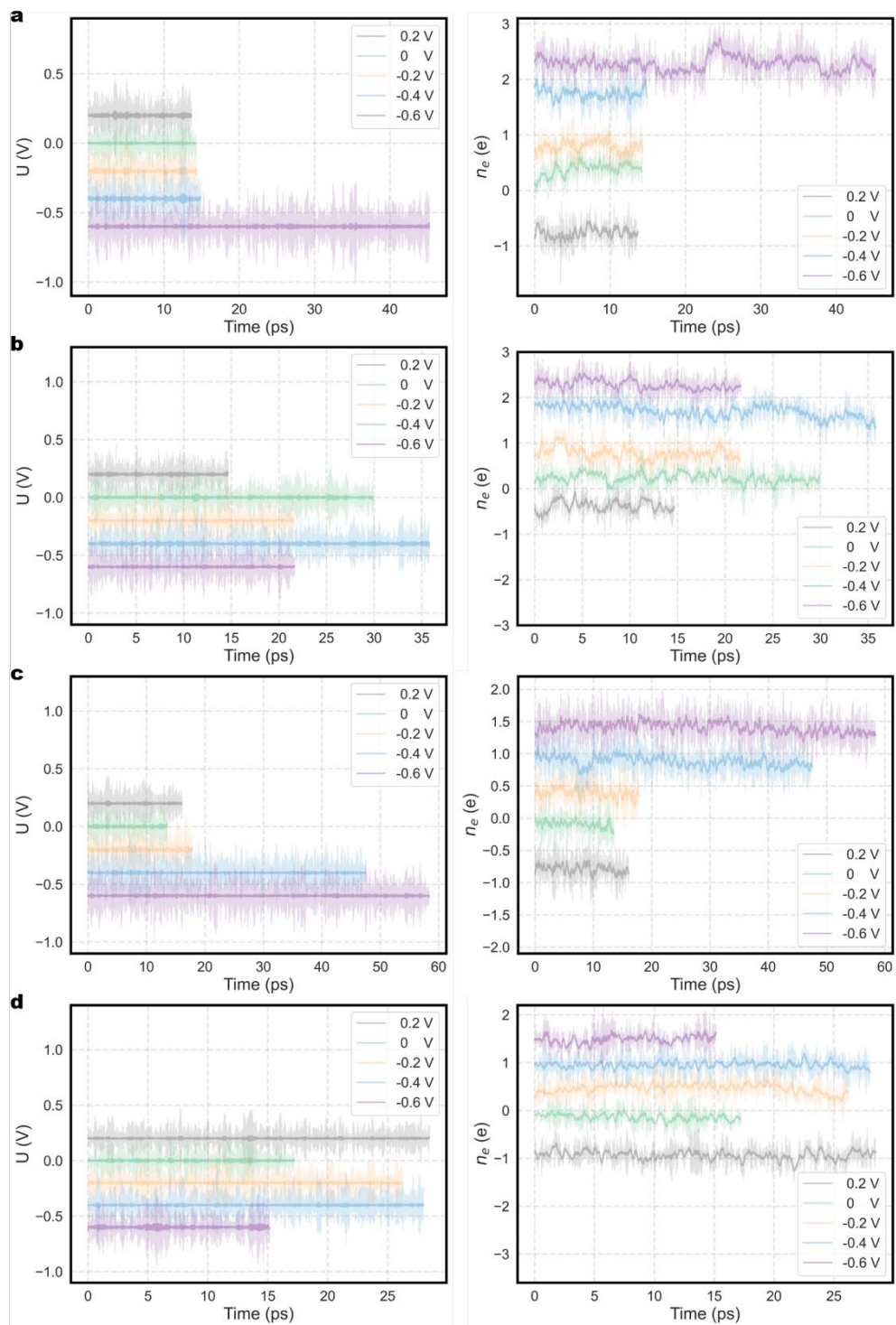

**Supplementary Figure 65. Applied potential and net electron count during the physical adsorption of $CO_2$.** Variations of applied potential (left) and net electron count (right) during the physical adsorption of $CO_2$ on **a**, a flat surface site and **b**, a protruding surface site at the Au(110)-water interface, as well as on **c**, a flat surface site and **d**, a protruding surface site at the Au(110)-water interface with 2 $K^+$ ions. The flat surface site refers to the Au site on the Au-water interface that remains relatively smooth after surface reconstruction, while the protruding surface site refers to the Au site that becomes raised compared to the original flat surface. The light-colored line represents the results based on each frame, while the dark-colored line is the average obtained with a 200 fs moving window.

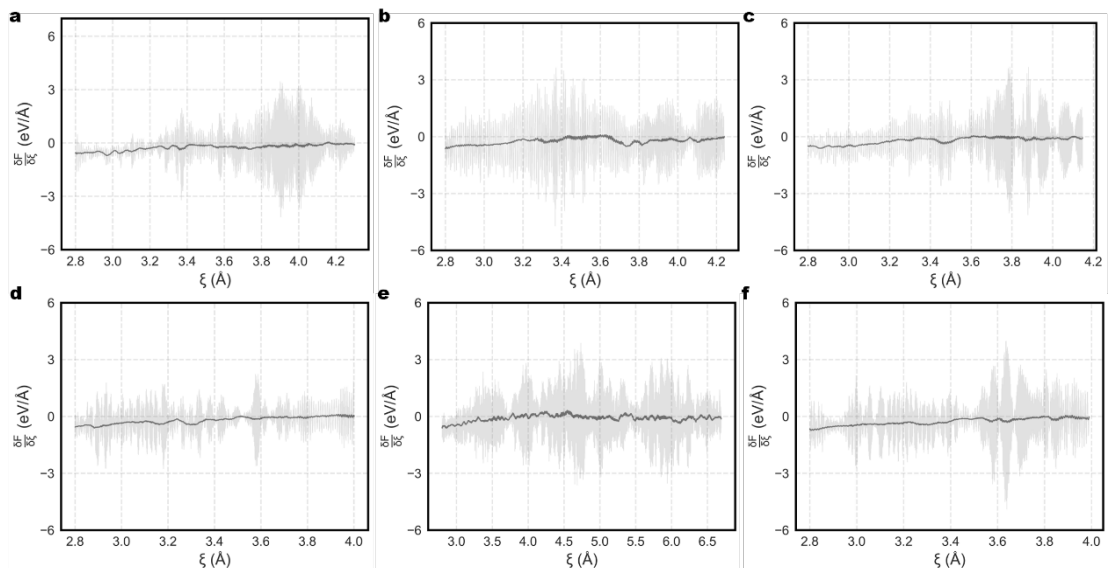

**Supplementary Figure 66. PMF for physical adsorption of $CO_2$ on the Au(111)-water interface.** The PMFs for physical adsorption of $CO_2$ on the Au(111)-water interfaces obtained using the slow-growth method under different conditions (**a**, constant charge, **b**, 0.2 V vs SHE, **c**, 0 V vs SHE, **d**, -0.2 V vs SHE, **e**, -0.4 V vs SHE, **f**, -0.6 V vs SHE). The light-colored line represents the results based on each frame, while the dark-colored line is the average obtained with a 200 fs moving window.

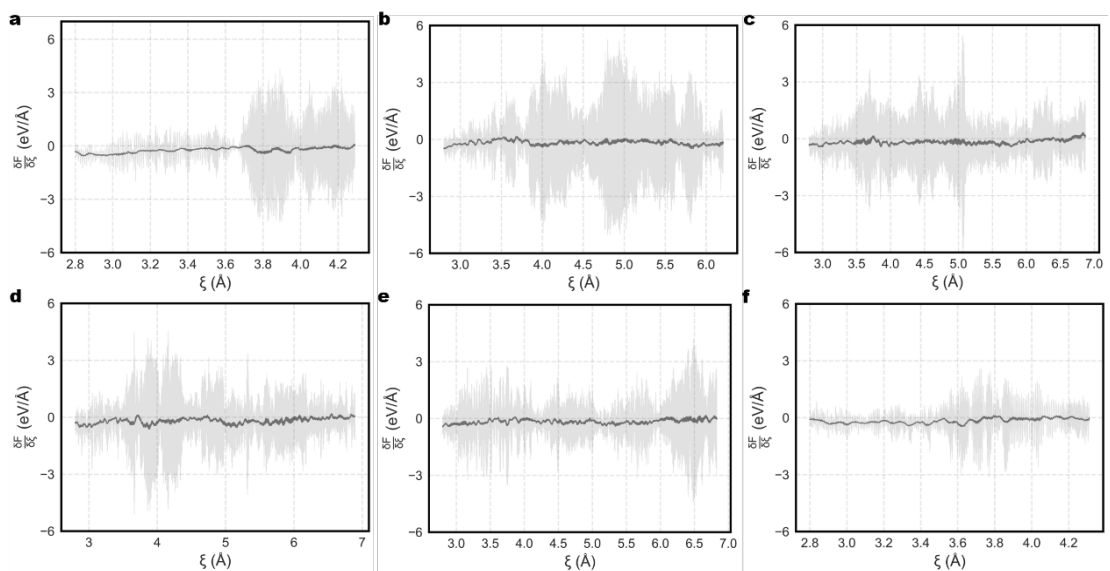

**Supplementary Figure 67. PMF for physical adsorption of $CO_2$ on the Au(111)-water interface with 2 $K^+$ ions.** The PMFs for physical adsorption of $CO_2$ on the Au(111)-water interfaces with 2 $K^+$ ions obtained using the slow-growth method under different conditions (**a**, constant charge, **b**, 0.2 V vs SHE, **c**, 0 V vs SHE, **d**, -0.2 V vs SHE, **e**, -0.4 V vs SHE, **f**, -0.6 V vs SHE). The light-colored line represents the results based on each frame, while the dark-colored line is the average obtained with a 200 fs moving window.

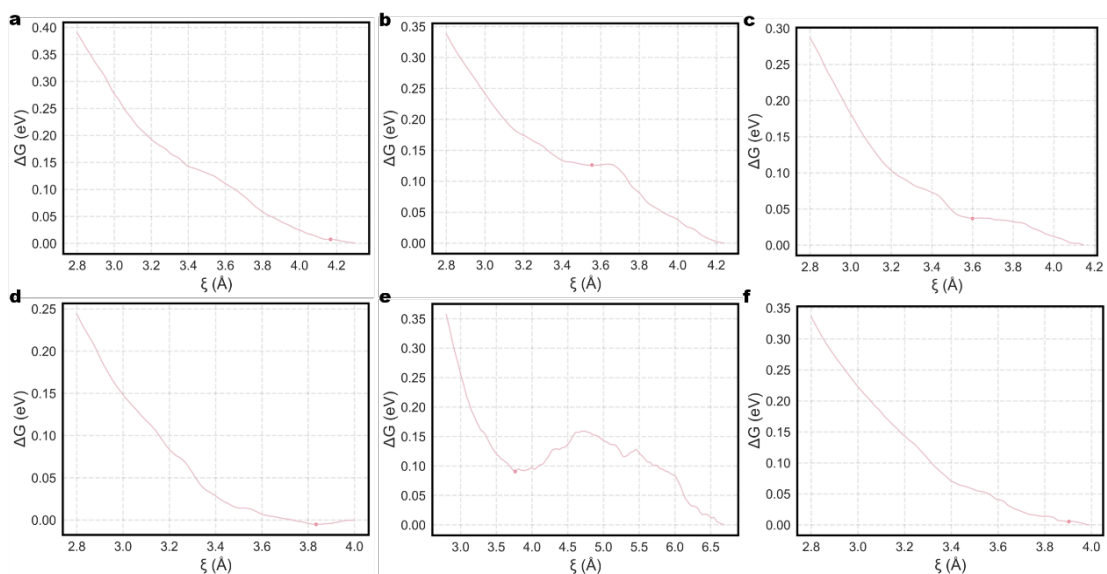

**Supplementary Figure 68. ΔG for physical adsorption of $CO_2$ on the Au(111)-water interface.** The ΔGs for physical adsorption of $CO_2$ on the Au(111)-water interfaces obtained using the slow-growth method under different conditions (**a**, constant charge, **b**, 0.2 V vs SHE, **c**, 0 V vs SHE, **d**, -0.2 V vs SHE, **e**, -0.4 V vs SHE, **f**, -0.6 V vs SHE). The pink dots indicate the identified stable physical adsorption positions of $CO_2$.

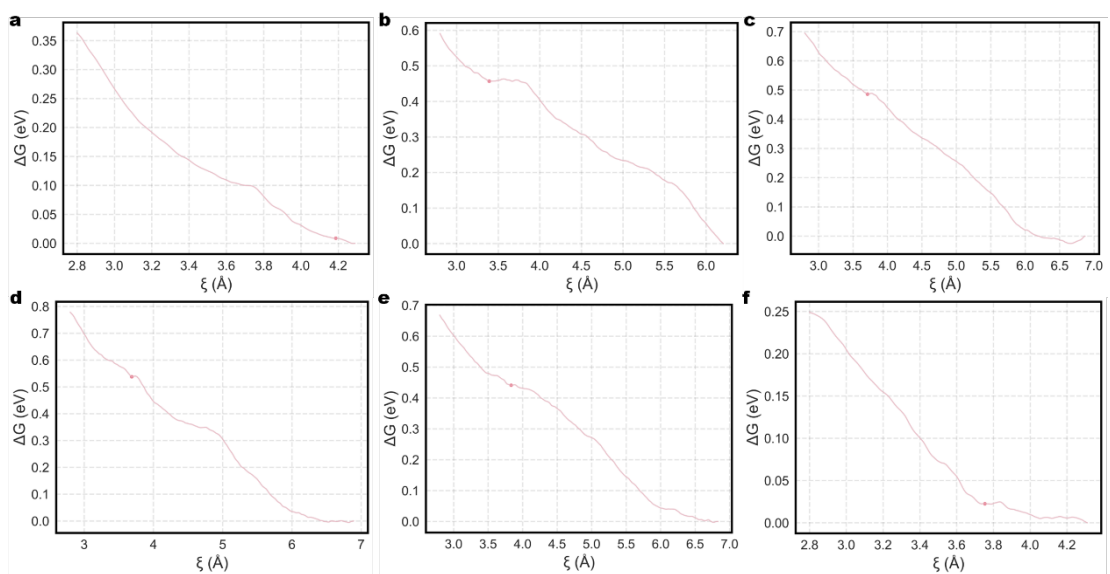

**Supplementary Figure 69. ΔG for physical adsorption of $CO_2$ on the Au(111)-water interface with 2 $K^+$ ions.** The ΔGs for physical adsorption of $CO_2$ on the Au(111)-water interfaces with 2 $K^+$ ions obtained using the slow-growth method under different conditions (**a**, constant charge, **b**, 0.2 V vs SHE, **c**, 0 V vs SHE, **d**, -0.2 V vs SHE, **e**, -0.4 V vs SHE, **f**, -0.6 V vs SHE). The pink dots indicate the identified stable physical adsorption positions of $CO_2$.

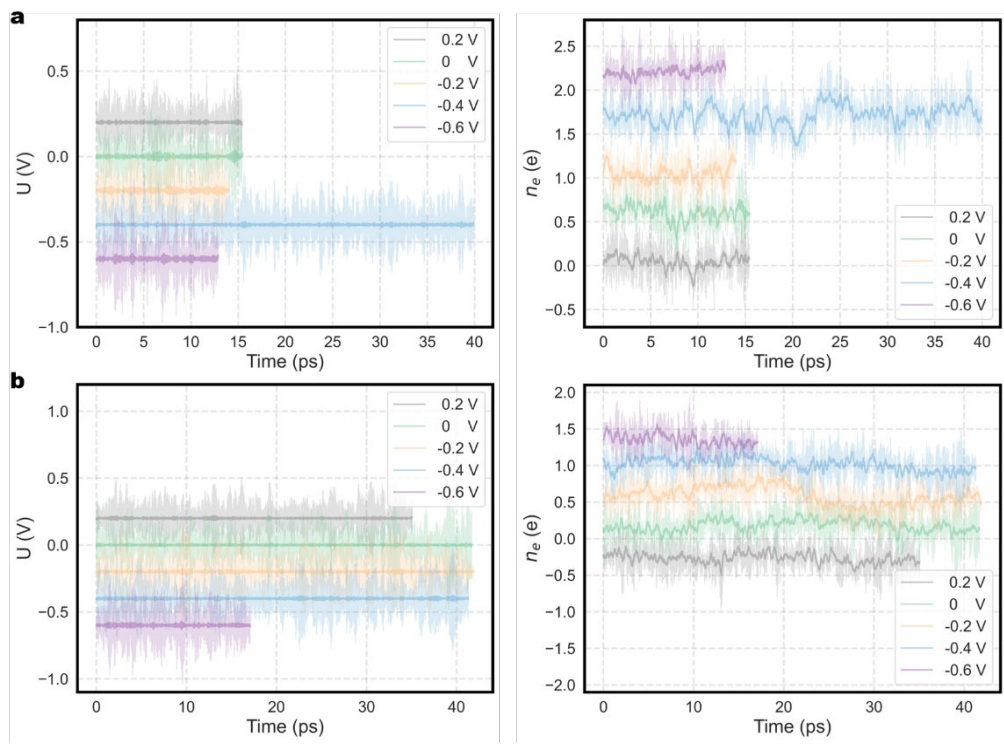

**Supplementary Figure 70. Applied potential and net electron count during the physical adsorption of $CO_2$.** Variations of applied potential (left) and net electron count (right) during the physical adsorption of $CO_2$ on **a**, the Au(111)-water interface and **b**, the Au(111)-water interface with 2 $K^+$ ions. The light-colored line represents the results based on each frame, while the dark-colored line is the average obtained with a 200 fs moving window.

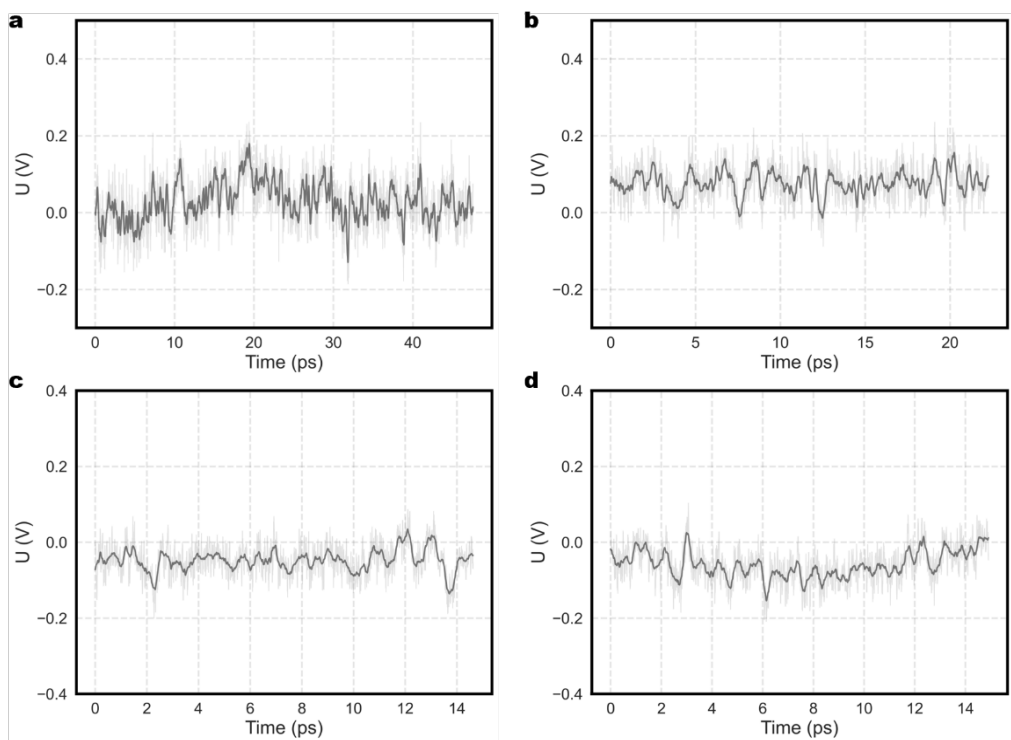

**Supplementary Figure 71. Applied potential during the physical adsorption of $CO_2$.** Variations of applied potential during the physical adsorption of $CO_2$ on **a**, a flat surface site and **b**, a protruding surface site at the Au(110)-water interface, as well as on **c**, a flat surface site and **d**, a protruding surface site at the Au(110)-water interface with 2 $K^+$ ions. The flat surface site refers to the Au site on the Au-water interface that remains relatively smooth after surface reconstruction, while the protruding surface site refers to the Au site that becomes raised compared to the original flat surface.

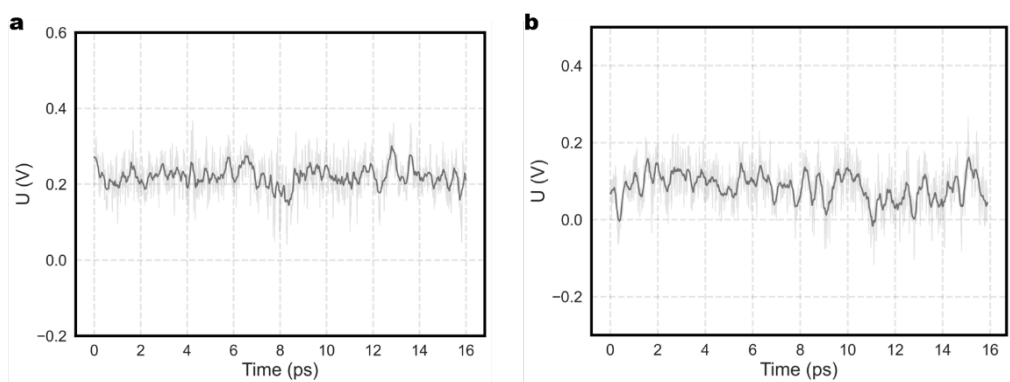

**Supplementary Figure 72. Applied potential during the physical adsorption of $CO_2$.** Variations of applied potential during the physical adsorption of $CO_2$ on **a**, the Au(110)-water interface and **b**, the Au(110)-water interface with 2 $K^+$ ions.

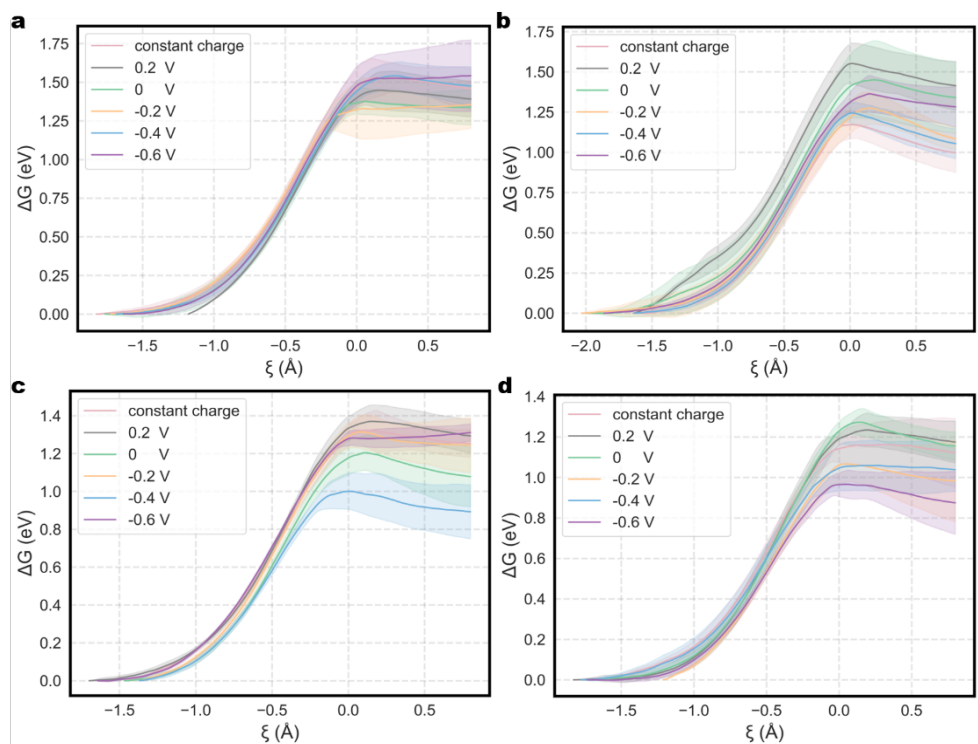

**Supplementary Figure 73. ΔG of the Volmer step at the Au(110)-water interface.** The ΔGs of the Volmer step for hydrogen atom adsorbed on **a**, a flat surface site and **b**, a protruding surface site at the Au(110)-water interface, as well as on **c**, a flat surface site and **d**, a protruding surface site at the Au(110)-water interface with 2 K$^+$ ions. The flat surface site refers to the Au site on the Au-water interface that remains relatively smooth after surface reconstruction, while the protruding surface site refers to the Au site that becomes raised compared to the original flat surface. The shaded area represents uncertainty.

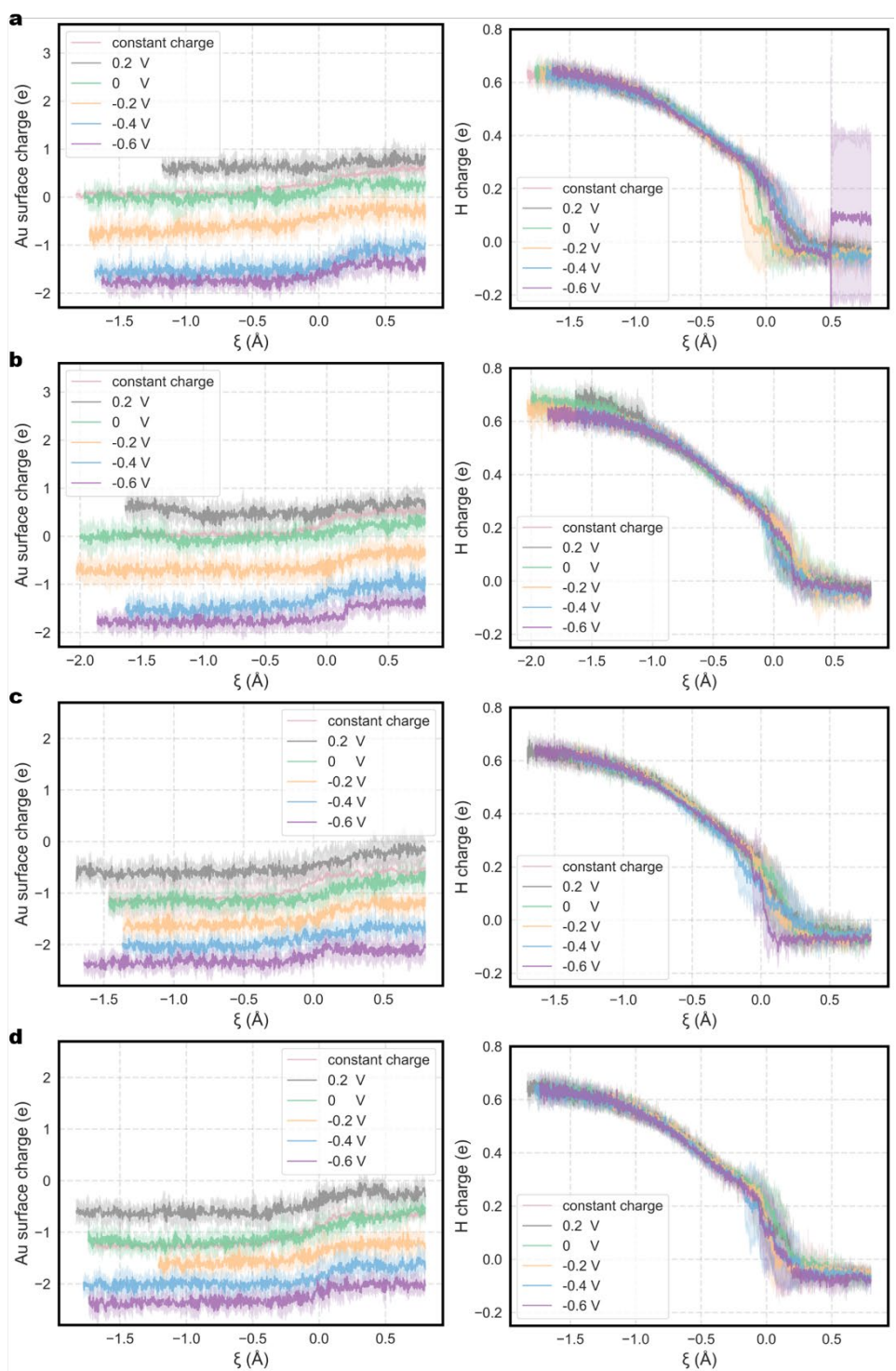

**Supplementary Figure 74. Change in the charge of the Au surface and adsorbed hydrogen atoms during the Volmer step as a function of the reaction coordinate.** Changes in the charge of the Au surface (left) and adsorbed hydrogen atom (right) for hydrogen atom adsorbed on **a**, a flat surface site and **b**, a protruding surface site at the Au(110)-water interface, as well as on **c**, a flat surface site and **d**, a protruding surface site at the Au(110)-water interface with 2 $K^+$ ions. The flat surface site refers to the Au site on the Au-water interface that remains relatively smooth after surface reconstruction, while the protruding surface site refers to the Au site that becomes raised compared to the original flat surface. The shaded area represents

uncertainty.

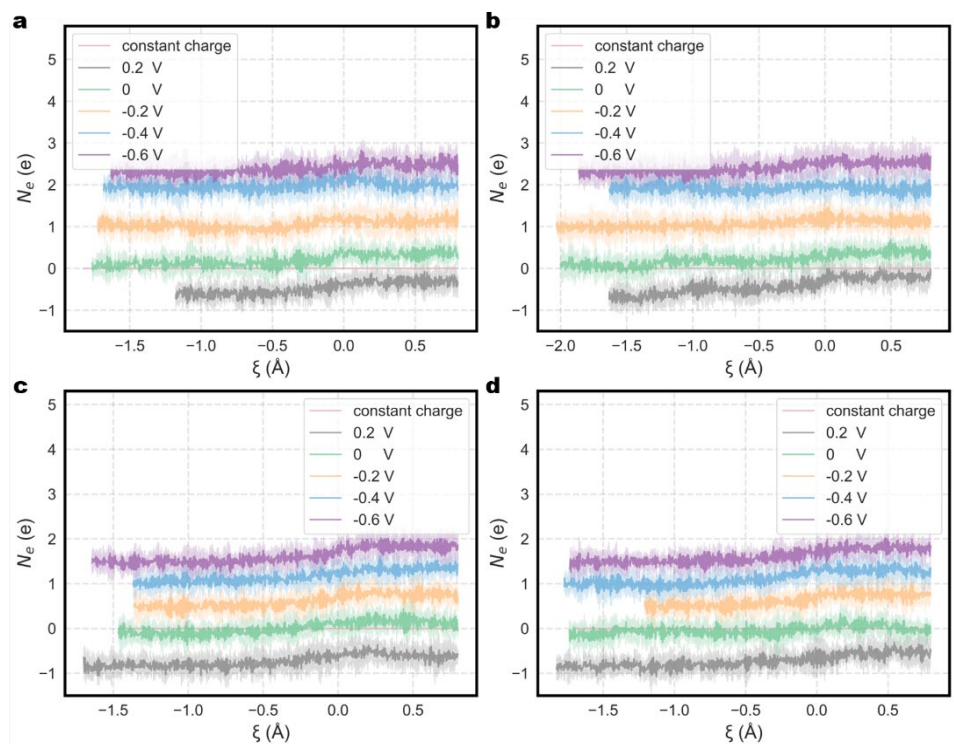

**Supplementary Figure 75. Change in the net electrons count of the Au(110)-water interface during the Volmer step as a function of the reaction coordinate.** The net electrons counts of the systems for hydrogen atom adsorbed on **a**, a flat surface site and **b**, a protruding surface site at the Au(110)-water interface, as well as on **c**, a flat surface site and **d**, a protruding surface site at the Au(110)-water interface with 2 $K^+$ ions. The flat surface site refers to the Au site on the Au-water interface that remains relatively smooth after surface reconstruction, while the protruding surface site refers to the Au site that becomes raised compared to the original flat surface. The shaded area represents uncertainty.

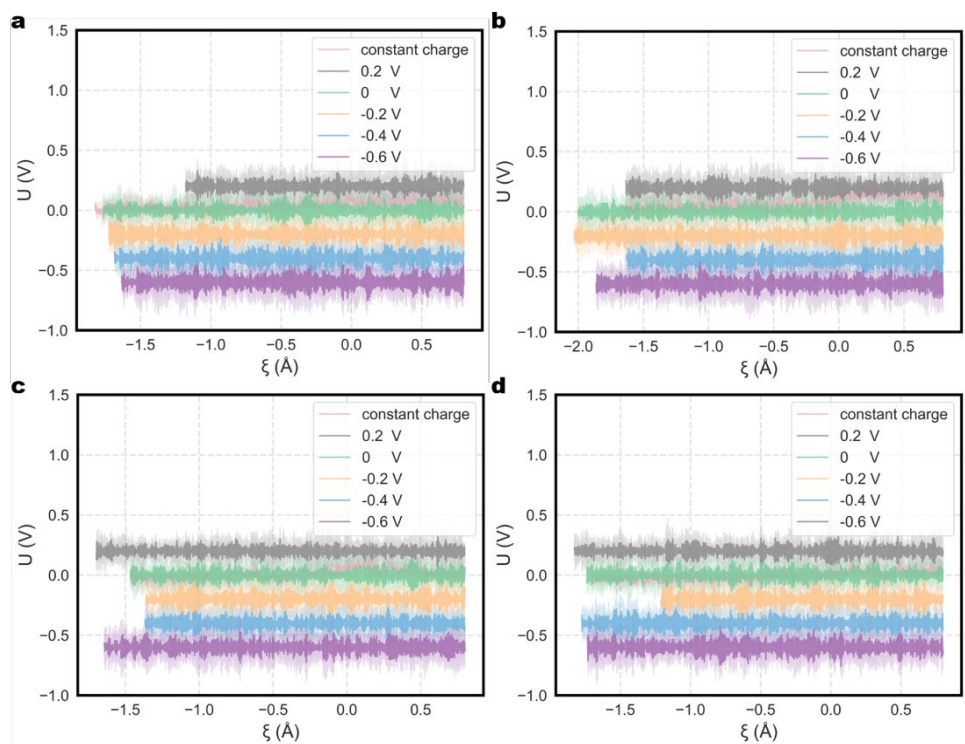

**Supplementary Figure 76. Change in the applied potential of the Au(110)-water interface during the Volmer step as a function of the reaction coordinate.** The applied potentials of the systems for hydrogen atom adsorbed on **a**, a flat surface site and **b**, a protruding surface site at the Au(110)-water interface, as well as on **c**, a flat surface site and **d**, a protruding surface site at the Au(110)-water interface with 2 $K^+$ ions. All data obtained using the slow-growth method under different applied potentials. The flat surface site refers to the Au site on the Au-water interface that remains relatively smooth after surface reconstruction, while the protruding surface site refers to the Au site that becomes raised compared to the original flat surface. The shaded area represents uncertainty.

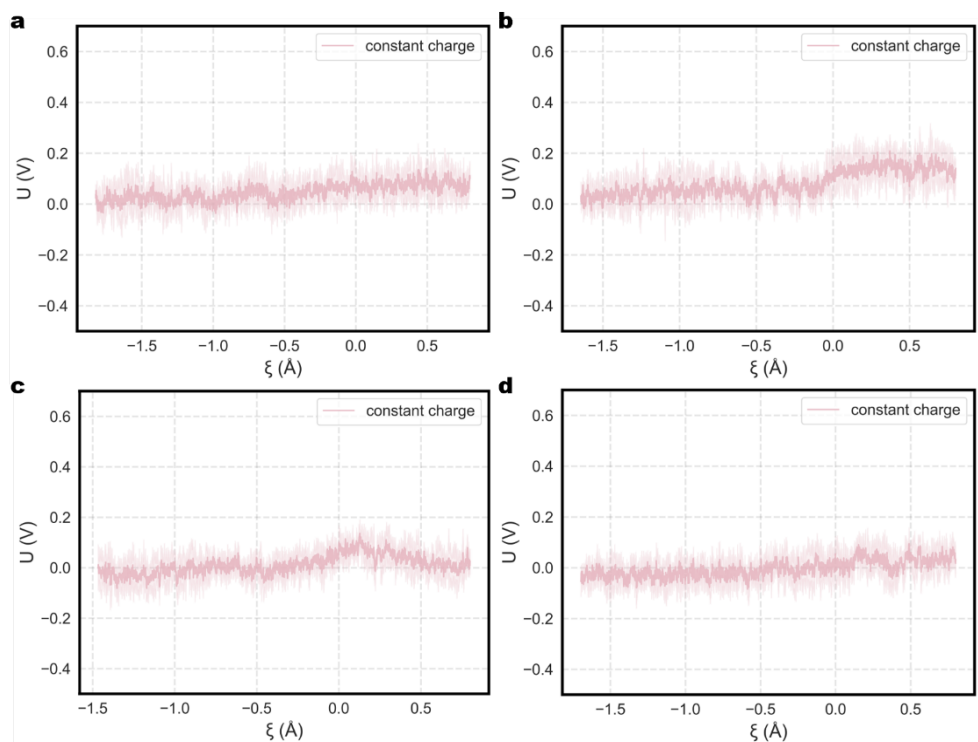

**Supplementary Figure 77. Change in the applied potential of the Au(110)-water interface during the Volmer step as a function of the reaction coordinate.** The applied potentials of the systems for hydrogen atom adsorbed on **a**, a flat surface site and **b**, a protruding surface site at the Au(110)-water interface, as well as on **c**, a flat surface site and **d**, a protruding surface site at the Au(110)-water interface with 2 $K^+$ ions. All data obtained using the slow-growth method under constant charge condition. The flat surface site refers to the Au site on the Au-water interface that remains relatively smooth after surface reconstruction, while the protruding surface site refers to the Au site that becomes raised compared to the original flat surface. The shaded area represents uncertainty.

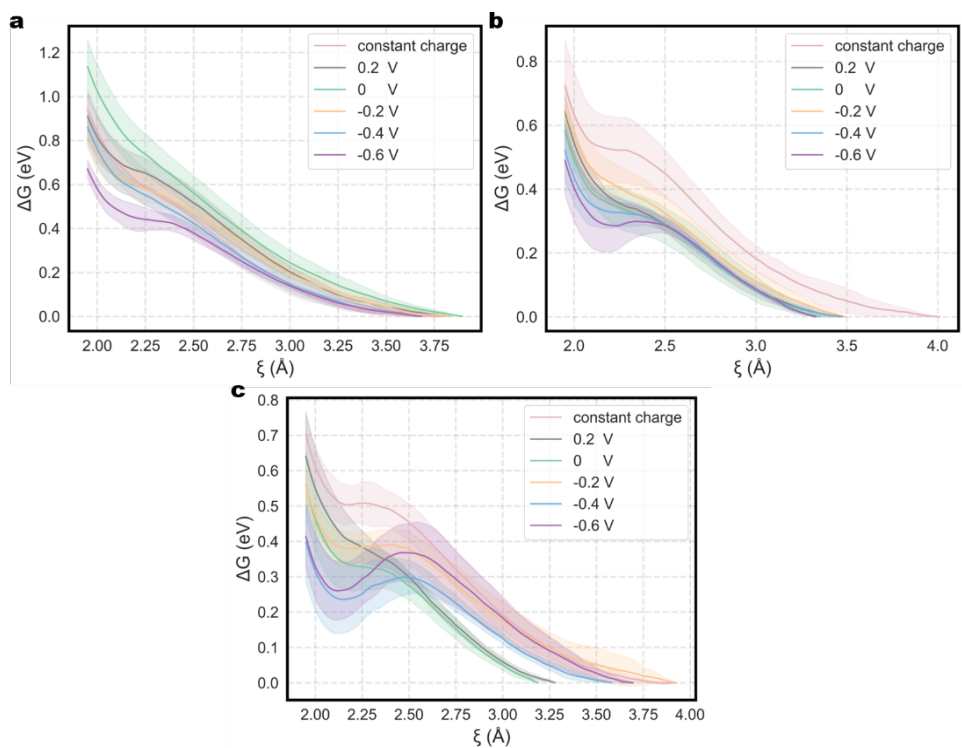

**Supplementary Figure 78. ΔG of chemical adsorption process for $CO_2$ adsorbed on the Au(110)-water interface.** ΔGs of chemical adsorption process for $CO_2$ adsorbed on **a**, a flat surface site and **b**, a protruding surface site at the Au(110)-water interface, as well as on **c**, a flat surface site at the Au(110)-water interface with 2 $K^+$ ions. The flat surface site refers to the Au site on the Au-water interface that remains relatively smooth after surface reconstruction, while the protruding surface site refers to the Au site that becomes raised compared to the original flat surface. The shaded area represents uncertainty.

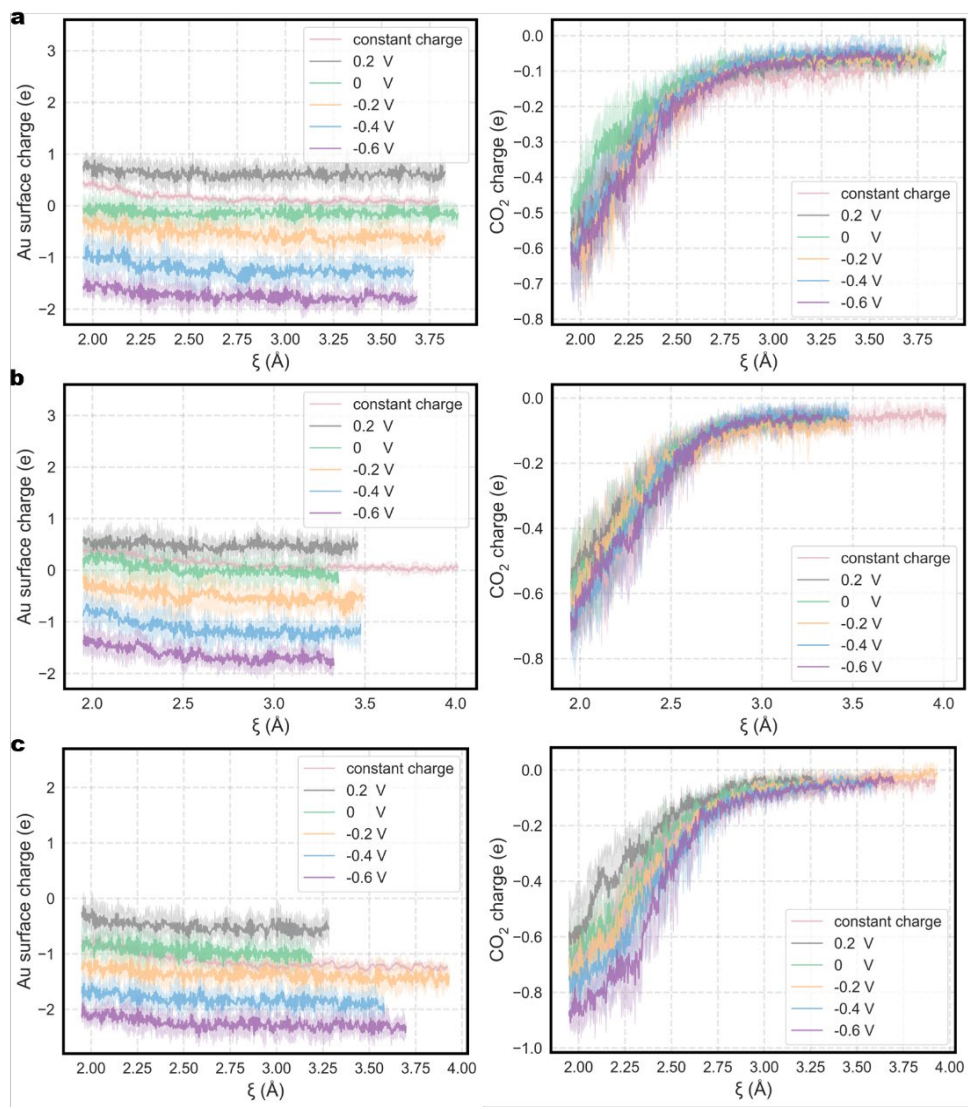

**Supplementary Figure 79. Changes in the charge of the Au surface and adsorbed CO₂ during the chemical adsorption process for as a function of the reaction coordinate.** Changes in the charge of the Au surface and adsorbed CO$_2$ for CO$_2$ adsorbed on **a**, a flat surface site and **b**, a protruding surface site at the Au(110)-water interface, as well as on **c**, a flat surface site at the Au(110)-water interface with 2 K$^+$ ions. The flat surface site refers to the Au site on the Au-water interface that remains relatively smooth after surface reconstruction, while the protruding surface site refers to the Au site that becomes raised compared to the original flat surface. The shaded area represents uncertainty.

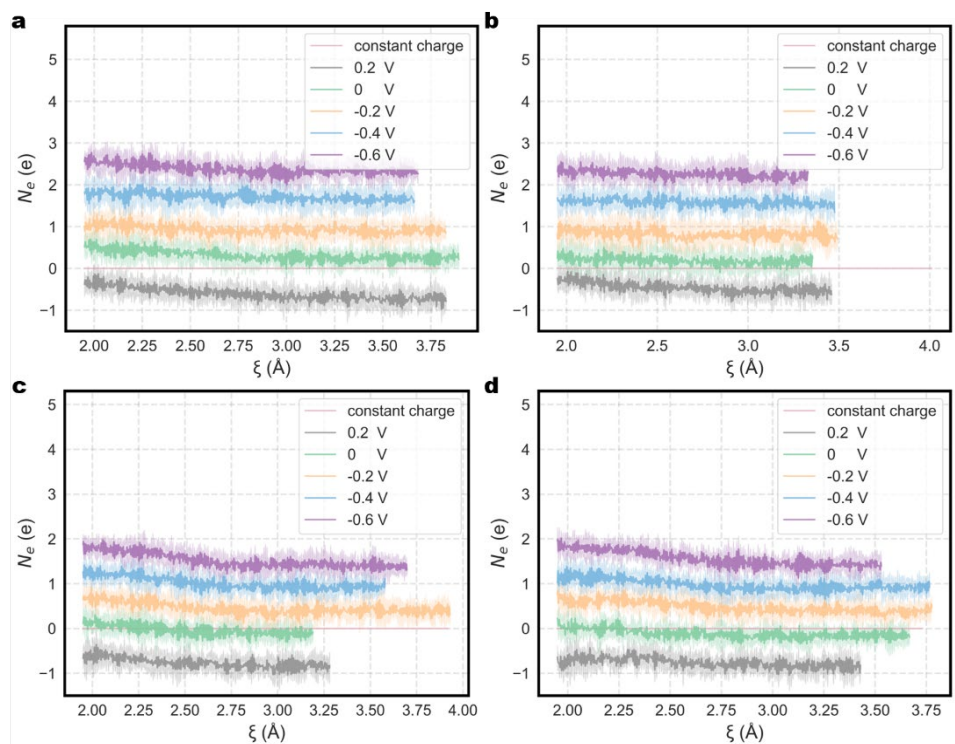

**Supplementary Figure 80. Change in the net electrons count of the Au(110)-water interface during the chemical adsorption process for as a function of the reaction coordinate.** The net electrons counts of the systems for $CO_2$ adsorbed on **a**, a flat surface site and **b**, a protruding surface site at the Au(110)-water interface, as well as on **c**, a flat surface site and **d**, a protruding surface site at the Au(110)-water interface with 2 $K^+$ ions. The flat surface site refers to the Au site on the Au-water interface that remains relatively smooth after surface reconstruction, while the protruding surface site refers to the Au site that becomes raised compared to the original flat surface. The shaded area represents uncertainty.

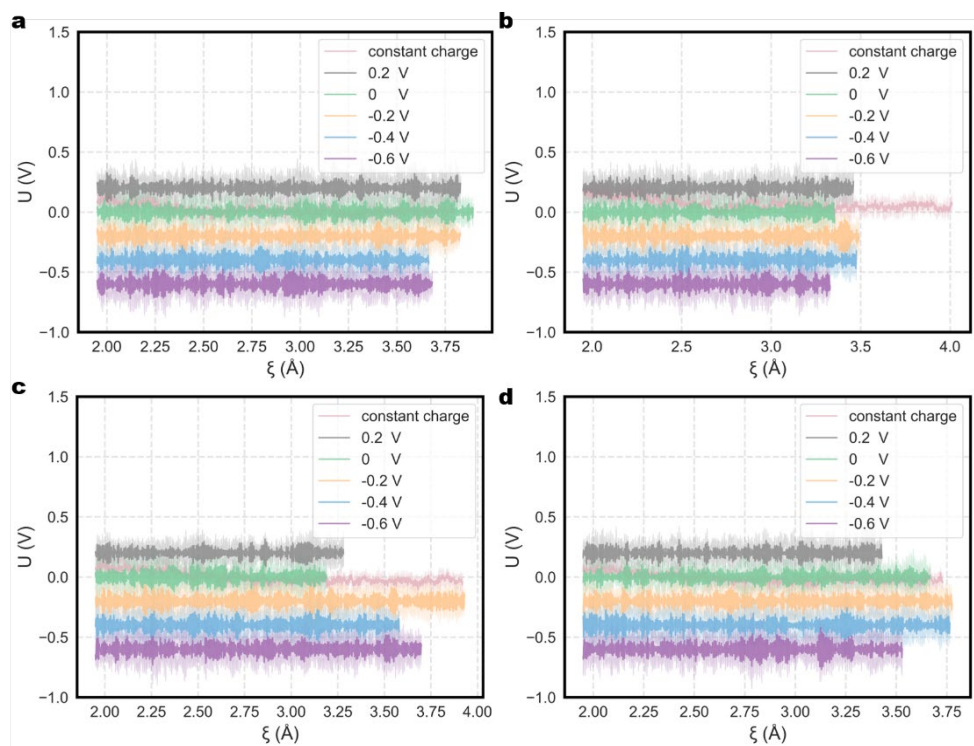

**Supplementary Figure 81. Change in the applied potential of the Au(110)-water interface during the chemical adsorption process for as a function of the reaction coordinate.** The applied potentials of the systems for $CO_2$ adsorbed on **a**, a flat surface site and **b**, a protruding surface site at the Au(110)-water interface, as well as on **c**, a flat surface site and **d**, a protruding surface site at the Au(110)-water interface with 2 $K^+$ ions. All data obtained using the slow-growth method under different applied potentials. The flat surface site refers to the Au site on the Au-water interface that remains relatively smooth after surface reconstruction, while the protruding surface site refers to the Au site that becomes raised compared to the original flat surface. The shaded area represents uncertainty.

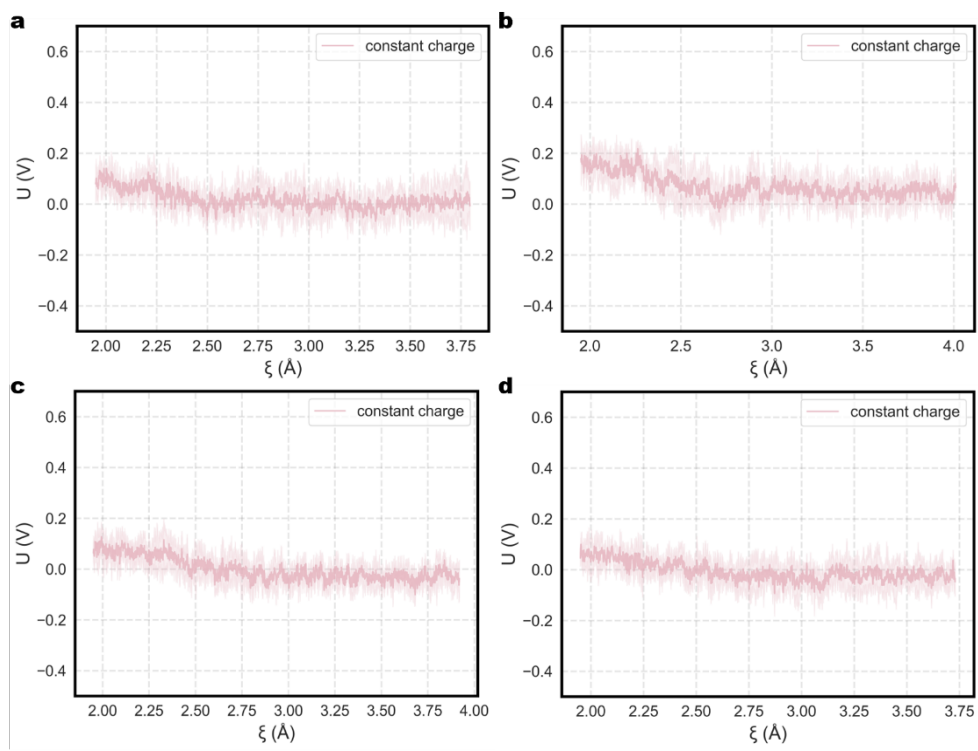

**Supplementary Figure 82. Change in the applied potential of the Au(110)-water interface during the chemical adsorption process for as a function of the reaction coordinate.** The applied potentials of the systems for $CO_2$ adsorbed on **a**, a flat surface site and **b**, a protruding surface site at the Au(110)-water interface, as well as on **c**, a flat surface site and **d**, a protruding surface site at the Au(110)-water interface with 2 $K^+$ ions. All data obtained using the slow-growth method under constant charge condition. The flat surface site refers to the Au site on the Au-water interface that remains relatively smooth after surface reconstruction, while the protruding surface site refers to the Au site that becomes raised compared to the original flat surface. The shaded area represents uncertainty.

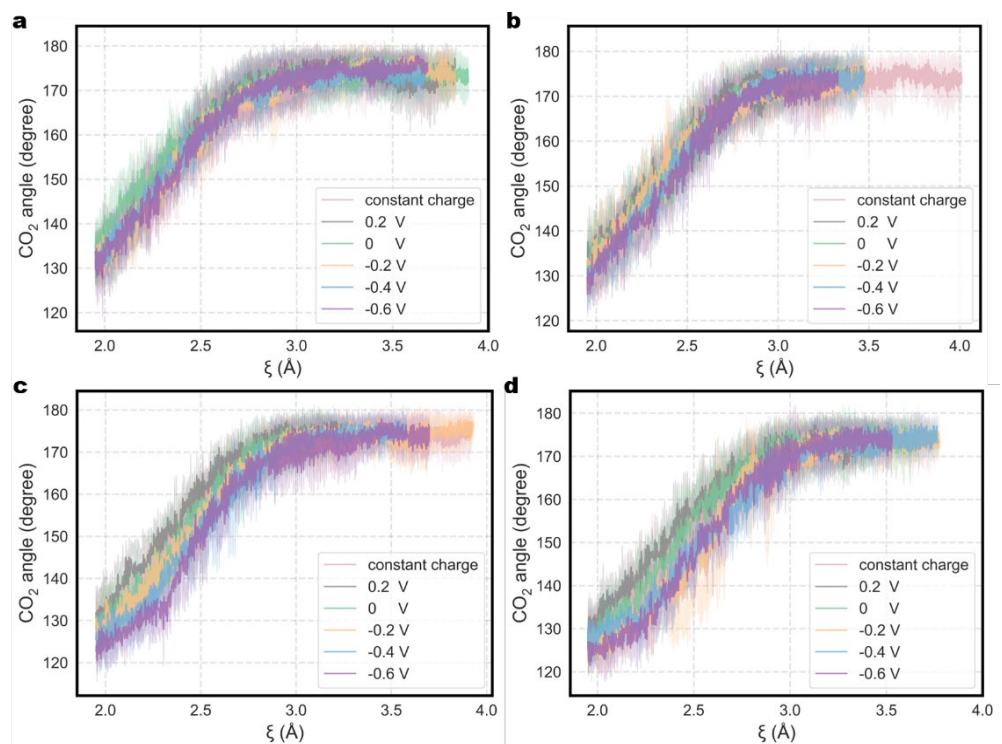

**Supplementary Figure 83. Change in CO$_2$ angle at the Au(110)-water interface during the chemical adsorption process for as a function of the reaction coordinate.** Changes in CO$_2$ angle for CO$_2$ adsorbed on **a**, a flat surface site and **b**, a protruding surface site at the Au(110)-water interface, as well as on **c**, a flat surface site and **d**, a protruding surface site at the Au(110)-water interface with 2 K$^+$ ions. The flat surface site refers to the Au site on the Au-water interface that remains relatively smooth after surface reconstruction, while the protruding surface site refers to the Au site that becomes raised compared to the original flat surface. The shaded area represents uncertainty.

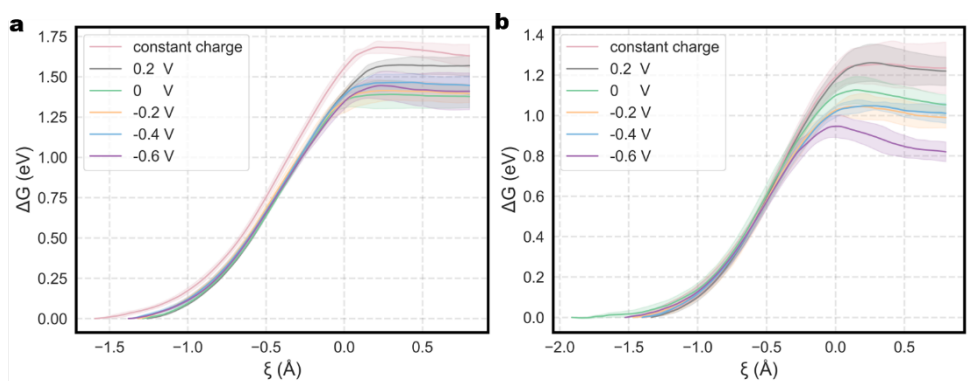

**Supplementary Figure 84. ΔG of the Volmer step at the Au(111)-water interface.** The ΔGs of the Volmer step for hydrogen atom adsorbed on **a**, the Au(111)-water interface and **b**, the Au(111)-water interface with 2 K$^+$ ions. The shaded area represents uncertainty.

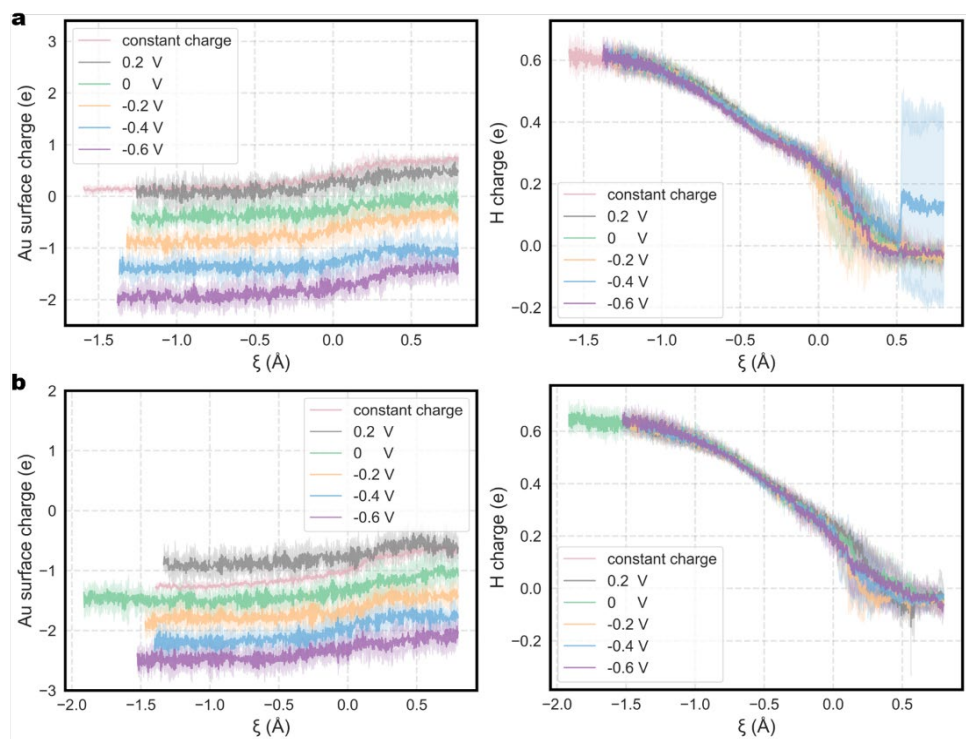

**Supplementary Figure 85. Change in the charge of the Au surface and adsorbed hydrogen atoms during the Volmer step as a function of the reaction coordinate.** Changes in the charge of the Au surface (left) and adsorbed hydrogen atom (right) for hydrogen atom adsorbed on **a**, the Au(111)-water interface, and **b**, the Au(111)-water interface with 2 $K^+$ ions. The shaded area represents uncertainty.

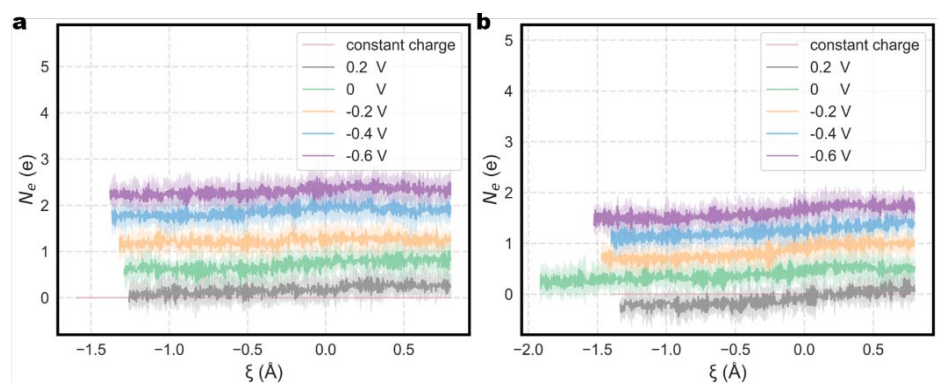

**Supplementary Figure 86. Change in the net electrons count of the Au(111)-water interface during the Volmer step as a function of the reaction coordinate.** The net electrons counts of the systems for hydrogen atom adsorbed on **a**, the Au(111)-water interface, and **b**, the Au(111)-water interface with 2 $K^+$ ions. The shaded area represents uncertainty.

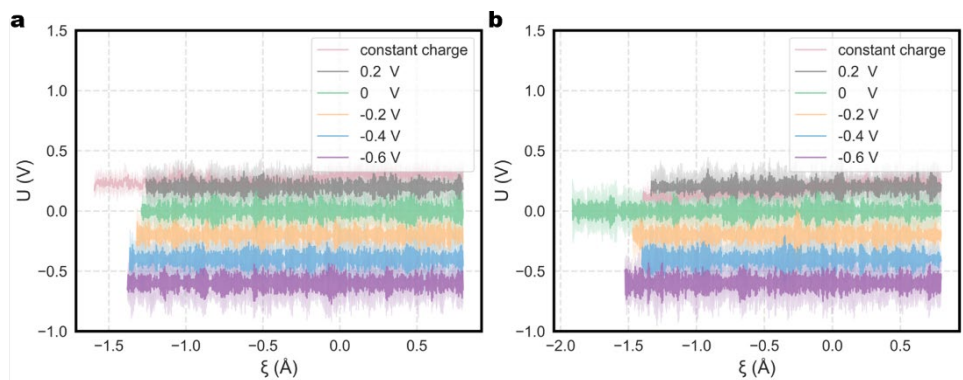

**Supplementary Figure 87. Change in the applied potential of the Au(111)-water interface during the Volmer step as a function of the reaction coordinate.** The applied potential of the systems for hydrogen atoms adsorbed on **a**, the Au(111)-water interface, and **b**, the Au(111)-water interface with 2 $K^+$ ions. All data obtained using the slow-growth method under different applied potentials. The shaded area represents uncertainty.

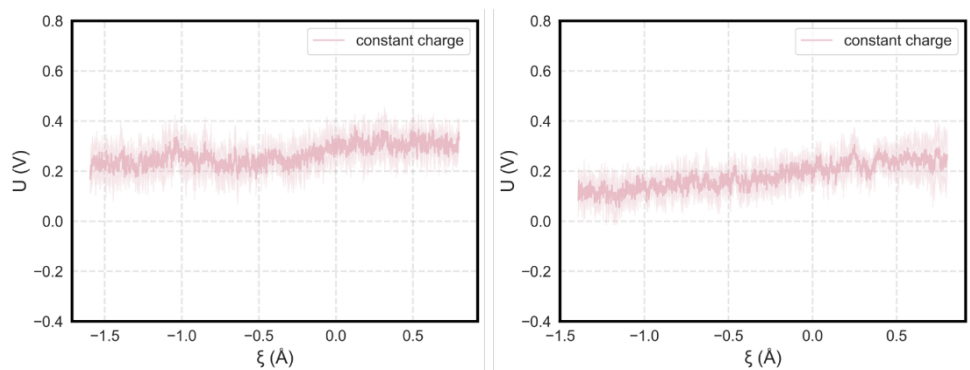

**Supplementary Figure 88. Change in the applied potential of the Au(111)-water interface during the Volmer step as a function of the reaction coordinate.** The applied potential of the systems for hydrogen atoms adsorbed on **a**, the Au(111)-water interface, and **b**, the Au(111)-water interface with 2 $K^+$ ions. All data obtained using the slow-growth method under constant charge condition. The shaded area represents uncertainty.

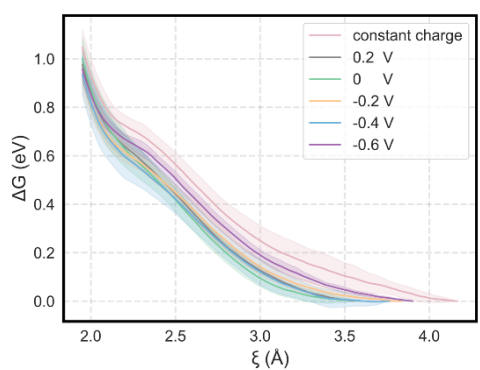

**Supplementary Figure 89. Chemical adsorption process for $CO_2$ adsorbed on the Au(111)-water interface.** The ΔG of chemical adsorption process for $CO_2$ adsorbed on the Au(111)-water interface. The shaded area represents uncertainty.

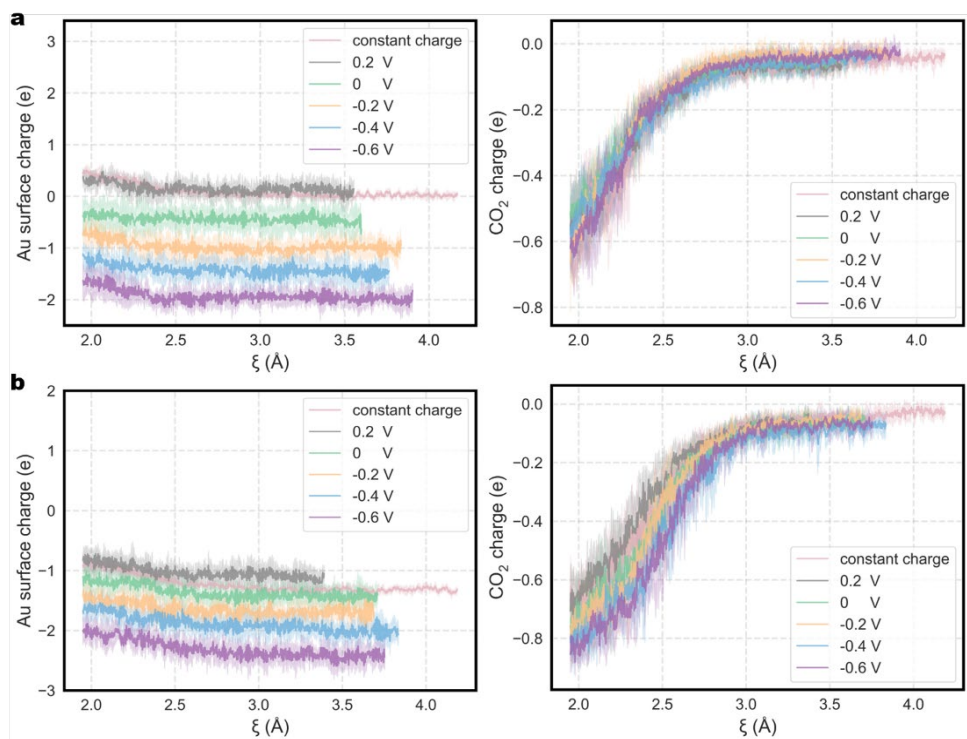

**Supplementary Figure 90. Changes in the charge of the Au surface and adsorbed $CO_2$ during the chemical adsorption process for as a function of the reaction coordinate.** The charge of the Au surface and adsorbed $CO_2$ for $CO_2$ adsorbed on **a**, the Au(111)-water interface and **b**, the Au(111)-water interface with 2 $K^+$ ions. The shaded area represents uncertainty.

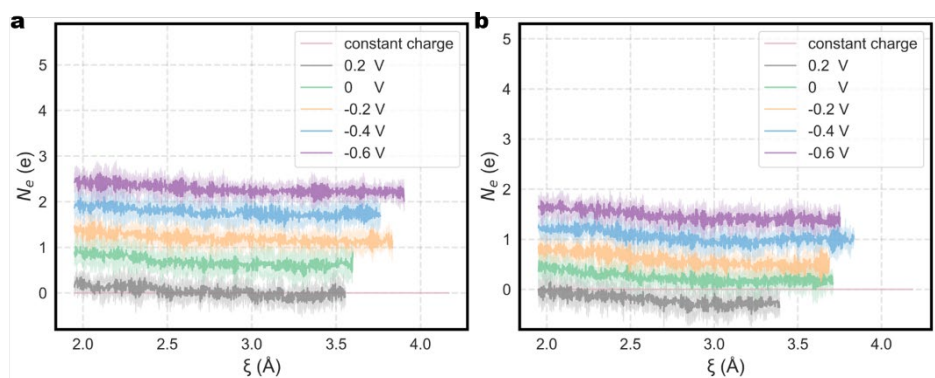

**Supplementary Figure 91. Change in the net electrons count of the Au(111)-water interface the chemical adsorption process for as a function of the reaction coordinate.** The net electrons counts of the systems for $CO_2$ adsorbed on **a**, the Au(111)-water interface and **b**, the Au(111)-water interface with 2 $K^+$ ions. The shaded area represents uncertainty.

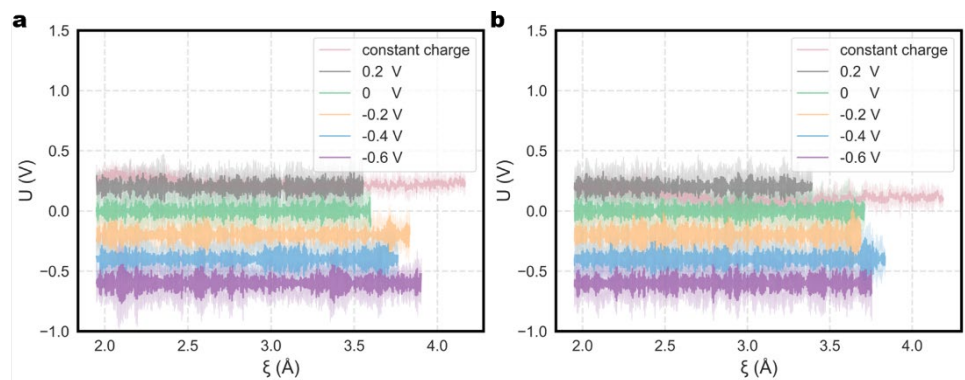

**Supplementary Figure 92. Change in the applied potential of the Au(111)-water interface the chemical adsorption process for as a function of the reaction coordinate.** The applied potential of the systems for $CO_2$ adsorbed on **a**, the Au(111)-water interface and **b**, the Au(111)-water interface with 2 $K^+$ ions. All data obtained using the slow-growth method under different applied potentials. The shaded area represents uncertainty.

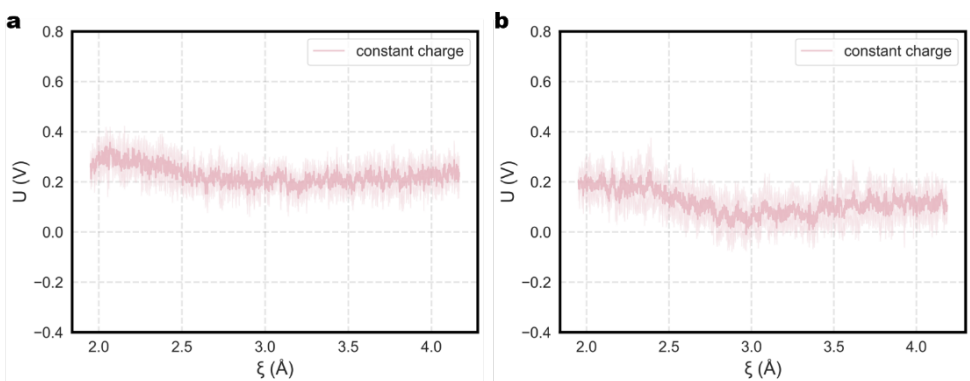

**Supplementary Figure 93. Change in the applied potential of the Au(111)-water interface the chemical adsorption process for as a function of the reaction coordinate.** The applied potential of the systems for $CO_2$ adsorbed on **a**, the Au(111)-water interface and **b**, the Au(111)-water interface with 2 $K^+$ ions. All data obtained using the slow-growth method under constant charge condition. The shaded area represents uncertainty.

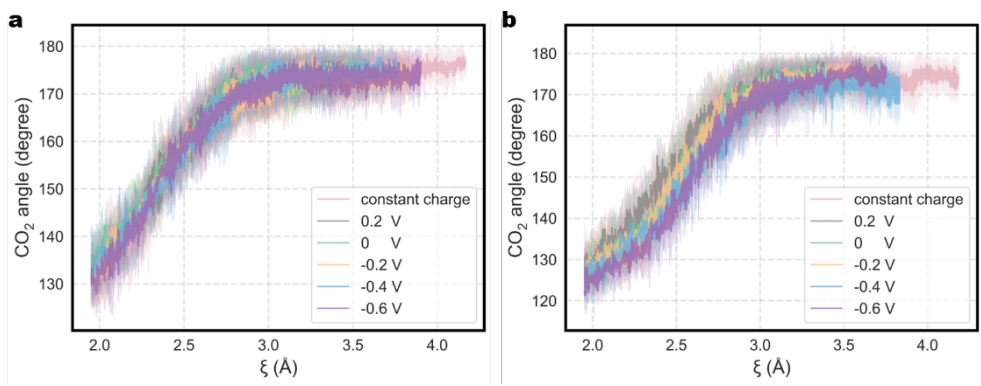

**Supplementary Figure 94. Change in CO$_2$ angle at the Au(111)-water interface during the chemical adsorption process for as a function of the reaction coordinate.** Changes in CO$_2$ angle for CO$_2$ adsorbed on **a**, the Au(111)-water interface and **b**, the Au(111)-water interface with 2 K$^+$ ions. The shaded area represents uncertainty.

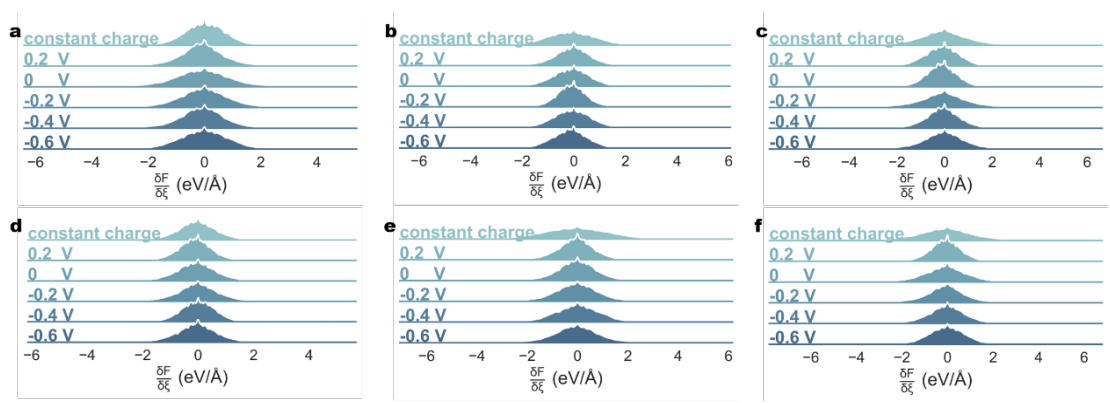

**Supplementary Figure 95. Comparison of probability distribution profile for the PMF at the initial position of the reaction coordinate.** Probability distribution profiles of the PMF for CO$_2$ adsorbed on **a**, a flat surface site and **b**, a protruding surface site at the Au(110)-water interface. Probability distribution profiles of the PMF for CO$_2$ adsorbed on **c**, a flat surface site and **d**, a protruding surface site at the Au(110)-water interface with 2 K+ ions. Probability distribution profiles of the PMF for CO$_2$ adsorbed on **e**, the Au(111)-water interface and **f**, the Au(111)-water interface with 2 K+ ions. All data are derived from the structural relaxation of molecular dynamics during the 20 ps before the reaction begins, using the Blue Moon method.

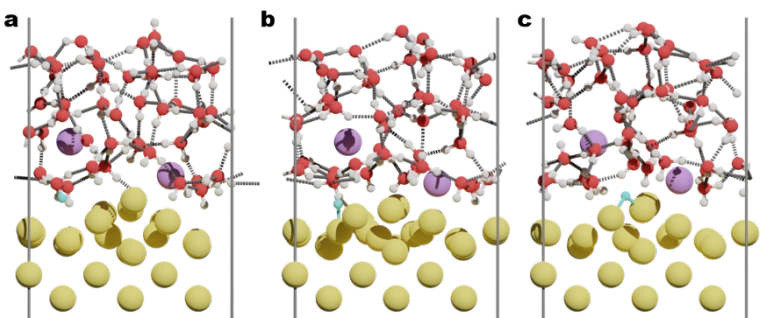

**Supplementary Figure 96. Key snapshot of the Volmer step occurring at the Au(110)-water interface with 2 K+ ions at -0.6 V vs SHE.** The images show the Volmer step where hydrogen atoms in water adsorb onto a flat site. **a**, Structure at the beginning of the Volmer step. **b**, Structure after hydrogen atom adsorption. **c**, Structure as the reaction coordinate progresses. The flat surface site refers to the Au site on the Au-water interface that remains relatively smooth after surface reconstruction. The yellow, white, and red spheres represent Au, H, and O atoms, respectively. The light blue and light green spheres represent the hydrogen atom set to react and the displaced hydrogen atom, respectively.

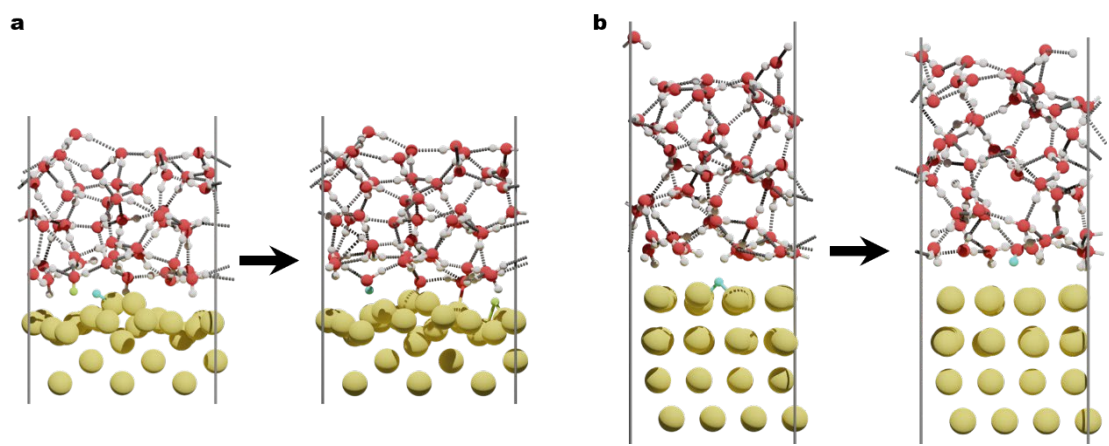

**Supplementary Figure 97. Incorrect process in the Volmer step. a**, The Volmer step occurs at the Au(110)-water interface at -0.6 V vs SHE. During the reaction, the hydrogen set to react desorbs and displaces the hydrogen from a neighboring water molecule. The displaced hydrogen atom then adsorbs onto a new Au site. **b**, The Volmer step occurs at the Au(111)-water interface at -0.4 V vs SHE. During the reaction, the hydrogen set to react desorbs and is attracted by the oxygen atom of a neighboring water molecule, forming a new bond. The yellow, white, and red spheres represent Au, H, and O atoms, respectively. The light blue and light green spheres represent the hydrogen atom set to react and the displaced hydrogen atom, respectively.

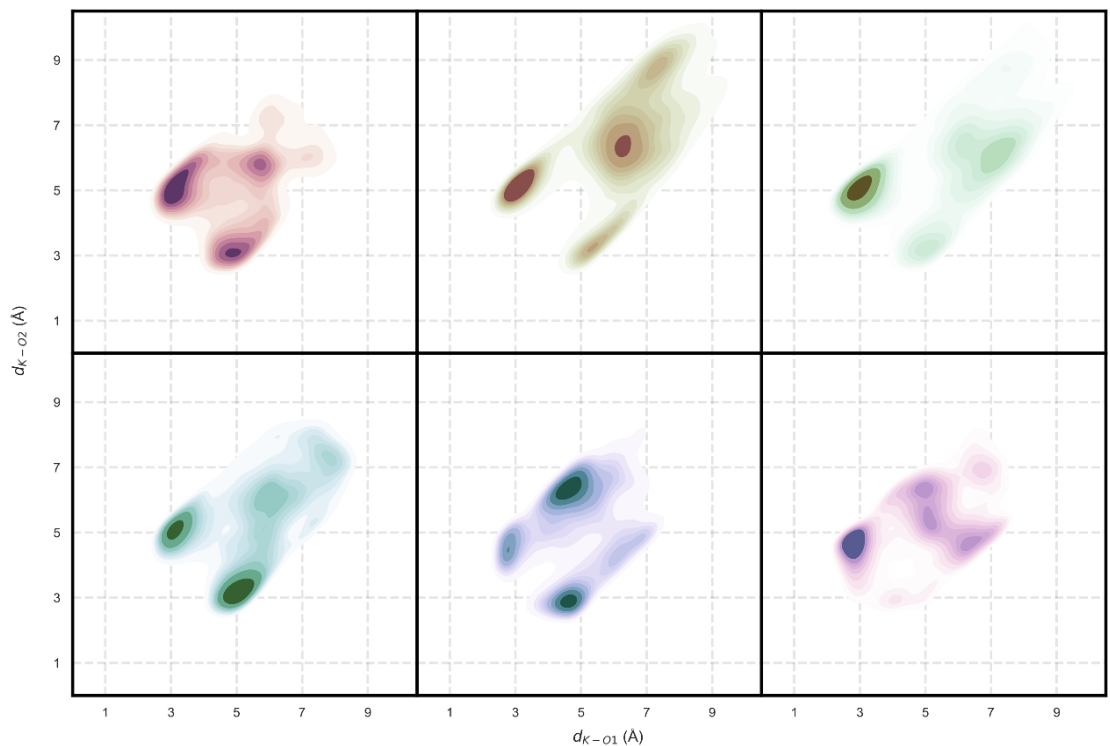

**Supplementary Figure 98. Probability distribution profile for distance distribution between K$^+$ ions and two oxygens atoms of CO$_2$ at the Au(110)-water.** Probability distribution profiles for distance distribution between K$^+$ ions and two oxygens atoms of CO$_2$ at the Au(110)-water. obtained using the slow-growth method under different conditions (**a**, constant charge, **b**, 0.2 V vs SHE, **c**, 0 V vs SHE, **d**, -0.2 V vs SHE, **e**, -0.4 V vs SHE, **f**, -0.6 V vs SHE).

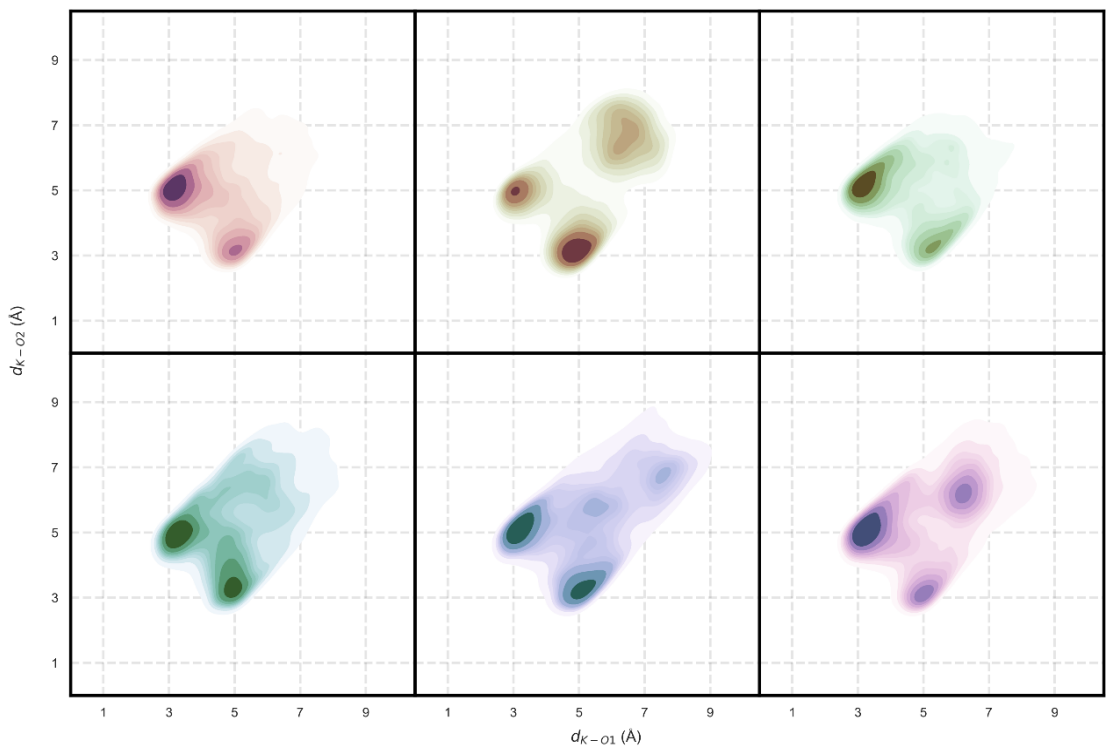

**Supplementary Figure 99. Probability distribution profile for distance distribution between $K^+$ ions and two oxygens atoms of $CO_2$ at the Au(111)-water.** Probability distribution profiles for distance distribution between $K^+$ ions and two oxygens atoms of $CO_2$ at the Au(111)-water obtained using the slow-growth method under different conditions (**a**, constant charge, **b**, 0.2 V vs SHE, **c**, 0 V vs SHE, **d**, -0.2 V vs SHE, **e**, -0.4 V vs SHE, **f**, -0.6 V vs SHE).

**Supplementary Table 1.** Maximum uncertainties and thresholds for energy, forces, corrected Fermi level, and atomic charge in the initial dataset analyses.

| Property | Maximum uncertainty | Threshold |
|---|---|---|
| **Energy** | 0.00172 | 0.0015 |
| **Force** | 0.165 | 0.15 |
| **Corrected $E_{fermi}$** | 0.0234 | 0.023 |
| **Atomic charge** | 0.108 | 0.1 |

**Supplementary Video 1.** Dynamics of the Au(110)-water interface with $2K^+$ ions and a $CO_2$ molecule at 300 K and an applied potential of −0.4 V vs. SHE.

**Supplementary Video 2.** Dynamics of the Au(110)-water interface with $2K^+$ ions and a $CO_2$ molecule at 300 K and an applied potential of −0.6 V vs. SHE.

**Supplementary Video 3.** Dynamics of the neutral (zero net charge) Au(110)-water interface at 300 K under canonical ensemble conditions.

**Supplementary Video 4.** Dynamics of the Au(110)-water interface at 300 K and an applied

potential of 0.2 V vs. SHE.

**Supplementary Video 5.** Dynamics of the Au(110)-water interface at 300 K and an applied potential of −0.2 V vs. SHE.

**Supplementary Video 6.** Dynamics of the Au(110)-water interface at 300 K and an applied potential of −0.6 V vs. SHE.

**Supplementary Video 7.** Dynamics of the neutral (zero net charge) Au(110)-water interface with $2K^+$ ions at 300 K under canonical ensemble conditions.

**Supplementary Video 8.** Dynamics of the Au(110)-water interface with $2K^+$ ions at 300 K and an applied potential of 0.2 V vs. SHE.

**Supplementary Video 9.** Dynamics of the Au(110)-water interface with $2K^+$ ions at 300 K and an applied potential of −0.2 V vs. SHE.

**Supplementary Video 10.** Dynamics of the Au(110)-water interface with $2K^+$ ions at 300 K and an applied potential of −0.6 V vs. SHE.

**Supplementary Video 11.** Dynamics of the neutral (zero net charge) Au(111)-water interface with $2K^+$ ions at 300 K under canonical ensemble conditions.

**Supplementary Video 12.** Dynamics of the Au(111)-water interface with $2K^+$ ions at 300 K and an applied potential of 0.2 V vs. SHE.

**Supplementary Video 13.** Dynamics of the Au(111)-water interface with $2K^+$ ions at 300 K and an applied potential of −0.2 V vs. SHE.

**Supplementary Video 14.** Dynamics of the Au(111)-water interface with $2K^+$ ions at 300 K and an applied potential of −0.6 V vs. SHE.

**Supplementary Video 15.** Dynamics of the Au(110)-water interface at 300 K and an applied potential of −0.6 V vs. SHE, with the Au(110) surface colored by Bader atomic charges.

**Supplementary Video 16.** Dynamics of the Au(110)-water interface with $2K^+$ ions at 300 K and an applied potential of −0.6 V vs. SHE, with the Au(110) surface colored by Bader atomic charges.

**Supplementary Video 17.** Dynamics of the Au(111)-water interface at 300 K and an applied potential of −0.6 V vs. SHE, with the Au(111) surface colored by Bader atomic charges.

**Supplementary Video 18.** Dynamics of the Au(111)-water interface with 2K$^+$ ions at 300 K and an applied potential of −0.6 V vs. SHE, with the Au(111) surface colored by Bader atomic charges.

**Reference**

1. Yang K., Liu J. & Yang B. Mechanism and Active Species in NH$_3$ Dehydrogenation under an Electrochemical Environment: An Ab Initio Molecular Dynamics Study. *ACS Catal.* **11**, 4310-4318 (2021).
2. Hu X. *et al.* Understanding the role of axial O in CO$_2$ electroreduction on NiN$_4$ single-atom catalysts via simulations in realistic electrochemical environment. *J. Mater. Chem. A* **9**, 23515-23521 (2021).
3. Zhao X. & Liu Y. Unveiling the Active Structure of Single Nickel Atom Catalysis: Critical Roles of Charge Capacity and Hydrogen Bonding. *J. Am. Chem. Soc.* **142**, 5773-5777 (2020).
4. Qin X., Vegge T. & Hansen H. A. Cation-Coordinated Inner-Sphere CO$_2$ Electroreduction at Au-Water Interfaces. *J. Am. Chem. Soc.* **145**, 1897-1905 (2023).